\documentclass{article}
\usepackage{graphicx}

\usepackage{euscript}
\usepackage{enumitem}
\usepackage{amsmath}   
\usepackage[compress]{cite}
\usepackage{amssymb}   
\usepackage{bm} 
\usepackage{dcolumn}
\usepackage{color}
\usepackage{mathrsfs}
\usepackage{amsfonts}
\usepackage{varioref}
\usepackage{textcomp}
\usepackage{float}
\usepackage{graphicx}
\usepackage{caption}
\usepackage{subcaption}
\RequirePackage[colorlinks,citecolor=blue,urlcolor=magenta,linkcolor=blue]{hyperref}
\allowdisplaybreaks

\labelformat{section}{Section #1} 
\labelformat{subsection}{Section #1} 
\labelformat{subsubsection}{Section #1}
\labelformat{subsubsubsection}{Section #1}
\labelformat{equation}{Eq.~(#1)} 
\labelformat{figure}{Fig.~#1} 
\labelformat{subfigure}{Fig.~\thefigure#1} 
\labelformat{table}{Tab.~#1} 
\labelformat{appendix}{Appendix #1}
\title{Lorentzian quantum cosmology with torsion}
\author{Vikramaditya Mondal\footnote{vikramaditya.academics@gmail.com}~ and Sumanta Chakraborty\footnote{tpsc@iacs.res.in}$~^{1}$
\\
$^{1}${\small{School of Physical Sciences}}\\
{\small{Indian Association for the Cultivation of Science, Kolkata-700032, India}}}
\begin{document}
  
\maketitle
\begin{abstract}
    We evaluate the Lorentzian gravitational path integral in the presence of non-vanishing torsion with the application of the Picard-Lefschetz theory for minisuperspaces corresponding to a number of phenomenological bouncing cosmological models as well as for the inflationary paradigm. It turns out that the semi-classical wave function derived from the saddle points of the path integral formalism coincides with the solutions of the Wheeler-DeWitt equation. Intriguingly, our analysis showed that the relative probability, derived using these semi-classical wave functions favors universes with smaller values of torsion. Moreover, we find that in the inflationary case, non-zero values of a certain parity-violating component of the torsion enhance the power in the large physical length scales, which can have important observational implications. On the other hand, in the case of bouncing models, the power spectrum is characterized by an initial region of growth, an intermediate oscillatory region, and then again a final region of growth. The shape of the power spectrum in the initial and intermediate regions is sensitive to the abundance of the bounce-enabling matter and torsion, along with the initial wave function of the universe, while the final size modifies the behavior of the power spectrum in the smaller length scales.
\end{abstract}

\maketitle
\section{Introduction}

The most successful theory in explaining the gravitational interaction at various scales, starting from the merger of binary black holes, to the accelerated expansion of the universe is Einstein's theory of general relativity. Formulated on (pseudo-)Riemannian manifolds, general relativity is free from torsion and is equipped with a unique metric-compatible, symmetric, and non-tensorial Levi-Civita connection \cite{Einstein:1916vd}. Shortly afterward, \'Elie Cartan explored the possibility of extending Riemannian geometry to include a more general situation in which one can have non-vanishing torsion (formally known as $U_{4}$ or Riemann-Cartan geometry), thereby generalizing the Einstein's theory of gravitation to the so-called Einstein-Cartan theory of gravitation \cite{Cartan1923,Cartan1925}, which is the simplest generalization of GR with the inclusion of torsion (for a review, see \cite{RevModPhys.48.393}). There have been numerous studies involving implications of torsion in the classical aspects of gravitational physics, for a small sample of such works, see \cite{Chakraborty:2018qew,Banerjee:2018yyi,Ivanov:2016xjm,Dey:2017fld,Blagojevic:2006jk,Barnich:2016rwk,Hehl:1974cn,Shapiro:2001rz,Shie:2008ms,Poplawski:2010kb}.  Recently, several studies \cite{PhysRevD.103.104008,PhysRevD.104.026002,Alexander_2021,Albertini:2022yny,Isichei:2022uzl,Gielen:2022yez,martini2023covariant} consider an underlying Riemann-Cartan geometry and inclusion of torsion in the framework of quantum cosmology --- a field of study which proposes that the early universe, as a whole, can be considered as a quantum mechanical system and is thus characterized by a wave function. The derivation of such a wave function, which is generally calculated assuming the framework of Riemannian geometry, was reworked in the above-mentioned studies to represent the wave function in terms of connection with (or without) the presence of torsion along with possible consequences. 

In the classical theory of gravity, there are two different formalisms --- (a) the second order formalism, wherein the metric tensor is considered fundamental and its variation in the gravitational action leads to the field equations, (b) the first order formalism, wherein the connection and the metric are both varied independently. However, if fermionic degrees of freedom are absent, a constraint relation, coming from the variation of the (gravity+matter) action with respect to the connection, implies that the torsion components must vanish. Subsequently, this implies that the first and second order formalisms are equivalent as far as the classical theory is concerned (for example, see \cite{Freedman:2012zz,Krasnov:2020lku,Feng:2021lfa,Chakraborty:2020yag}). 

However, in quantum theory, the above equivalence is not guaranteed. Including torsion can lead to various avenues based on how the quantum nature of the torsional degree of freedom is incorporated. We, here, consider the views of \cite{PhysRevD.103.104008}, which proposes that in the quantum theory, the constraint implying a vanishing momentum for torsion and the secondary constraint implying the vanishing of torsion itself cannot be imposed together, thanks to the uncertainty principle in the quantum domain. One way to proceed, then, is to quantize the theory ignoring the secondary constraint at first and computing a (kinematical) wave function of the universe in which torsion appears as a label, almost like the spatial curvature index in standard cosmology. Thereafter, to incorporate the quantum fluctuations of torsion one can form a wave packet (a ``beam") of universes with different torsion labels having a Gaussian (or, other suitable) distribution over the torsion space around the classical value of the torsion (which is zero). The last step can then be considered equivalent to imposing the secondary constraint. The authors show, an advantage of this way of quantizing the universe in the presence of torsion is that it immediately cures the problem of infinite norms encountered in quantum cosmology, and probabilities from the wave function can now be defined in a conventional manner as in particle quantum mechanics. For our purpose, we shall primarily focus on deriving the kinematical torsionful wave function of the universe, in which the quantum fluctuations of torsion have not yet been incorporated, using the Lorentzian gravitational path integral in the minisuperspace. What becomes apparent in our treatment, as we shall see in the following, is that not all values of the relevant torsion parameter are acceptable in the quantum theory due to a strict bound following from the stability analysis of perturbations around saddle point geometries. Therefore, one has to be careful in choosing a distribution in the torsion space. Although, in principle, beams of universes can still be constructed as suggested in \cite{PhysRevD.103.104008}. 

Given this interesting development in the quantum mechanical treatment of cosmologies based on Riemann-Cartan geometries in the presence of torsion, several pertinent questions arise naturally and are in need of thorough investigations. We motivate the present paper based on some of these questions and these are briefly described in the following.
\begin{itemize}

\item The main question we wish to address in the present work is --- can the techniques of Lorentzian Quantum Cosmology be consistently translated to derive a torsionful version of the Hartle-Haking wave function for inflationary and bouncing scenarios? Hartle and Hawking defined their no-boundary wave function as a Euclidean path integral over all regular geometries that start from zero size (``no-boundary") \cite{PhysRevD.28.2960}. Mathematically, this no-boundary condition has been a challenge to implement, due to various issues ranging from the conformal factor problem to the choice of a suitable contour in the path integral. A rigorous approach based on the Lorentzian path integral was recently developed in \cite{PhysRevD.95.103508}, which ultimately failed to provide a no-boundary wave function, because of the uncontrolled growth of perturbations around the no-boundary saddle point geometry \cite{PhysRevLett.119.171301,PhysRevD.97.023509,PhysRevD.100.063517}. Subsequently, it was suggested in \cite{PhysRevLett.122.201302,PhysRevD.100.123543,PhysRevD.106.023511} that it is possible to achieve controlled perturbations with alternative boundary conditions for which the trade-off is that only the saddle point geometry starts with initial zero size complying with the no-boundary proposal, whereas the other off-shell geometries that also contribute to the path integral, do start with all possible initial sizes. It seems, as of now, this is the best possible way to implement the no-boundary proposal consistently (for a recent review, see \cite{Lehners:2023yrj}). Moreover, in \cite{PhysRevD.103.106008}, we showed that the no-boundary proposal defined in the way of \cite{PhysRevD.100.123543} can also be extended to a large class of phenomenological bouncing scenarios. Thus it only seems natural to ask, how do all of these results change when we have some non-vanishing torsion components in the gravitational action? The bulk of this paper is concerned with demonstrating the subtle changes that are to be made along the way in the Lorentzian path integral approach to compute the kinematical torsionful version of the no-boundary wave function of the universe in the inflationary and bouncing scenarios. Moreover, we shall also discuss various boundary conditions, in both inflationary and bouncing scenarios, leading to saddle points, which are stable. We note that recently the techniques of Lorentzian Quantum Cosmology has also been generalized to the case of Gauss-Bonnet gravity \cite{Narain_2021,Narain_2022,ailiga2023lorentzian}, which we shall not consider in the present article.

We start with a general setup in \ref{sec:torsion_in_minisuperspace} describing the minisuperspace models that we shall be quantizing, first, by constructing and solving the Wheeler-DeWitt equation, as in the canonical approach in \ref{sec:Wheeler_DeWitt_quantization} and then by the path integral quantization in \ref{sec:PIquantization}.

\item The most important aspect that comes next is the stability of these saddle points in the path integral approach under perturbations. It is indeed possible to have saddle points with intriguing behavior, possibly mimicking the Hartle-Hawking no-boundary wave function. However, the stability of these saddle points under external perturbation is what makes certain boundary conditions and saddle points to stand out. The stability analysis also has implications in two distinct aspects --- (a) the power spectrum and (b) the possibility for cross-talk between differently curved FRW universes \cite{PhysRevD.103.104008}. The importance of determining the power spectrum cannot be over-emphasized, since it relates the initial quantum phase of the universe to the observable regime. In particular, we wish to explore how the presence of torsion affects the scale invariance of the inflationary power spectrum and also the corresponding changes in the context of bounce. In general, bouncing models do not predict scale-invariant power spectrum (see, however \cite{Brandenberger:2016vhg,Raveendran:2018why,Raveendran:2017vfx,PhysRevD.100.083523}), but the conclusion may change in the presence of torsion. Regarding the cross-talk between universes with positive, negative, and zero spatial curvature, the stability argument can play an important role. Since the saddle points will depend on the effective curvature of the three-space, which has contributions from the spatial curvature index appearing in the metric and torsion, but with an opposite sign. The stability of these saddle points also tells us a lot about possible choices of the three-space curvature. In particular, we will show that stability demands that such cross-talks between different three-space curvatures should not exist. These issues have been more elaborated in \ref{sec:quantized_perturbations}. 

\item The final question we wish to address in this work has been harped upon several times, and from various different perspectives, namely why is the spacetime torsion of the observable universe small? The explanations range from extra dimensions to modified theories of gravity \cite{Will:1993ns,Hehl:1974cn,Bohmer:2017dqs,PhysRevLett.89.121101,PhysRevD.90.107901}. In this work, we wish to provide another explanation for the same, but from the initial quantum phase of the universe. The hope is, the relative probability of two universes with and without torsion, as derived from the wave functions using saddle points of the Lorentzian path integral approach, will be small. In other words, we wish to explore, if the probability of a universe having larger values of torsion is significantly small compared to the probability of a universe having zero torsion. This will provide a quantum explanation of why our universe has a very small torsion --- since such universes are more favored during the initial quantum phase of the universe. 

\end{itemize}
Finally, we would like to mention that in dealing with torsion there is scope for possible ambiguities. We show below that one can construct different models of torsion, in which the torsion parameter and the scale factor can have non-trivial coupling. We have worked with the simplest case, wherein the torsion parameter remains uncoupled to the scale factor and behaves as a parameter. Moreover, there can be two types of torsion degrees of freedom in the cosmological setting one of them being a parity-odd component (see, \cite{TSAMPARLIS197927,PhysRevD.103.104008}). These two torsion degrees of freedom can be treated differently in the quantum theory, leading to further ambiguities. We shall discuss all of these issues below.

The paper is organized as follows: We start by introducing the basics of the mini-superspace model with torsion in \ref{sec:torsion_in_minisuperspace}. The canonical quantization method and the solutions of the Wheeler-DeWitt equation have been derived in \ref{sec:Wheeler_DeWitt_quantization} for both inflationary and bouncing scenarios. Subsequently, we performed the Lorentzian path integral and determined the relevant saddle points for both inflationary and bouncing scenarios with different boundary conditions in \ref{sec:PIquantization}. Then we have discussed the stability of these saddle points under scalar perturbations and have determined the associated power spectrum for the scalar field in the background of both inflationary and bouncing cosmologies in \ref{sec:quantized_perturbations}. Then we conclude with a discussion of the results obtained and the future directions of exploration. Detailed calculations regarding the first order formalism with torsion in the minisuperspace have been presented in \ref{AppA} and \ref{AppB}, respectively.

\emph{Notations and Conventions:} Throughout this paper, we shall use natural units, i.e., we shall set the fundamental constants $c=1=\hbar$ unless mentioned otherwise. The four-dimensional spacetime is spanned by the Greek indices $\mu,\nu,\ldots$ and the four-dimensional local Minkowski spacetime will be spanned by lowercase Roman indices $a,b,\ldots$, whereas the indices $i,j, \ldots$ at places may refer to spatial components of vectors or tensors.  
\section{Torsion in the minisuperspace}
\label{sec:torsion_in_minisuperspace}
In this section, we provide all the necessary ingredients for incorporating torsion in the context of minisuperspace quantum cosmology. We first provide our setup and compute the Palatini action, which in turn provides us the Hamiltonian for the system necessary for canonical quantization. Subsequently, we discuss these setups in two distinct scenarios, firstly in the case of inflationary cosmology, and then we consider a number of bouncing models of the universe. 
\subsection{General setup}

In the minisuperspace approximation, quantization is restricted to only a limited number of degrees of freedom by invoking the symmetries exhibited by the classical cosmological spacetime, namely homogeneity, and isotropy in the present case. As a consequence, the spacetime metric is characterized by two functions of time, the scale factor $q(t)$ and the lapse function $N(t)$.  Therefore, to start with, we consider the following general parametrized metric for the minisuperspace, which reads,
\begin{align}\label{minimetric}
{\rm d}s^2=-\frac{N^2(t)}{q^{2p}(t)}{\rm d}t^2+q^{2s}(t)\left(\frac{{\rm d}r^2}{1-\mathcal{K} r^2} + r^2 \left({\rm d}\theta^2 + \sin^2\theta {\rm d}\phi^2\right)\right)~.
\end{align}
Here, $p$ and $s$ are taken to be real and rational numbers. For the moment, we keep these two numbers as independent parameters, but later we shall constrain them so that the action for the system is quadratic in $q(t)$, and as a result, the quantization becomes straightforward. One immediately recognizes from \cite{PhysRevD.39.2206} that for an inflationary minisuperspace model, $s$ and $p$ both can be chosen to be $1/2$, however, for different bouncing scenarios the values of $s$ and $p$ will depend on the context (clear from \cite{PhysRevD.103.106008,Rajeev:2021yyl}). We shall discuss this point in detail below.

Since we are interested in the Einstein-Cartan theory, it is better to make a transition to the first order formalism, involving tetrads, such that, $g_{\mu \nu}=e^{a}_{\mu}e^{b}_{\nu}\eta_{ab}$, with $\eta_{ab}=\textrm{diag}(-1,1,1,1)$. For the above minisuperspace metric, the four basis tetrads have the following expressions,
\begin{align} 
\boldsymbol{e}^{0}=\frac{N}{q^{p}}\boldsymbol{d}t~,
\qquad
\boldsymbol{e}^{1}=\frac{q^{s}}{K(r)}\boldsymbol{d}r~,
\qquad
\boldsymbol{e}^{2}=q^{s}r\boldsymbol{d}\theta~,
\qquad 
\boldsymbol{e}^{3}=q^{s}r\sin \theta \boldsymbol{d}\phi~,
\label{tetrad}
\end{align}
where, the bold-faced quantities denotes the 1-forms, and the function $K(r)$ reads, $K(r)=\sqrt{1-\mathcal{K}r^2}$. In the above expressions, the quantity $\mathcal{K}$ determines the spatial curvature index of the spacetime and is a constant. In particular, it can take values $\mathcal{K}=0,\pm 1$, corresponding to a flat, closed, or open universe, respectively, when torsion is absent. For brevity, we have introduced the following notation, $q^{a}\equiv \left(q(t)\right)^{a}$. 

Given the above basis 1-forms $\boldsymbol{e}^{a}$, the spin-connections $\boldsymbol{\omega}^a_{\,\,\, b}$ can be determined from the following Cartan structure equation,
\begin{align}\label{cartan_first_structure}
\boldsymbol{T}^{a}=\boldsymbol{d}\boldsymbol{e}^a+\boldsymbol{\omega}^a_{\,\,\, b} \wedge \boldsymbol{e}^b~, 
\end{align}
where, $\boldsymbol{T}^{a}$ is the torsion two-form. In the absence of torsion, the spin-connections $\boldsymbol{\omega}^a_{\,\,\, b}$ are determined by the tetrads alone, while for nonzero torsion, the spin-connections are dependent on the tetrads and as well as on the components of the torsion field. Since torsion is also a geometrical entity, it must respect the symmetries of the minisuperspace, in our case, homogeneity and isotropy. Under these symmetries, the torsion two-form has the following expression \cite{TSAMPARLIS197927},
\begin{align}
\boldsymbol{T}^0=0~; 
\qquad 
\boldsymbol{T}^i=\mathcal{T}(t) \boldsymbol{e}^0 \wedge \boldsymbol{e}^i + \mathcal{C}(t) \epsilon^i_{\,\,\, jk} \boldsymbol{e}^j \wedge \boldsymbol{e}^k~,
\label{torsion}
\end{align}
where, $\mathcal{T}(t)$ and $\mathcal{C}(t)$ are arbitrary functions of the coordinate time $t$. Thus, given the above structure of the torsion two-form and the Cartan structure equation in \ref{cartan_first_structure}, we obtain the following expression for the components of the spin connection (see \ref{AppA} for derivation),
\begin{align}
\boldsymbol{\omega}^{0i}&=q^{-3s+1} \mathcal{B} (t) \boldsymbol{e}^i~, 
\label{connection1}
\\ 
\boldsymbol{\omega}^{12}&=-\frac{K(r)}{r} q^{-s} \boldsymbol{e}^2 -  q^{d} c(t) \boldsymbol{e}^3~, 
\\ 
\boldsymbol{\omega}^{13}&=-\frac{K(r)}{r} q^{-s} \boldsymbol{e}^3 + q^{d} c(t) \boldsymbol{e}^2~, 
\\ 
\boldsymbol{\omega}^{23}&=-\frac{\cot \theta}{r} q^{-s} \boldsymbol{e}^3 - q^{d} c(t) \boldsymbol{e}^{1}~.
\label{connection2}
\end{align}
Here, we have introduced two new functions of time $\mathcal{B}(t)$ and $c(t)$, such that,
\begin{align}
c(t) &\equiv \frac{\mathcal{C}(t)}{q^{d}}~, 
\label{eq:torsionCcomponent}
\\ 
\mathcal{B}(t) &\equiv \left \{-\mathcal{T}(t)+ sq^{p-1}\frac{\dot{q}}{N} \right\}q^{3s-1}~, 
\label{eq:torsionTcomponent}
\end{align}
with $d$ being a real and rational number which we shall constrain later. Thus instead of the functions $\mathcal{T}(t)$ and $\mathcal{C}(t)$, we may use the newly defined function $\mathcal{B}(t)$ and $c(t)$, such that the torsion tensor becomes,
\begin{align} 
\boldsymbol{T}^0=0~;\qquad \boldsymbol{T}^i=\left(\frac{s}{N}q^{p}\frac{\dot{q}}{q} - q^{-3s+1} \mathcal{B} (t)\right)\boldsymbol{e}^0 \wedge \boldsymbol{e}^i + q^d c(t) \epsilon^i_{\,\,\, jk} \boldsymbol{e}^j \wedge \boldsymbol{e}^k~.
\end{align}
As evident from the above expression, $c(t)$ corresponds to the completely anti-symmetric part of the torsion tensor and can be related to the pseudo-scalar degree of freedom. On the other hand, $\mathcal{B}(t)$ contains the parity-even part of the torsion tensor. Having derived the connections in terms of the scale factor $q$, lapse function $N$, and the two torsion parameters, such that one of them is contained in $\mathcal{B}$ and the other, completely antisymmetric part is given by $c$, we shall now determine the curvature two-form, which has the following definition,
\begin{align} 
\boldsymbol{R}^{ab} \equiv {\bm d} \boldsymbol{\omega}^{a b} + \boldsymbol{\omega}^a_{\,\,\, c} \wedge\boldsymbol{\omega}^{cb}~.
\end{align}
Explicitly, using the connections from \ref{connection1} to \ref{connection2}, it turns out that only the temporal-spatial and purely spatial components of the curvature two-form are non-zero and they read (for a derivation, see \ref{AppA}),
\begin{align} 
\boldsymbol{R}^{0i}&=\frac{q^{-3s+p}}{N}\left(q \dot{\mathcal{B}}(t) + \left(1-2s\right) \dot{q} \mathcal{B}(t) \right) \boldsymbol{e}^0 \wedge \boldsymbol{e}^i + q^{-3s+d+1} \mathcal{B} (t) c(t) \epsilon^i_{\,\,\, jk}\boldsymbol{e}^j \wedge \boldsymbol{e}^k ~,
\\ 
\boldsymbol{R}^{ij}&= q^{-2s} \left(\mathcal{K} + \mathcal{B}^2(t) q^{-4s+2} - c^2(t) q^{2(s+d)} \right)\boldsymbol{e}^i \wedge \boldsymbol{e}^j - \frac{q^{p+d}}{N}\left(\dot{c}(t) + \left(s+d\right) \frac{\dot{q}}{q} c(t)\right) \epsilon^{ij}_{\,\,\,\,k}\boldsymbol{e}^0 \wedge \boldsymbol{e}^k~. 
\end{align}
Therefore, the symmetry of the spacetime reduces the number of degrees of freedom in the spacetime metric to two (scale factor $q$, lapse function $N$) from ten. Similarly, the torsion tensor, which in general, can have 24 independent components, reduces to two degrees of freedom (the parity-even part residing in $\mathcal{B}$ and the parity-odd part $c$). In the minisuperspace, the gravitational action must be expressed in terms of the reduced variables, in this case, the action must be expressed in terms of the two degrees of freedom of the metric and the two degrees of freedom of the torsion. Further, since we are working with first order formalism, we should use the Palatini action, which takes the form,
\begin{align} 
\kappa \mathcal{A}_{\rm Palatini}=\frac{1}{2!}\int \epsilon_{abcd}\bigg(\boldsymbol{e}^a\wedge \boldsymbol{e}^b \wedge \boldsymbol{R}^{cd}
- \frac{\Lambda}{6} \boldsymbol{e}^a \wedge \boldsymbol{e}^b \wedge \boldsymbol{e}^c \wedge \boldsymbol{e}^d \bigg) ~,
\end{align}
where $\kappa=16 \pi G_{\rm N}$, with $G_{\rm N}$ being the Newton's gravitational constant. We also assume that the cosmological constant is either positive or, zero. Note that, $\epsilon_{abcd}$ is the completely antisymmetric Levi-Civita symbol in the locally flat spacetime and we have used the following convention: $\epsilon_{abcd}\bm{e}^{a}\wedge \bm{e}^{b}\wedge \bm{e}^{c}\wedge \bm{e}^{d}=4!\sqrt{-g}d^{4}x$. Using the expressions for the curvature two-form and the tetrad one-forms, we can compute the Palatini action for the minisuperspace cosmology which reads (for a derivation, see \ref{AppB}),
\begin{align} 
\mathcal{A}_{\rm Palatini}=\frac{3 V_{3}}{8\pi G_{\rm N}}\int {\rm d}t \left[\left(q \dot{\mathcal{B}}(t) + \left(1-2s\right) \dot{q} \mathcal{B}(t) \right) + q^{s-p} N \left(\mathcal{K} + \mathcal{B}^2(t) q^{-4s+2} - c^2(t) q^{2(s+d)} \right) -\frac{\Lambda}{3}q^{3s-p} N \right] ~, 
\end{align}
where, $V_{3}$ is the volume of the three-space. As promised, the action depends on the four degrees of freedom, and on the three parameters, $s$, $p$, and $d$, respectively.

The quantity inside the integral of the above action is the Lagrangian of the (gravity+torsion) system in the cosmological setup within the context of first order formalism. Since, any Lagrangian can be expressed as, $L=\mathsf{p}\dot{q}-H$, where $H$ is the Hamiltonian associated with the Lagrangian $L$, the above action can be rewritten, up to a boundary term, in the following form:
\begin{align}
\label{eq:PalatiniAction}
\mathcal{A}_{\rm Palatini}=\frac{3 s V_{3}}{4\pi G_{\rm N}}\int {\rm d} t \left[-\dot{q}\mathcal{B} - N \left(- \frac{q^{s-p}}{2s}\mathcal{K} - \frac{q^{-3s-p+2}}{2s} \mathcal{B}^2 + \frac{q^{2d+3s-p}}{2s} c^2 + \frac{\Lambda}{6 s}q^{3s-p} \right) \right] + \frac{3V_{3}}{8 \pi G_{\rm N}} q \mathcal{B}\Big|_{\rm Boundary}~,
\end{align}
allowing us to identify the momentum conjugate to $q$ as $\mathsf{p}=-\mathcal{B}$. The resulting Lagrangian, as evident from the above action, is directly in the form $(\mathsf{p}\dot{q}-N\mathcal{H})$, where the factor $N$ before the Hamiltonian arises due to non-trivial lapse function in the spacetime metric, and hence we can identify the corresponding Hamiltonian `density' $\mathcal{H}$\footnote{The Hamiltonian of the system is given by $H=(3sV_{3}/4\pi G_{\rm N})N \mathcal{H}$, which is simply equal to the Hamiltonian density times an overall multiplicative factor, proportional to the volume of three space. However, in what follows we shall call the quantity $\mathcal{H}$ as the Hamiltonian of the system since no further confusion is likely to arise.} as,
\begin{align}\label{hamiltonian1}
\mathcal{H}=-\frac{q^{-3s-p+2}}{2s} \mathcal{B}^2 - \frac{q^{s-p}}{2s}\mathcal{K} + \frac{q^{2d+3s-p}}{2s} c^2 + \frac{\Lambda}{6 s}q^{3s-p}~.
\end{align}
The variation of the Hamiltonian with respect to $q$ and $\mathcal{B}$ yields the dynamical equations for $\mathcal{B}$ and $q$, respectively. On the other hand, the torsional degree of freedom $c$ has no conjugate momentum, which in turn demonstrates that the purely anti-symmetric part of torsion is non-dynamical, at least classically. Thus in what follows, we can take the torsional degree of freedom $c$ to be a constant, not evolving with time. We will observe that a similar conclusion holds true in the quantum domain as well. 

It is to be emphasized that, so far we have used the Palatini form of the action in the tetrad formalism, so that both the metric degree of freedom $q$ and the connection degree of freedom $\mathcal{B}$ are treated as independent. As a consequence we have only one torsion degree of freedom $c$, while the other gets hidden inside $\mathcal{B}$ and does not play any role. However, had we used the metric form of the action, the connection $\mathcal{B}$ needs to be expressed in terms of the metric degree of freedom and derivative thereof, see \ref{eq:torsionTcomponent}. Thus we will now have two torsion degrees of freedom, namely $c$ and $\mathcal{T}$. In particular, all the $c^2$ terms in the expressions, need to be replaced by $\mathcal{T}^2-c^2$. In what follows, we shall continue with the Palatini formalism for the rest of our analysis, which can be changed to the metric formalism by simply performing the above replacement.

Further simplification can be achieved by fixing the powers of $q$ in the above expression for the Hamiltonian. Since our aim is to quantize the system and determine the wave function of the universe using path integral formalism, we want the Hamiltonian to involve terms that are at most of the quadratic order of the dynamical variables. We notice that the torsional parameters are already quadratic and hence we simply need to fix the powers of the scale factor appropriately. If we transform the first term of the above Hamiltonian, in \ref{hamiltonian1} to a quadratic one in the dynamical variable, it requires imposing the following condition: $2-3s-p=0$. This in turn determines the exponent $p$ to be: $p=-3s+2$ and hence the Hamiltonian can be re-written as
\begin{align}\label{hamiltonian2}
\mathcal{H}= - \frac{1}{2s} \mathcal{B}^2 - \frac{q^{4s-2}}{2s}\mathcal{K}  + \frac{q^{2d+6s-2}}{2s} c^2 + \frac{\Lambda}{6 s}q^{6s-2}~. 
\end{align}
The two parameters $s$ and $d$ are still free. Further constraints can be imposed on these parameters, depending on whether we are considering inflationary or bouncing scenarios. In the inflationary scenario, one considers the cosmological constant $\Lambda$ to be dominant over any other matter fields existing in the universe. However, in the bouncing scenario, at least near the bounce, the cosmological constant is irrelevant, and instead one needs to add phenomenological and suitably parametrized two-component fluid, leading to bouncing cosmology. We discuss these two cases in the following.

\subsection{Inflationary cosmology} 
\label{sec:inflationary_minisuperspace}

As emphasized earlier, in an inflationary scenario, the cosmological constant is dominant and there is no additional matter content in the universe, except for the spacetime torsion and spatial curvature index ($\mathcal{K}$), arising from the geometric sector. In this case, the quadratic nature of the Hamiltonian in \ref{hamiltonian2} requires $s=(1/2)$, since there is no other way to make both the spatial curvature index term and the cosmological constant term to be at most of the quadratic order in the scale factor $q$ (this can also be seen from \cite{PhysRevD.39.2206}). Given the above value for the parameter $s$, if we also make the following choice: $d=-3s+1$, then the Hamiltonian, in \ref{hamiltonian2} becomes
\begin{align}\label{hamiltonianinf}
\mathcal{H}_{\rm inflation}=-\mathcal{B}^2 - \mathcal{K}  + c^2 + \frac{\Lambda}{3}q~. 
\end{align}
We would like to emphasize that the above Hamiltonian coincides with the one suggested in \cite{PhysRevD.103.104008}. For the above choices of the parameters, the time coordinate $t$, defined in \ref{minimetric}, gets related to the cosmological time coordinate $t_{\rm co}$, by the relation: $dt_{\rm co}=(N/\sqrt{q})dt$, and the scale factor $a(t_{\rm co})$ in the cosmological coordinate gets related to $q(t)$ by, $a=\sqrt{q}$. Note that this demands $q$ to be a positive definite quantity in the classical theory \footnote{However, note that in the quantum theory, say in path integral quantization this restriction is lifted and the limit in the quantum fluctuation integrals is chosen to be the whole range of $(-\infty,\infty)$, see \cite{PhysRevD.39.2206}. However, for a different approach to quantum cosmology, wherein, $q$ is restricted to only a positive range, see \cite{Jia:2022nda}.}. Therefore, the Hubble parameter, defined in the cosmological coordinate, in terms of the scale factor $a(t_{\rm co})$ reads, 
\begin{align}
H(t)\equiv\frac{1}{a(t_{\rm co})}\dfrac{da(t_{\rm co})}{dt_{\rm co}}=\frac{\dot{q}}{2N\sqrt{q}}~.
\end{align}
From the equation \ref{eq:torsionTcomponent}, we see that the canonically conjugate momentum ($\mathsf{p}=-\mathcal{B}$) is related to the Hubble rate offset by the torsion component $\mathcal{T}(t)$. On the other hand, given the Hamiltonian \ref{hamiltonianinf}, we can calculate the equations of motion
\begin{align}
\dot{q}&=N\left(\frac{\partial \mathcal{H}_{\rm inflation}}{\partial \mathsf{p}}\right)=-N\left(\frac{\partial \mathcal{H}_{\rm inflation}}{\partial \mathcal{B}}\right)=2 N \mathcal{B}~,
\\
\dot{\mathsf{p}} & = - \dot{\mathcal{B}} = - N\left(\frac{\partial \mathcal{H}_{\rm inflation}}{\partial q}\right) = - N\frac{\Lambda}{3}~.
\end{align}
Consistency of the first Hamilton's equation with \ref{eq:torsionTcomponent} demands $\mathcal{T}(t)=0$ on-shell. Now, as we shall perform path integral quantization in the phase space, the result of the path integral will be approximated by the classical solution, and hence for all intents and purposes, $\mathcal{T}(t)$ can be set to zero. The only non-trivial contribution, in the quantum theory, from the torsional sector arises from the purely anti-symmetric degree of freedom, namely $c$. As in \cite{PhysRevD.103.104008}, we leave this torsion component as a parameter while quantizing the theory to obtain the kinematical torsionful wave function of the universe.
\subsection{Bouncing scenario} 
\label{sec:bouncing_minisuperspace}

For bouncing models of cosmology, near the bounce, the matter content of the universe is more important than the cosmological constant, and hence we can safely set $\Lambda=0$. Instead of the cosmological constant, here we shall introduce a two-component fluid, like in \cite{PhysRevD.103.106008,Rajeev:2021yyl}, such that the background classical spacetime undergoes a bounce. In the presence of the bounce-enabling fluid, the overall (gravity+torsion+matter) action has to be obtained by adding the matter action to the previously computed action for the (gravity+torsion) sector. The action for the matter sector reads,
\begin{align} 
\mathcal{A}_{\rm M}= -\int {\rm d}^4x \sqrt{-g} \rho= V_{3}\int{\rm d}t\, N U(q)~,
\end{align}
where $\rho(q)$ is the energy density of the matter sector, including that of the bounce-enabling matter. Moreover, for later convenience, we have introduced the potential function $U(q)$, defined as,
\begin{align}
U(q)\equiv -q^{3s-p}\rho(q)=-q^{6s-2}\rho(q)~.
\end{align}
Therefore, the total action involving the metric, torsional, and matter degrees of freedom altogether becomes,
\begin{align}
\label{eq:NewPalatiniAction}
\mathcal{A}_{\rm total}=\frac{3 s V_{3}}{4\pi G_{\rm N}}\int {\rm d} t \left[-\dot{q}\mathcal{B} - N \left(- \frac{q^{4s-2}}{2s}\mathcal{K} - \frac{1}{2s} \mathcal{B}^2 + \frac{q^{2d+6s-2}}{2s} c^2 -\frac{4\pi G_{\rm N}}{3s}U \right) \right]~,
\end{align}
where we have ignored the boundary contribution. Therefore, the Hamiltonian for the system now reads,
\begin{align}
\label{eq:bouncehamilton}
\mathcal{H}=-\frac{1}{2s} \mathcal{B}^2-\frac{q^{4s-2}}{2s}\mathcal{K} + \frac{q^{2d+6s-2}}{2s} c^2 -\frac{4\pi G_{\rm N}}{3s}U~.
\end{align}
The variation of the above Hamiltonian with respect to the conjugate momentum $\mathsf{p}$, as well as the variation with respect to the scale factor $q$ yields Hamilton's equations of motion as,
\begin{align}
\dot{q}&=N\left(\frac{\partial \mathcal{H}}{\partial \mathsf{p}}\right)=-N\left(\frac{\partial \mathcal{H}}{\partial \mathcal{B}}\right)=N(\mathcal{B}/s)~,
\\
\dot{\mathsf{p}} & = -\dot{\mathcal{B}} = -N\left(\frac{\partial \mathcal{H}}{\partial q}\right)=N\left(\frac{4s-2}{2s}\right)q^{4s-3}\mathcal{K}-N\left(\frac{2d+6s-2}{2s}\right)q^{2d+6s-3}c^2+\frac{4\pi G_{\rm N}}{3s}N\left(\frac{\partial U}{\partial q}\right)~.
\end{align}
Again, we see that consistency of the first Hamilton's equation with \ref{eq:torsionTcomponent} demands $\mathcal{T}(t)=0$ on-shell, leaving the purely anti-symmetric degree of freedom $c$ as the only non-trivial torsion degree of freedom. Note that there is an additional equation given by $\mathcal{H}=0$, corresponding to the Hamiltonian constraint of general relativity.\par
As in the context of the inflationary paradigm, for the bouncing scenario as well, one would prefer the Hamiltonian to be quadratic in the scale factor $q$, in order to facilitate the path integral formulation. For that, we need to provide the behavior of the energy density of the fluid $\rho$ with the scale factor $q$. As is customary for phenomenological bouncing models of cosmology, we introduce a two-component fluid with generic power law fall-off behaviors, such that,
\begin{align}
\rho=\rho_{0}\left[-\frac{\Omega}{a^{m}}+\frac{1}{a^{n}}\right]~.
\end{align}
Here, $\rho_{0}$ is a constant, related to the energy scale close to the time of bounce. Moreover, $\Omega$ is the relative fractional abundance of the bounce-enabling exotic matter compared to the normal matter component. We shall now relate the powers of the scale factor $n$ and $m$ with $s$, such that we have a wide range of fluid material that can contribute to the potential $U(q)$ at most quadratically in the scale factor $q$.

In order to determine the effect of the energy density of the fluid on the Hamiltonian \ref{eq:bouncehamilton}, we use the relation $a=q^s$, and express the potential $U$ in terms of the scale factor $q$ as,
\begin{align}
U=\rho_{0}\left[\Omega q^{6s-2-ms}-q^{6s-2-ns}\right]~.
\end{align}
Since the first non-trivial effect arises when the potential is linear in the scale factor $q$, we will consider this linear model of the potential throughout this work. The main motivation is, the linear model is the simplest one, which captures the non-trivial dynamics of the bounce-enabling matter. Thus, requiring that the form of the potential be $U=\rho_{0}\left(\Omega-q\right)$, so that for small $q$, the bounce-enabling matter starts to become important and for large $q$, the normal matter dominates. Given the above structure of the potential, we obtain the following conditions: $s=\{3/(6-n)\}$ and $m=\{(6s-2)/s\}=6-2\{(6-n)/3\}=\{(6+2n)/3\}$. Therefore, the energy density of the matter contained in the bouncing scenario becomes,
\begin{align}\label{bounce_fluid_scale}
\rho=\rho_{0}\left[\frac{1}{a^{n}}-\frac{\Omega}{a^{\frac{6+2n}{3}}}\right]; \quad s=\frac{3}{6-n}~,
\end{align}
where it is assumed that $n<6$. The other terms in the Hamiltonian \ref{eq:bouncehamilton} correspond to the spatial curvature associated with the metric degrees of freedom and torsion. The scale factor dependence of the spatial curvature index term becomes, $q^{2n/(6-n)}$, so that for $n=0$, $n=2$ and $n=3$, the scale factor attached to the spatial curvature $\mathcal{K}$ becomes independent of $q$, linear in $q$ and quadratic in $q$ respectively. Hence, if we want to keep the spatial curvature term  $\mathcal{K}$, arising from the metric alone, in the Hamiltonian, we need to restrict ourselves to the above values of $n$. To avoid any such restrictions, so that a broader class of bouncing models can be considered, we will take $\mathcal{K}=0$ for the bouncing scenario.  

Finally, the scale factor associated with the torsion degree of freedom $c^{2}$ has the following behavior: $q^{2d+6s-2}$, such that we can choose the parameter $d$ in order for the power of the scale factor to be at most of the quadratic order. All of these different choices of the parameter $d$, lead to inequivalent models of the parity-odd torsion. Treatment of a system with Hamiltonian of quadratic order in the scale factor $q$ can become complicated in the path integral approach, see e.g., \cite{PhysRevD.103.106008,Rajeev:2021yyl} and make the physics obscure. Thus, we will consider two convenient choices for the parameter $d$ which are 
\begin{align}
d^{(\mathsf{0}),(\mathsf{L})} = \begin{cases} 1-3s; \quad &\text{leading to the term}~c^2~,
\\
\frac{3(1-2s)}{2}; \quad &\text{leading to the term}~q c^2~.
\end{cases}
\end{align}
Thus the possible Hamiltonians in the bouncing model of cosmology will be given by, 
\begin{align}
\mathcal{H}_{\rm bounce}^{(\mathsf{0})}&=-\frac{1}{2s}\mathcal{B}^2+\frac{1}{2s}c^2-\frac{4\pi G_{\rm N}\rho_{0}}{3s}\left(\Omega-q\right)~.
\label{eq:bouncehamilton1}
\\
\mathcal{H}_{\rm bounce}^{(\mathsf{L})}&=-\frac{1}{2s}\mathcal{B}^2+\frac{1}{2s}qc^2-\frac{4\pi G_{\rm N}\rho_{0}}{3s}\left(\Omega-q\right)~.
\label{eq:bouncehamilton2}
\end{align}
From a cursory comparison between the inflationary and bouncing Hamiltonians \ref{hamiltonianinf} and \ref{eq:bouncehamilton1}, it may appear that the exotic and normal fluid components in the bouncing Hamiltonian correspond to the `curvature' and the `cosmological constant', if one identifies $(4\pi G_{\rm N}/3s)\rho_{0}\Omega$ with $\mathcal{K}$ and $(4\pi G_{\rm N}/3s)\rho_{0}$ with $\Lambda/3$, respectively. However, such an appearance is deceptive, since in the cosmic time coordinate frame the energy densities of these two fluids in the bouncing scenario satisfy the scaling law given in \ref{bounce_fluid_scale}, which is very different from the scaling of the curvature and the cosmological constant. The only exception is when $n=0$. Similarly, the `energy density' corresponding to torsion component $c$ scales as $a^{-2(n+3)/3}$, a similar scaling to that of the bounce-enabling matter.

Note that the classical bouncing scale is obtained by setting $\dot{q}=0$, which in turn demands $\mathcal{B}=0$ (on-shell). Thus the Hamiltonian constraint $\mathcal{H}=0$ yields the scale factor $q_{\rm B}$ at which the bounce occurs, for both the $q$-independent and linear in $q$ models,
\begin{align}\label{scale_bounce}
q_{\rm B}^{(\mathsf{0})}=\Omega-\frac{3c^{2}}{8\pi G_{\rm N}\rho_{0}}~;
\qquad
q_{\rm B}^{(\mathsf{L})}=\Omega\left(1+\frac{3c^{2}}{8\pi G_{\rm N}\rho_{0}}\right)^{-1}~.
\end{align}
Thus, in order to have a finite and positive scale factor at which bounce happens, we must have non-zero $\Omega$, and also the contribution of the torsion to the energy budget must be less than the contribution from the bounce-enabling matter. Note that in the following we shall only work with the Hamiltonian \ref{eq:bouncehamilton1}.

It is to be emphasized that, in the quantum domain, both the classical degrees of freedom $q$ and $c$ must be promoted to operators, $\hat{q}$ and $\hat{c}$, respectively. Even then there is no operator ordering ambiguity due to the terms of the form $qc^{2}$, since these two operators are not canonical conjugates and hence they commute with each other, i.e., $[\hat{q},\hat{c}]=0$.

In what follows we will first quantize the minisuperspace model of cosmology with torsion in the Wheeler-DeWitt approach and then shall discuss how the semi-classical results derived from the Wheeler-DeWitt approach arise from the path integral computation as well. The path integral approach will provide us with an idea about the stability of the semi-classical geometry and the behavior of perturbations around the same, which in turn will help us to determine the power spectrum of these perturbations. We discuss each of these aspects in the subsequent sections.  

\section{Quantization using the Wheeler-DeWitt equation}
\label{sec:Wheeler_DeWitt_quantization}
We shall now study the quantization of the minisuperspace models described above. Like any other quantum system, the quantization of cosmological models can be approached from two distinct directions, namely, (a) the canonical quantization, and (b) the path integral quantization. Generally, the equivalence between these approaches is expected, however, often subtle differences come into play, in the sense that certain issues are more apparent in one approach than in the other. For example, recently, the issue of the stability of the wave function of the universe obtained with different initial conditions has been greatly debated, along with the merits of the path integral approach versus the canonical approach in dealing with the issues of stability under perturbations \cite{PhysRevD.95.103508,PhysRevLett.119.171301,PhysRevD.97.023509,Lehners:2018eeo,PhysRevD.100.123543,PhysRevLett.122.201302,PhysRevD.98.066003,Feldbrugge:2018gin,PhysRevD.99.066010,PhysRevD.99.043526,PhysRevD.100.043544}. As far as the present article is concerned, for the quantization of the minisuperspace model(s), first, we shall solve the Schr\"odinger-like Wheeler-DeWitt equation(s) to obtain the wave function(s) of the unperturbed universe with the Hartle-Hawking no-boundary condition. Next, we shall show that these solutions can be obtained from the path integral approach as well. Finally, we shall deal with scalar perturbations around the saddle point geometry, that closely approximates the path integral for the background spacetime. We shall derive the wave function for such scalar degrees of freedom and consequently derive corresponding power spectra. We shall perform these calculations for the inflationary and bouncing scenarios and highlight differences between these two types of cosmologies wherever appropriate.

Now, if one assumes the connection component $\mathcal{B}$ to be an independent variable, alongside the tetrads, one can then have a wave function of the universe characterized by connection $\Phi(\mathcal{B})$. However, due to canonical conjugacy between $q$ and $\mathcal{B}$, this \textit{connection wave function} is related, by means of Fourier transform, to a wave function $\Psi(q)$ defined in the coordinate space representation. Solutions of the WDW equation in different representations can be obtained from the path integral approach by choosing suitable boundary terms, as we shall show in the \ref{sec:PIquantization}.

\subsection{Inflationary scenario}
\label{sec:inflationary_scenario_WDW}

In this section, we shall deal with the canonical quantization of the minisuperspace model described in \ref{sec:inflationary_minisuperspace}, which corresponds to an inflationary universe with torsion. Note that the canonical quantization of such a scenario has already been thoroughly discussed in the literature, e.g. in \cite{PhysRevD.103.104008}. However, we present the analysis here as well for the sake of completeness and also for the fact that this discussion will flesh out the notations and conventions we are going to follow for the rest of our paper.

As discussed in the previous sections, the exponents of the scale factors in the Palatini action for the inflationary cosmology can be so chosen that the integrand becomes quadratic in the degrees of freedom and the corresponding Hamiltonian for the system is given by \ref{hamiltonianinf} with $\mathsf{p}=-\mathcal{B}$ being identified as the canonically conjugate momentum to $q$. Thus for the Palatini action, the (negative of) the connection component $\mathcal{B}$ is the canonical conjugate momentum of the metric variable $q$. Therefore, following the canonical quantization scheme, where all the conjugate variables are promoted to operators and satisfy the usual Heisenberg algebra, here also, we promote the variables $q$, and $\mathcal{B}$ to operators and impose fundamental commutation relations between them, such that,
\begin{align}
\label{eq:commutation_relation}
[\hat{q},\hat{\mathcal{B}}] = -i \mathfrak{h}~.
\end{align}
Following \cite{PhysRevD.103.104008}, we have defined the effective Planck's constant $\mathfrak{h}$ as
\begin{align}
\mathfrak{h} \equiv \frac{8\pi G_{\rm N} \hbar}{3V_{3}}=\frac{\ell_{\rm Pl}^2}{3V_{3}}~.
\end{align}
The last equality follows from the definition of the Planck length, $\ell_{\rm Pl} = \sqrt{8 \pi G_{\rm N} \hbar}$. Note that in our convention we have ignored the coefficient in front of the action $\frac{3V_{3}}{8 \pi G_{\rm N}}$ while defining the conjugate momentum and then redefined the Planck's constant for quantization. Alternatively, one could have defined the conjugate momentum along with the pre-factor $(3V_{3}/8 \pi G_{\rm N})$, and then imposing the standard commutation relation involving $\hbar$ alone would, essentially, result into the same algebra. However, defining the conjugate momentum without the pre-factor and then defining an effective Planck's constant for the quantization seems notation-wise convenient and we shall stick to that convention. For a physical interpretation of such an effective Planck's constant, see \cite{Barrow:2020coo}. However, it is not necessary for us to commit to any particular interpretation at this point. 

Further note that the Heisenberg evolution equation for the torsion yields, $(d\hat{c}/dt)=0$, since $\hat{c}$ commutes with all the other operators, including the Hamiltonian itself. Thus even in the quantum domain, we can treat the torsion as a constant valued operator.   

In the momentum representation, the operator $\hat{q}$ must be represented as a differential operator constructed from the conjugate momentum $\mathcal{B}$. As in ordinary quantum mechanics, in order to satisfy the fundamental commutation relation in \ref{eq:commutation_relation}, one represents the operator $\hat{q}$ as follows,
\begin{align}
    \hat{q} \to + i\mathfrak{h}\frac{\partial}{\partial \mathsf{p}} = -i\mathfrak{h}\frac{\partial}{\partial \mathcal{B}}~.
\end{align}
Therefore, the quantum Hamiltonian constraint or the Wheeler-DeWitt equation, $\hat{\mathcal{H}}_{\rm inflation}\Phi^{\rm I}(\mathcal{B})=0$, for the inflationary universe takes the following form in the momentum representation
\begin{align}
\label{eq:WDW_inflation}
\left(-i\mathfrak{h}\frac{\Lambda}{3}\frac{\partial}{\partial \mathcal{B}} - \mathcal{K}_{c} - \mathcal{B}^2 \right)\Phi^{\rm I}(\mathcal{B})=0~,
\end{align}
where we have defined an effective curvature scale: $\mathcal{K}_c \equiv \mathcal{K} - c^2$, and the superscript `${\rm I}$' in the wave function $\Phi^{\rm I}(\mathcal{B})$ indicates the fact that we are here concerned with an inflationary scenario. Two comments are in order---(a) we see that the parity-odd torsion component `$c$' acts, in this particular case of an inflationary scenario, as a source for spatial curvature and produces an effective spatial curvature $\mathcal{K}_c$; (b) the other torsion scalar $\mathcal{T}$ is hidden inside the connection component $\mathcal{B}$, and as a result, essentially, its information is erased in the quantum theory, in the phase space. This fact, therefore, suggests an in-equivalent treatment of the two torsion components $\mathcal{T}$ and $c$ in the quantum theory.

The Wheeler-DeWitt equation in the momentum space, presented in \ref{eq:WDW_inflation}, is a first order ordinary differential equation, whose most general solution reads,
\begin{align}
\label{eq:WDW_solution_inflation}
    \Phi^{\rm I}(\mathcal{B}) = \phi_{0}\exp \left[i\frac{3}{\mathfrak{h}\Lambda}\left(\frac{\mathcal{B}^3}{3}+\mathcal{K}_{c}\mathcal{B}\right)\right]~,
\end{align}
where ${\phi_0}$ is a constant of integration. For the moment, we shall keep the constant of integration arbitrary and shall fix it later using the Hartle-Hawking no-boundary condition.


The wave function in the coordinate space representation, denoted by $\Psi^{\rm I}(q)$ can either be obtained by identifying the operator $\hat{\mathcal{B}}$ with $i\mathfrak{h}(\partial/\partial q)$ and subsequently solving the quantum Hamiltonian constraint or by means of a Fourier transformation of the wave function $\Phi^{\rm I}(\mathcal{B})$ in the momentum space as, 
\begin{align}
\Psi^{\rm I}(q)=\int \frac{{\rm d}\mathcal{B}}{\sqrt{2\pi \mathfrak{h}}} \Phi^{\rm I}(\mathcal{B}) e^{-\frac{i}{\mathfrak{h}}\mathcal{B}q}~.
\end{align}
The result of the above Fourier transformation with the integration range for the conjugate momentum being along whole the real line \footnote{On the other hand, if one restricts to only outgoing modes, as expected in Vilenkin's tunneling proposal, the range of integration is changed to only the half infinite line. For details, see \cite{PhysRevD.102.044034}.}, such that $\mathcal{B}\in (-\infty,\infty)$, is well known and has a closed form in terms of the Airy function ${\rm Ai}(x)$ that turns out to be,
\begin{align}
\label{eq:Fourier_wave_function}
    \Psi^{\rm I}(q) = \phi_{0}\sqrt{\frac{2\pi}{\mathfrak{h}}}\left(\frac{\mathfrak{h} \Lambda}{3}\right)^{\frac{1}{3}}
   \text{Ai}\left[\left(\frac{3}{\mathfrak{h} \Lambda}\right)^{\frac{2}{3}}\left(\mathcal{K}_{c} - \frac{\Lambda}{3} q\right)\right]~.
\end{align}
Now, we can determine the integration constant by using the boundary condition corresponding to Hartle-Hawking no-boundary proposal, which is $\Psi^{\rm I}\to 1$ as $q\to0$ (see, \cite{PhysRevD.31.1777,PhysRevD.37.888}). This condition implies
\begin{align}
\label{eq:HH_wave_function_WDW}
    \Psi^{\rm I}(q) = \frac{
   \text{Ai}\left[\left(\frac{3}{\mathfrak{h} \Lambda}\right)^{\frac{2}{3}}\left(\mathcal{K}_{c} - \frac{\Lambda}{3} q\right)\right]}{\text{Ai}\left[\left(\frac{3}{\mathfrak{h} \Lambda}\right)^{\frac{2}{3}}\mathcal{K}_{c}\right]}~.
\end{align}
Intriguingly, the above wave function exactly corresponds to the original Hartle-Hawking no-boundary wave function, when torsion is absent, i.e., in the limit $c\to 0$. Thus, it immediately follows that the wave functions derived in \ref{eq:WDW_solution_inflation} and \ref{eq:HH_wave_function_WDW} represent the no-boundary state of an inflationary universe in the presence of spacetime torsion. Moreover, in the semi-classical limit, with $\mathfrak{h}\rightarrow 0$, the above wave function, in the classically allowed domain $q > \left(3/\Lambda\right)  \mathcal{K}_{c}$, reduces to the standard Cosine function with an exponential prefactor outside, as fit for the no-boundary wave function. Explicitly, this prefactor reads, $\exp\left[+\left(2/\mathfrak{h}\Lambda\right)\mathcal{K}_{c}^{3/2}\right]$, and we shall see that in the path integral approach, we recover this exact prefactor when the saddle point geometry is required to have an initial zero size, i.e., the saddle point geometry corresponds to a Hartle-Hawking no-boundary geometry.

\subsection{Bouncing scenarios}

The generalization of the above quantization procedure to bouncing models of cosmology is straightforward. In the case of a bouncing scenario, we start with a (gravity+torsion+matter) action, as detailed in \ref{sec:bouncing_minisuperspace}. The bouncing models have an extra parameter $n$, whose different choices correspond to different classical bouncing scenarios. For example, as discussed above, the choice $n=3$ (or, equivalently $s=1$) corresponds to a matter bounce scenario. Moreover, recall that we have chosen all the other free parameters in the metric, such that the Hamiltonian becomes at most of the quadratic order in either the conjugate momentum or the scale factor $q$. The resulting Hamiltonians corresponding to various possible bouncing scenarios have been presented in \ref{eq:bouncehamilton1} and \ref{eq:bouncehamilton2}, respectively. In what follows, we shall consider the case where the scale factor and the torsional degree are decoupled, i.e., the Hamiltonian \ref{eq:bouncehamilton1}.



Quantization of the theory proceeds, as in the previous scenario, by imposing the fundamental commutation relation between the `coordinate' and the canonically conjugate momentum operator as,
\begin{align}
[\hat{q},\hat{\mathcal{B}}] = - i \mathfrak{h}_{s}~,
\end{align}
where, the effective Planck's constant $\mathfrak{h}_{s}$ depends on the model of bouncing cosmology through the parameter $s$ (or, equivalently through $n$) as,
\begin{align}
\mathfrak{h}_{s} \equiv \frac{4\pi G_{\rm N}\hbar}{3sV_{3}}~.
\end{align}
Note that the effective Planck's constant in the inflationary scenario can be obtained by choosing $s=(1/2)$ in the above expression. Here also, we can make the identification, $\hat{q} \to -i\mathfrak{h}_{s} \frac{\partial}{\partial \mathcal{B}}$, in order to write down the quantum Hamiltonian constraint or, the Wheeler-DeWitt equation, which becomes,
\begin{align}
\left(-i\mathfrak{h}_{s} \sigma \frac{\partial}{\partial \mathcal{B}}- \mathcal{B}^2 + c^2 - \sigma \Omega \right)\Phi^{\rm B}(\mathcal{B}) = 0~,
\end{align}
where the superscript `B' denotes that we are considering the bouncing models of our universe. Moreover, we have also introduced the following notation for future convenience,
\begin{align}
\sigma \equiv 2s\mathfrak{h}_s V_{3} \rho_{0} = \frac{8 \pi G_{\rm N} \rho_{0}}{3}~,
\end{align}
where $\sigma$ corresponds to the constant Hubble parameter (squared), had a torsionless universe been filled with fluid with constant energy density $\rho_{0}$. The general solution of the above first order differential equation in the momentum space has the following form
\begin{align}
\label{eq:WDW_solution_bounce1}
\Phi^{\rm B}(\mathcal{B}) = \phi'_{0} \exp \left[\frac{i}{\mathfrak{h}_{s} \sigma }\left(\frac{\mathcal{B}^3}{3}+\left(\sigma \Omega -c^2\right)\mathcal{B}\right)\right]~,
\end{align}
where, $\phi'_{0}$ is again a constant of integration. However, unlike the case of an inflationary universe discussed in \ref{sec:inflationary_scenario_WDW}, where we have used a boundary condition corresponding to the no-boundary proposal to determine the integration constant, for the present case of a bouncing scenario, no such proposal exists. We shall leave this integration constant unfixed, for the moment, only to return to it during the discussion of path integral quantization of the corresponding problem in \ref{sec:PI_quantization_bouncing_scenario}. The wave function in the coordinate space representation can be obtained, again, by Fourier transformation, which results into
\begin{align}\label{eq:bounce_wavefunction}
    \Psi^{\rm B}(q) = \phi'_{0} \sqrt{\frac{2 \pi}{\mathfrak{h}_s}} \left(\mathfrak{h}_s \sigma\right)^{\frac{1}{3}}  \text{Ai}\left[\left(\frac{1}{\mathfrak{h}_s \sigma}\right)^{\frac{2}{3}}\left(\sigma 
   \Omega -c^2 - q \sigma \right)\right]~.
\end{align}
The above wave function, in the absence of torsion, is consistent with what we had obtained for the analog of the no-boundary wave function in the bouncing cosmologies in \cite{PhysRevD.103.106008}. Therefore, the coordinate space wave function $\Psi^{\rm B}(q)$ derived above describes the quantum state for bouncing scenarios in the presence of torsion. Further, in the semi-classical limit $\mathfrak{h}_{s}\to 0$, and in the classically allowed domain $q > \Omega-\frac{c^2}{\sigma}$, the bouncing wave function also takes a Cosine form, and a relevant pre-factor depending on the choice of $\phi'_{0}$ (to be discussed in \ref{sec:PI_quantization_bouncing_scenario}).

\section{Path integral quantization}
\label{sec:PIquantization}

The alternative to obtaining the quantum state of the universe by solving the Wheeler-DeWitt equation corresponds to the Path integral quantization procedure. We first discuss the general setup and subsequently, we shall concentrate on the specific cases of inflationary and bouncing scenarios.

\subsection{General setup}

As in the case of a point particle, where both the canonical as well as the path integral formalism for quantization can be developed, in an analogous manner here also the transition amplitude between any two equal-time hypersurfaces of the universe can be determined using the path integral formalism. Due to the assumption of homogeneity and isotropy, each of these equal-time hypersurfaces can be characterized by simply the scale factors corresponding to such hypersurfaces. In some cases, however, it might be beneficial to characterize a hypersurface by the Hubble rate or by a combination of the scale factor and the Hubble rate at that hypersurface \cite{PhysRevD.100.123543,PhysRevLett.122.201302,PhysRevD.106.023511}. Formally, one evaluates such a quantum transition amplitude by summing over all possible histories between two given initial and final 3-hypersurfaces separated by finite proper time interval, with each history being weighted by $e^{i\mathcal{A}}$, where $\mathcal{A}$ is the total (gravity+matter) action for the system. Subsequently one integrates over all possible time intervals separating the boundary hypersurfaces as well in order to exhaust the time reparametrization invariance. The gravitational action, in general, assumes the following form \cite{PhysRevD.25.3159}:
\begin{align}
\mathcal{A}_{\rm gravity}=\int \left(\pi^{ij} \dot{h}_{ij} - N \mathcal{H} - N^{i} \mathcal{H}_{i}\right) {\rm d}^3 x {\rm d}t ~,
\end{align}
where, $h_{ij}$ is the 3-metric on these spatial hypersurfaces characterized by constant scale factors, and $\pi^{ij}$ is the canonically conjugate momenta to $h_{ij}$. Furthermore, $N$ and $N^i$ are the lapse and the shift functions, respectively; and $\mathcal{H}$ and $\mathcal{H}^i$ turn out to be Hamiltonian and momenta constraints when the action is varied with respect to $N$ and $N^i$, respectively, as the lapse and shift functions appear as Lagrange multipliers in the gravitational action. Due to the presence of constraints in the system, it is necessary to include ghosts and gauge-fixing conditions in the path integral quantization so that the over-counting of histories that are equivalent to each other due to the gauge symmetry can be prevented. In the case of minisuperspace model(s) the gravitational path integral simplifies enormously. After the gauge conditions $\dot{N}=0=N^i$ are enforced and ghost degrees of freedom are integrated out, the path integral for the transition amplitude between the initial and the final boundary hypersurfaces ${}^{(3)}\Sigma_{\rm i}\left(\mathsf{p},q\right)$ and ${}^{(3)}\Sigma_{\rm f}\left(\mathsf{p},q\right)$ can be written formally as follows \cite{PhysRevD.25.3159,PhysRevD.38.2468},
\begin{align}
\label{eq:PI_definition}
    \left\langle {}^{(3)}\Sigma_{\rm f}\left(\mathsf{p},q\right) \bigg| {}^{(3)}\Sigma_{\rm i}\left(\mathsf{p},q\right) \right\rangle =  \int (t_{\rm f}-t_{\rm i}) {\rm d} N  \int_{{}^{(3)}\Sigma_{\rm i},t_{\rm i}}^{{}^{(3)}\Sigma_{\rm f},t_{\rm f}} \mathcal{D}[\mathsf{p}] \mathcal{D} [q] ~ \exp \left[\frac{i}{\hbar} \int_{t_{\rm i}}^{t_{\rm f}} {\rm d} t \left(\mathsf{p}\dot{q} - N \mathcal{H}\left(\mathsf{p},q\right)\right)\right]~,
\end{align}
where, $q$ is the minisuperspace `coordinate' and $\mathsf{p}$ is the conjugate momentum, with $\mathcal{H}(\mathsf{p},q)$ being the Hamiltonian, and $\mathcal{D}[\mathsf{p}]\mathcal{D}[q]$ corresponding to the Liouville path integral measure. Without any loss of generality, we can choose the coordinate times for the initial and final hypersurfaces, such that $0=t_{\rm i} \leq t \leq t_{\rm f} = 1$ and hence the final integral becomes an integral over the lapse function alone. Since the proper time separating the two boundary hypersurfaces can be defined as $\tau\equiv \int_{t_{\rm i}=0}^{t_{\rm f}=1} N {\rm d}t $, thanks to the gauge fixing condition $\dot{N}=0$, the proper time becomes $\tau=N$. As a result, the final integration in \ref{eq:PI_definition} reduces to an ordinary integral over the lapse function $N$ which is then interpreted as integrating over all possible proper time separation between the boundary hypersurfaces. In the literature, it has been shown that the lapse function integral over the whole infinite range $(-\infty,\infty)$, indeed, produces a solution of the quantum Hamiltonian constraint or, the Wheeler-DeWitt equation \cite{PhysRevD.38.2468}. \textcolor{red}{}. We notice that given the torsionful action in \ref{eq:PalatiniAction}, insofar as one considers torsion parameter $c$ to be only a label like spatial curvature index $\mathcal{K}$, the path integral is structurally similar to what has been already discussed in \cite{PhysRevD.38.2468,PhysRevD.39.2206} in great detail. Therefore, we shall not, here, repeat the process of quantum fluctuation integration, and rather assume that in the semi-classical limit $\hbar \to 0$, we can approximate the wave function of the system from the path integral in the following manner
\begin{align} 
\Psi \sim \int {\rm d}N\int \mathcal{D} \mathcal{B} \mathcal{D}q e^{\frac{i}{\hbar}\left(\mathcal{A}_{\rm Palatini}+\mathcal{A}_{\rm matter}\right)} \propto \int_{\mathcal{C}} {\rm d} N f(N) e^{\frac{i}{\hbar}\mathcal{A}_{\rm cl}(\bar{q},N)} ~,
\end{align}
where, $\mathcal{A}_{\rm cl}$ is the total classical action evaluated for the classical trajectory $\bar{q}$ (the solution of the dynamical Einstein equations), and $f(N)$ carries the contribution from the quantum fluctuation integrals. In the case of Dirichlet boundary conditions imposed on the path integral, the function $f(N)$ will be proportional to $1/\sqrt{N}$, while for Dirichlet-Neumann mixed boundary conditions the function $f(N)$ reduces to a simple numerical factor \textcolor{red}{(see, \cite{PhysRevD.39.2206,PhysRevD.102.086011})}. Since our interest will never be on the Dirichlet boundary conditions alone and we consider the semiclassical limit $\hbar\to0$ when performing the lapse integral, for all intents and purposes we shall ignore the prefactor $f(N)$. Further, to perform the oscillatory integral over the lapse function, we shall extend the domain of the lapse function to the complex plane. In which case the part of the integrand, namely $e^{{\rm Re}[i\mathcal{A}_{\rm cl}]}$, will provide the dominant contribution and we are interested in its behavior. Therefore, we shall focus most of our attention on finding the appropriate complex integration contour $\mathcal{C}$, in accordance with the Picard-Lefschetz theory, for which the convergence of the above oscillatory integration becomes most apparent and the integration is approximated by means of the steepest descent method. Moreover, we shall also compute the saddle points of the above action and find out the behavior of scalar perturbations around the saddle point geometry in order to analyze the stability of these saddle point geometries.

In order to generate solutions in different representations, that is either in connection or coordinate space representations, we shall have to carefully choose the initial and final hypersurfaces in the path integral problem. This will require us to add different boundary terms to the action, as we shall see below.

\subsection{Inflationary scenario}

Let us start by computing the inflationary minisuperspace path integral to determine the wave function in the momentum representation, so that it can be compared with the one obtained in \ref{eq:WDW_solution_inflation} using the Wheeler-DeWitt equation. To do this, as the wave function is characterized by the conjugate momentum, we shall have to choose a Neumann condition on the final boundary in the path integral, i.e., we need to sum over paths for which the momentum at the final hypersurface is fixed to a certain value. For the initial boundary, we can either choose a Dirichlet condition or a Robin boundary condition. However, we will demonstrate that choosing an initial Dirichlet condition leads to a saddle point geometry, which is unstable under perturbations. Stable saddle point(s) can be achieved, in this case, only by using Robin boundary conditions.

On the other hand, generating a solution in the coordinate representation requires us to sum over paths for which at the final hypersurface, the Dirichlet boundary condition is being used. Then for the initial boundary, either a Neumann or a Robin condition will be necessary to obtain stable perturbations around the saddle point geometry. An initial Dirichlet boundary condition generically leads to an unstable saddle point geometry (see, \cite{PhysRevLett.119.171301,PhysRevD.97.023509,PhysRevD.100.063517}). In the following, we proceed to discuss the various boundary conditions we mentioned above.

\subsubsection{Momentum space wave function: Dirichlet-Neumann boundary condition} \label{sec:Inf_Diri_Neu}

\begin{figure}[ht!]
    \centering
    \includegraphics[width=\textwidth]{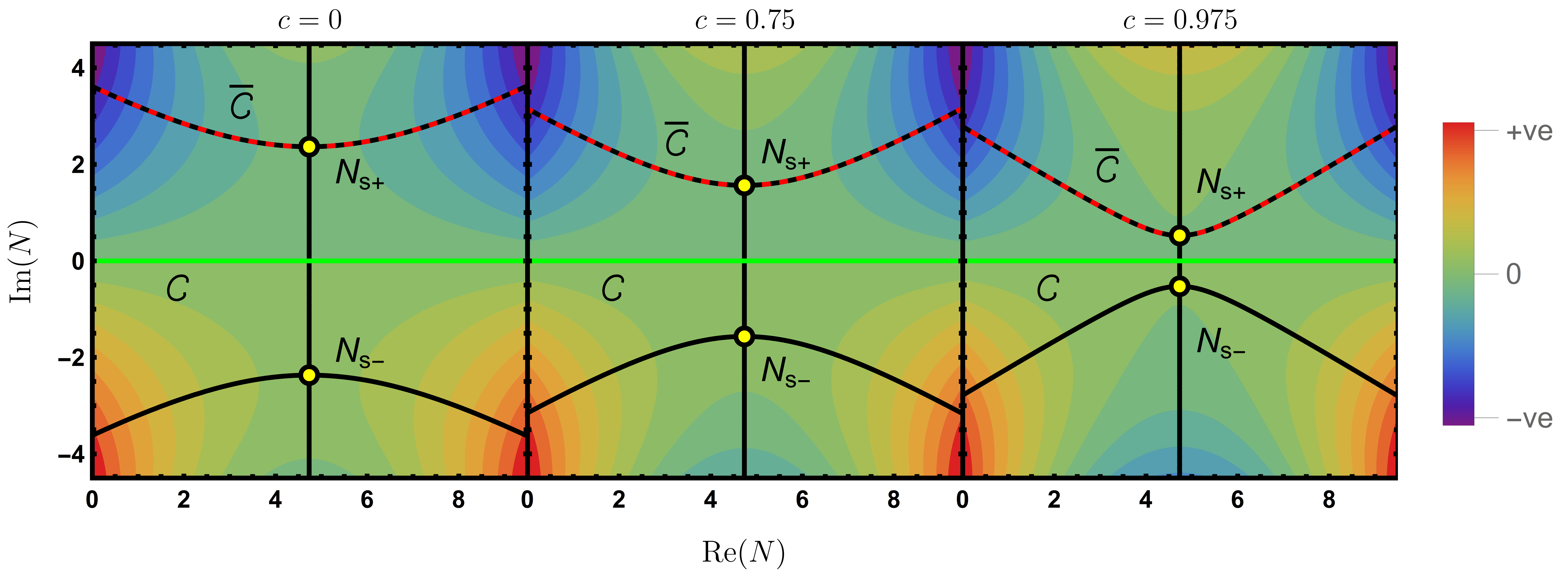}
    \caption{We have plotted the ${\rm Re}[i\mathcal{A}_{\rm cl}]$ in the plane of the complex lapse function for different choices of the torsion parameter $c$. The leftmost plot is for zero torsion, while the middle plot is for an intermediate value of the torsion and the rightmost plot is for a large value of the torsion. In all the plots, the asymptotic regions of convergence (${\rm Re}[i\mathcal{A}_{\rm cl}]<0$) have been shown in shades of blue, whereas the complementary regions, where (${\rm Re}[i\mathcal{A}_{\rm cl}]>0$), have been shown in shades of red. The two saddle points $(N_{\rm s\pm})$ have been presented as yellow dots. The black curves are the steepest descent/ascent flow lines passing through the saddle points. $\mathcal{C}$ corresponds to the original integration contour (green line), which is deformed into $\bar{\mathcal{C}}$ (red dashed line) such that it passes through the saddle point $N_{\rm s+}$ and runs along the Lefschetz thimbles starting and ending in the regions of convergence.}
    \label{fig:inf_Dirichlet_Neumann}
\end{figure}

In the semiclassical approximation, the path integral is dominated by the classical action, that is, the action evaluated at the classical path or, the trajectory. However, the classical solution must respect the boundary condition set in the problem, in this case, an initial Dirichlet condition and a final Neumann condition. As the classical solution is obtained from the variation of the action, to make the variational problem well-defined for the boundary conditions we want to impose, we shall have to add a boundary term to the action, such that the final action becomes,
\begin{align}\label{eq:inf_total_action} 
\mathcal{A}_{\rm final} =\frac{3V_{3}}{8\pi G_{\rm N}}\int_{0}^{1} {\rm d} t \left[-\dot{q}\mathcal{B} - N\left(-\mathcal{B}^2 - \mathcal{K}  + c^2 + \frac{\Lambda}{3}q\right)\right]+\frac{3V_{3}}{8\pi G_{\rm N}}\Bigg[q(1)\mathcal{B}(1)-q(0)\mathcal{B}(0)\Bigg]+\mathcal{A}_{\rm Boundary}~, 
\end{align}
where, we have used the Hamiltonian presented in \ref{hamiltonianinf}, derived in the inflationary scenario. Since we wish to impose an initial Dirichlet and final Neumann condition (the final Neumann condition is motivated by the fact that we are interested in a wave function in the momentum representation), the following boundary term should be added to the action $\mathcal{A}_{\rm Boundary} =(3 V_{3}/8\pi G_{\rm N}) q(0) \mathcal{B}(0)$. Such that the variation of the action $\mathcal{A}_{\rm final}$ reads,
\begin{align} 
\delta \mathcal{A}_{\rm final}&=\frac{3 V_{3}}{8\pi G_{\rm N}}\int_{0}^{1} {\rm d} t 
\left[\Big(-\dot{q}+2N\mathcal{B}\Big)\delta\mathcal{B} + \left(\dot{\mathcal{B}} - N\frac{\Lambda}{3}\right)\delta q - \left( - \mathcal{B}^2 - \mathcal{K}_{c} + \frac{\Lambda}{3} q \right) \delta N \right]
\nonumber
\\ 
&+\frac{3 V_{3}}{8\pi G_{\rm N}}\left[q(1)\delta \mathcal{B}(1)+\mathcal{B}(0)\delta q(0) \right]~.
\end{align}
As evident from the above variation, one needs to fix the momenta $\mathcal{B}$ at the final hypersurface and the scale factor $q$ at the initial hypersurface, which justifies the boundary term added to the action. We have defined the following quantity, $\mathcal{K}_{c}\equiv \mathcal{K}-c^{2}$, and hence this quantity captures the information regarding the torsional degree of freedom. In addition to spelling out the appropriate boundary contributions, the variation of the above action also yields the following equations of motion for the variables $q$, $\mathcal{B}$ and $N$ respectively,
\begin{align}
\dot{\mathcal{B}}&=N\frac{\Lambda}{3}~, 
\label{eq:inflation_dynamics1} 
\\
\dot{q}&=2 N \mathcal{B}~,
\label{eq:inflation_dynamics2}
\\
\mathcal{B}^2&+\mathcal{K}_{c}-\frac{\Lambda}{3}q=0~.
\label{eq:inflation_constraint}
\end{align}
where \ref{eq:inflation_dynamics1} and \ref{eq:inflation_dynamics2} captures the dynamics of the system, while \ref{eq:inflation_constraint} is a constraint equation. Further, the boundary conditions imposed on the initial and the final hypersurface require the following
\begin{align}\label{eq:inf_DirichletNeumann} 
\mathcal{B}(1)\equiv \mathcal{B}_{1}~;
\qquad 
q(0)\equiv q_{0}~. 
\end{align}
The path integral over the conjugate momentum and the scale factor is dominated by the classical solution $\bar{q}(t)$ of the dynamical equations presented in \ref{eq:inflation_dynamics1} and \ref{eq:inflation_dynamics2}, such that the solution satisfies the boundary conditions in \ref{eq:inf_DirichletNeumann}. The classical solution is obtained by taking a time derivative of \ref{eq:inflation_dynamics2} and then substituting $\dot{\mathcal{B}}$ from \ref{eq:inflation_dynamics1}, thereby obtaining a second-order differential equation for the scale factor. The solution consistent with the boundary conditions \ref{eq:inf_DirichletNeumann} has the following form
\begin{align}\label{eq:inf_class_sol_diri_neu} 
\bar{q}(t) & = N^2 \frac{\Lambda}{3} t(t - 2) + 2N\mathcal{B}_{1} t + q_{0}~. 
\end{align}
Evaluation of the final action, as in \ref{eq:inf_total_action}, at the classical solution $\bar{q}(t)$ yields
\begin{align}\label{eq:inflation_action_Dirichlet_Neumann} 
\mathcal{A}_{\rm cl}(N)\equiv \mathcal{A}_{\rm final}[N,\bar{q}] = & \frac{3 V_{3}}{8\pi G_{\rm N}} \left[ N^3 \frac{\Lambda^2}{27} -N^2 \frac{\Lambda}{3}\mathcal{B}_{1} + N \mathcal{B}_{1}^2 + N \mathcal{K}_{c} - N\frac{\Lambda}{3} q_{0} + \mathcal{B}_{1} q_{0}\right]~. 
\end{align}
Therefore, determining the wave function of our universe in the momentum space representation, given the scale factor on an initial hypersurface, amounts to calculating an ordinary integral of the exponential of the classical action over the lapse function, such that
\begin{align}\label{reduced_int_inf}
\Phi_{q_{0}}^{\rm I}(\mathcal{B}_{1})= \int_{-\infty}^{\infty} {\rm d} N e^{\frac{i}{\hbar}\mathcal{A}_{\rm cl}(N)} ~.
\end{align}
Above oscillatory integral can be evaluated using the method of steepest descent, in the semiclassical limit $\hbar\to0$. For this purpose, one extends the domain of the lapse function $N$, originally the entire real line, to the complex plane. As a result, the classical action becomes complex as well, $\mathcal{A}_{\rm cl} = {\rm Re}[\mathcal{A}_{\rm cl}] + i {\rm Im}[\mathcal{A}_{\rm cl}]$, and hence the integrand becomes, $\exp(i{\rm Re}[\mathcal{A}_{\rm cl}])\exp(-{\rm Im}[\mathcal{A}_{\rm cl}])$. As evident, the above integration will be convergent for large values of the lapse function, if we have ${\rm Im}[\mathcal{A}_{\rm cl}]>0$, as $|N|\to\infty$. Thus, the deformed contour of integration on the complex lapse function plane must start and end in those regions, where ${\rm Im}[\mathcal{A}_{\rm cl}]>0$. Explicitly, in the large lapse function limit ($|N|\to\infty$), these regions are such that ${\rm Arg}\left[N-\left(3/\Lambda\right)\mathcal{B}_{1}\right] \in \left(0,\pi/3\right)\cup\left(2\pi/3,\pi\right)\cup\left(4\pi/3,5\pi/3\right)$ and have been depicted in \ref{fig:inf_Dirichlet_Neumann} with shades of dark blue color.

The next step is to employ the Picard-Lefschetz theory in order to identify the flow lines (also known as Lefschetz thimbles) starting from the saddle point(s) of $\mathcal{A}_{\rm cl}(N)$, along which the oscillatory part of the integrand, namely $\exp(i{\rm Re}[\mathcal{A}_{\rm cl}])$ freezes to a constant and the real part $\exp(-{\rm Im}[\mathcal{A}_{\rm cl}])$ becomes an exponentially decaying function. Then, if one continuously deforms the contour running along the real line towards the Lefschetz thimble(s), the above oscillatory integral becomes convergent and will be dominated by the integrand evaluated at the saddle point(s). In the present context, there are two saddle points of $\mathcal{A}_{\rm cl}(N)$ in the lapse plane, both of which can be determined by solving the equation $(\partial \mathcal{A}_{\rm cl}[\bar{q}(N)]/\partial N)_{N_{\rm s}}=0$, yielding
\begin{align}\label{eq:inf_Diri_Neu_saddle} 
N_{\rm s\pm} = \frac{3}{\Lambda}\left(\mathcal{B}_{1} \pm \sqrt{ - \mathcal{K}_{c} + \frac{\Lambda}{3}q_{0}} \right)~.
\end{align}
The location of these saddle points depends on the specific choices for the parameter values of $\mathcal{K}$, $c$, and $q_{0}$. If we desire the path integral to be dominated by a Hartle-Hawking type geometry, which corresponds to a closed ($\mathcal{K}=1$) universe starting from a zero size, setting $q_0=0$ is a reasonable choice. Moreover, a closed universe is possible only if the squared value of the parity-odd torsion component is positive and cannot exceed $\mathcal{K}$, that is, $\mathcal{K}_{c}>0$. In that case, the saddle points are complex. The action evaluated at these saddle points reads
\begin{align} 
\mathcal{A}_{\rm cl}(N_{\rm s\pm}) = & \frac{3V_{3}}{8 \pi G_{\rm N} \Lambda} \left(\mathcal{B}_{1}^3 + 3 \mathcal{K}_{c} \mathcal{B}_{1} \pm i 2\mathcal{K}_{c}^{\frac{3}{2}} \right)~. 
\end{align}
The only remaining bit corresponds to the choice of the saddle point. It is clear from \ref{fig:inf_Dirichlet_Neumann} that the real line contour $\mathcal{C}$ has to be deformed into a new contour ($\bar{\mathcal{C}}$) in the complex plane running along the Lefschetz thimble through $N_{\rm s+}$, to ensure the convergence of the integral. This is because the other contour passing through $N_{\rm s-}$ runs along the steepest ascent contours and the integral for this choice does not converge. As a result, the integral in \ref{reduced_int_inf} is dominated by the upper saddle point, located at $N_{\rm s+}$. Thus the semi-classical wave function in the momentum representation can be expressed as
\begin{align}
\label{eq:path_Dirichlet_Neumann_inflation}
    \Phi^{\rm I}(\mathcal{B}_{1}) \sim \exp\left(\frac{i}{\hbar}\mathcal{A}_{\rm cl}(N_{s+})\right) = \exp \left[i\frac{3}{\mathfrak{h}\Lambda}\left(\frac{\mathcal{B}^3_1}{3}+\mathcal{K}_{c}\mathcal{B}_1\right) -\frac{2}{\mathfrak{h}\Lambda}\mathcal{K}_{c}^{\frac{3}{2}}\right]~.
\end{align}
Therefore, from the path integral approach, we have obtained the connection wave function as in \ref{eq:WDW_solution_inflation}, but with the normalization factor $\phi_0$, which has an opposite sign in the exponent compared to what is expected from the Hartle-Hawking no-boundary condition. We shall show below (in \ref{sec:inf_perturbation_stability_Dir_Neu}) that the perturbation
around the saddle point geometry corresponding to $N_{\rm s+}$ is unstable, and hence this wave function, even though mathematically correct, is not physical and we must explore other boundary conditions.

Note that the thimbles are defined as contours on the plane of complex lapse function, for which the ${\rm Re}[\mathcal{A}_{\rm cl}]$ remains constant. For example, the thimbles relevant for the present context originating at the saddle point $N_{\rm s+}$ and flowing into the regions of convergence, satisfy the relation ${\rm Re}[\mathcal{A}_{\rm cl}]={\rm Re}[\mathcal{A}_{\rm cl}({N_{s+}})]$. Since $N_{\rm s+}$ is a saddle point of the action, it follows that slightly away from $N_{\rm s+}$, along the contour of steepest descent, we have
\begin{align}
{\rm Re} \left[\frac{1}{2}\left(\partial^2_{N}\mathcal{A}_{\rm cl}\right)_{N_{\rm s+}} \left(\delta N\right)^2\right] = 0~.
\end{align}
Expressing, $\delta N$ as $\delta N\equiv \boldsymbol{|} \delta N \boldsymbol{|} e^{i \alpha}$, where $\alpha$ is the angle subtended by the thimble with respect to the real line, near the saddle point, and using \ref{eq:inflation_action_Dirichlet_Neumann} and \ref{eq:inf_Diri_Neu_saddle} to express $(\partial^2_{N}\mathcal{A}_{\rm cl})$ at the saddle point, reduces the above condition to
\begin{align}
{\rm Re} \left[\boldsymbol{|} \delta N \boldsymbol{|}^2 i\frac{\Lambda}{3}\sqrt{\mathcal{K}_{c}} e^{2i\alpha}\right] = 0 ~,
\end{align}
Since we have $0<\mathcal{K}_{c}<1$, the above equation implies $\sin 2\alpha = 0$. Thus the thimble at the saddle point subtends an angle $\alpha = n \pi,~ n\in \mathbb{Z}$, where $n=0,1$ corresponds to the thimble originating at $N_{\rm s+}$ and flowing down to right and left, respectively. On the other hand, much away from the saddle point we have ${\rm Re} \left[\boldsymbol{|} \delta N \boldsymbol{|}^3 \left(\partial_{N}^3 \mathcal{A}_{\rm cl} \right)_{N_{\rm s+}} e^{3i\alpha}\right] = 0$. As $\left(\partial_{N}^3 \mathcal{A}_{\rm cl} \right)_{N_{\rm s+}}$ is a real constant, this equation immediately implies that the thimbles run to infinity at an angle $\left(4n \pm 1\right)\pi/6$. Visual inspection reveals thimble originating at $N_{\rm s+}$ and flowing down to the right runs to infinity at an angle $\pi/6$. This completes our study of the path-integral derivation of the momentum space wave function using the Dirichlet-Neumann boundary condition. We will now specialize to a different boundary condition, leading to a modified momentum space wave function with the correct sign of the exponential prefactor in the semi-classical limit.  

\subsubsection{Momentum space wave function: Robin-Neumann boundary condition}\label{sec:Robin_Neumann_inflation}

In the previous section, we obtained the wave function in the momentum representation, with Dirichlet-Neumann boundary conditions. It is possible to generate another wave function in the momentum representation by setting a Robin boundary condition on the initial hypersurface, instead of the Dirichlet boundary condition. For this purpose, we may consider the following boundary term
\begin{align} 
\mathcal{A}_{\rm Boundary} = - \frac{V_{3}}{8\pi G_{\rm N}}\sqrt{\frac{\Lambda}{3}} q^{\frac{3}{2}}(0)~, \end{align}
the inclusion of which in the action given in \ref{eq:inf_total_action}, yields upon variation the following
\begin{align} 
\delta \mathcal{A}_{\rm final}&=\frac{3V_{3}}{8\pi G_{\rm N}}\int_{0}^{1} {\rm d} t \left[ \left(-\dot{q} + 2N\mathcal{B} \right)\delta \mathcal{B} + \left(\dot{\mathcal{B}} - N\frac{\Lambda}{3}\right)\delta q - \left( - \mathcal{B}^2 - \mathcal{K}_{c} + \frac{\Lambda}{3} q \right) \delta N \right]  
\nonumber
\\ 
&\qquad +\frac{3V_{3}}{8\pi G_{\rm N}}\left[- q(0)\delta \left(\mathcal{B}(0) +\sqrt{\frac{\Lambda}{3}} \sqrt{q(0)}\right) + q(1) \delta \mathcal{B}(1)\right]~. 
\end{align}
\begin{figure}[ht!]
    \centering
    \includegraphics[width=\textwidth]{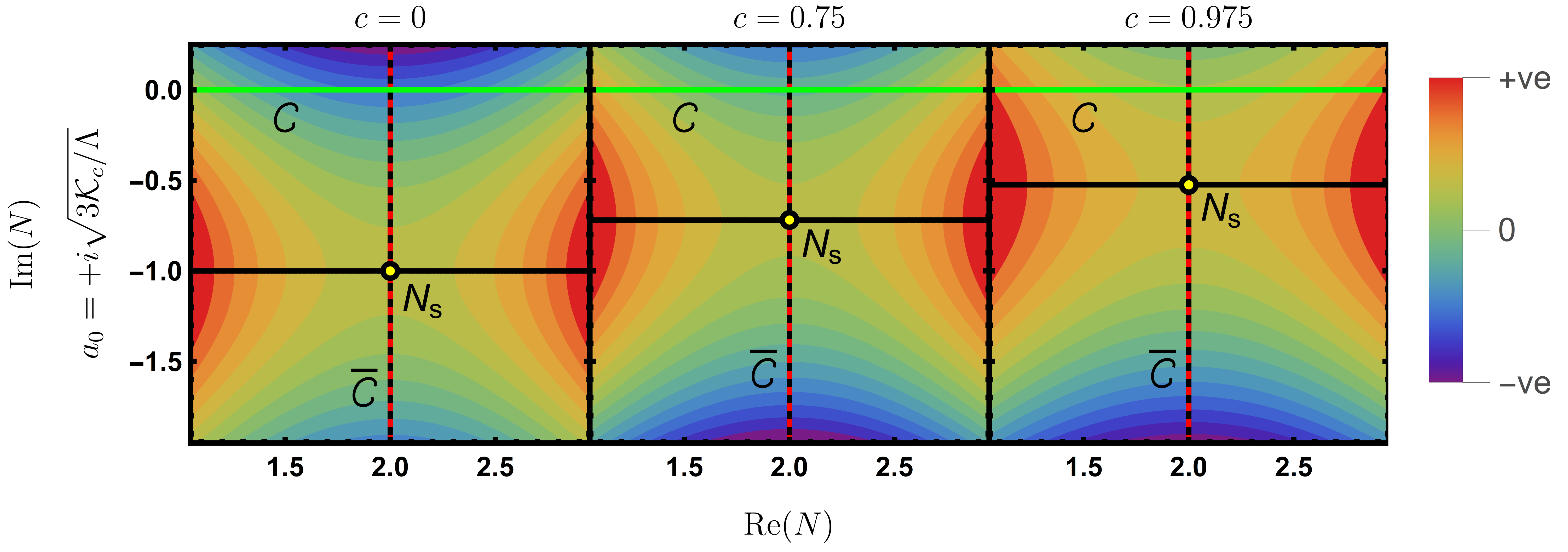}
    \caption{Contours of ${\rm Re}[i\mathcal{A}_{\rm cl}]$ in the plane of the complex lapse function have been presented, for different choices of the torsion parameter $c$. In all of these plots, the regions for which ${\rm Re}[i\mathcal{A}_{\rm cl}]$ is positive(negative) are shown in shades of dark red(blue). The regions of convergence lie with the wedges with asymptotic bounds ${\rm Arg}\left[N-\left(3/\Lambda\right)\left(\mathcal{B}_{1}-i\sqrt{\mathcal{K}_{c}}\right)\right] \in \left(\pi/4,3\pi/4\right)\cup\left(5\pi/4,7\pi/4\right)$, for the choice $a_{0}=+i\sqrt{3\mathcal{K}_{c}/\Lambda}$. The only saddle point $N_{\rm s}$ for this choice of $a_{0}$ is located on the lower half of the complex plane. The black lines correspond to the steepest ascent/descent flow lines passing through the saddle point. The original contour of integration is represented by $\mathcal{C}$ and corresponds to the green line, on the other hand, $\bar{\mathcal{C}}$ depicts the actual contour of integration passing through the only saddle point $N_{\rm s}$ (red dashed line).}
    \label{fig:inf_Robin_Neumann}
\end{figure}
As evident from the variation, for consistency, we need to fix the conjugate momentum on the final surface (characterized by $t_{f}=1$), corresponding to the Neumann boundary condition, and on the initial hypersurface (characterized by $t_{i}=0$) we fix a combination of $\mathcal{B}$ and $q$, leading to a Robin boundary condition. When explicitly spelled out, these conditions read
\begin{align}
\label{eq:inf_boundary_conditions_Robin}
\mathcal{B}(1)\equiv\mathcal{B}_{1}~; 
\qquad 
\mathcal{B}(0) +\sqrt{\frac{\Lambda}{3}} \sqrt{q(0)}\equiv\sqrt{\frac{\Lambda}{3}} a_{0}~, 
\end{align}
where $a_{0}$ is, in general, a complex number, for which we shall choose an appropriate value later. For an interesting interpretation of the above boundary conditions, see \cite{PhysRevD.106.023511}. Again, the phase space path integral will be dominated by the solution of the dynamical Einstein equations (\ref{eq:inflation_dynamics1} and \ref{eq:inflation_dynamics2}) satisfying the boundary conditions \ref{eq:inf_boundary_conditions_Robin} and its explicit form is the following
\begin{align} \label{eq:class_sol_robin_neumann}
\bar{q}(t) = {\frac{\Lambda}{3}}N^2 t^2 +2N t \left(\mathcal{B}_{1} - {\frac{\Lambda}{3}} N\right) 
+ {\frac{3}{\Lambda}}{\left[ \sqrt{\frac{\Lambda}{3}} \left(a_{0}+ \sqrt{\frac{\Lambda}{3}} N\right)- \mathcal{B}_{1}\right]^2}~.
\end{align}
Like before, in this case, as well, we can identify the saddle point(s) of the action evaluated on the classical trajectory, by solving the equation $(\partial \mathcal{A}_{\rm cl}[\bar{q}(N)]/\partial N)_{N_{\rm s}}=0$. This yields, a single saddle point given by
\begin{align} \label{eq:saddle_robin_neumann}
N_{\rm s}=\frac{a_0 \sqrt{\frac{\Lambda}{3}} \left(2 \mathcal{B}_{1} - a_{0} \sqrt{\frac{\Lambda}{3}}\right) + \mathcal{K}_c}{2 a_{0} \left(\frac{\Lambda}{3}\right)^\frac{3}{2}}~.
\end{align}
If we wish to have a wave function that corresponds to the Hartle-Hawking no-boundary proposal, then it is natural to impose the boundary condition that the mini-superspace geometry at the above saddle point must start from a zero size. This condition can be imposed by setting $\bar{q}(t=0)\big|_{N_{\rm s}} = 0$, which in turn, fixes the parameter $a_{0}$ to have the following values
\begin{align}
a_{0} = \pm i\sqrt{\frac{3}{\Lambda}}\sqrt{\mathcal{K}_{c}}~.
\end{align}
Therefore, the classical action evaluated at the saddle point, but with different choices for the initial Robin boundary condition, characterized by $a_{0}$, reads
\begin{align} 
\mathcal{A}_{\rm cl}\left[N_{\rm s};a_{0}=\pm i\sqrt{3\mathcal{K}_{c}/\Lambda}\right] = & \frac{3V_{3}}{8 \pi G_{\rm N} \Lambda} \left(\mathcal{B}_{1}^3 + 3 \mathcal{K}_{c} \mathcal{B}_{1} \mp 2i \mathcal{K}_{c}^{\frac{3}{2}}\right)~. 
\end{align}
We shall show below (in \ref{sec:inf_perturbation_stability_Dir_Neu}) that the perturbation
around the saddle point geometry corresponding to the choice $a_{0}=-i\sqrt{3\mathcal{K}_{c}/\Lambda}$ is unstable, and hence this choice is not physical. We shall proceed with the positive imaginary choice for $a_{0}$. To find out the wave function in the momentum space, we need to figure out which contour on the complex lapse function plane should $\mathcal{C}$, which is the original contour along the real lapse function axis, be deformed to. As evident from \ref{fig:inf_Robin_Neumann}, there are only two possible contours that pass through the saddle point and along which the ${\rm Re}[\mathcal{A}_{\rm cl}]$ is a constant. The vertical one among these two contours corresponds to the steepest descent contour, as it starts and ends in regions where ${\rm Im}[\mathcal{A}_{\rm cl}]$ remains positive. Given this deformation contour $\bar{\mathcal{C}}$ (red dashed line), the momentum space wave function can be immediately obtained, since in the semiclassical limit, the wave function is given by the classical action at the saddle point. Therefore, the momentum space wave function assumes the following form
\begin{align}
\Phi^{\rm I}\left(\mathcal{B}_{1};a_{0}=+ i\sqrt{3\mathcal{K}_{c}/\Lambda}\right) & \sim \exp\left(\frac{i}{\hbar}\mathcal{A}_{\rm cl}\left[N_{s},a_{0}=+ i\sqrt{3\mathcal{K}_{c}/\Lambda}\right]\right) 
=\exp \left[i\frac{3}{\mathfrak{h}\Lambda}\left(\frac{\mathcal{B}^3_1}{3}+\mathcal{K}_{c}\mathcal{B}_1\right) +\frac{2}{\mathfrak{h}\Lambda}\mathcal{K}_{c}^{\frac{3}{2}}\right]~.
\end{align}
Thus the choice, $a_0 = + i\sqrt{3\mathcal{K}_{c}/\Lambda}$, leads to an exponential factor in the front, growing with a decrease in $\Lambda$, as per the no-boundary wave function. Moreover, for the choice $a_0 = + i\sqrt{3\mathcal{K}_{c}/\Lambda}$, it follows from \ref{eq:saddle_robin_neumann} that the saddle point is given by the expression $N_{\rm s}=\left(3/\Lambda\right)\left(\mathcal{B}_{1} - i\sqrt{\mathcal{K}_{c}}\right)$. Then it is clear that the position of the saddle point depends on the torsion parameter. As the torsion parameter $c$ increases, $\mathcal{K}_{c}$ decreases, and hence the saddle point approaches the real line (as $\mathcal{B}_{1}$ is real), which can be readily verified from \ref{fig:inf_Robin_Neumann}. It is also to be noted performing a similar calculation as before, the angle of the thimble at the saddle point can be calculated. It turns out that the thimble runs at a constant angle $(2n+1)\pi/2~,~n\in\mathbb{Z}$ throughout, where $n=0,1$ may correspond to the relevant thimbles.

We shall now derive the coordinate space wave function using the path integral method. 

\subsubsection{Coordinate space wave function: Neumann-Dirichlet boundary condition}

Having derived the wave functions in the momentum representation, in the previous sections, with various boundary conditions, here we wish to determine the wave function in the coordinate representation. As evident, this will require the use of boundary conditions, different from those we employed earlier. To be specific, we shall be using a Dirichlet condition (fixed coordinate) for the final boundary hypersurface. Whereas, for the initial boundary hypersurface we have three possibilities --- (a) a Dirichlet condition, (b) a Neumann condition, or (c) a Robin condition. The issue with an initial Dirichlet condition (along with a final Dirichlet condition as well) is that the perturbations become unstable (see \cite{PhysRevLett.119.171301,PhysRevD.97.023509}). We do not expect this result to change even with the inclusion of torsion. Thus we straight away move to discuss the scenario, where we have set Dirichlet and Neumann conditions on the final and initial boundary hypersurfaces, respectively.

In order for the Neumann-Dirichlet boundary condition to be consistent with the classical variational problem, described by the action in \ref{eq:inf_total_action}, we must use the following boundary term,
\begin{align}
\mathcal{A}_{\rm Boundary}=-\frac{3V_{3}}{8\pi G_{\rm N}} q(1) \mathcal{B}(1)~.
\end{align}
Besides getting the same equations of motion, as in \ref{eq:inflation_dynamics1} --- \ref{eq:inflation_constraint}, we will have to fix $q$ at the final hypersurface ($t_{\rm f}=1$) and the conjugate momentum $\mathcal{B}$ at the initial hypersurface ($t_{\rm i}=0$). This allows us to impose the initial Neumann and final Dirichlet boundary conditions as follows
\begin{align}
\label{eq:inf_neu_diri_bdycon}
q(1) = q_{1}~;
\qquad 
\mathcal{B}(0) = \pi_{0}~.
\end{align}
The solution to the classical dynamical equations, as in \ref{eq:inflation_dynamics1} and \ref{eq:inflation_dynamics2},  consistent with the above boundary conditions becomes
\begin{align}
\label{eq:class_sol_inf_neu_diri}
\bar{q}(t) = \frac{\Lambda}{3} N^2 t^2 + 2 N \pi_{0} t + q_{1} - \frac{\Lambda}{3} N^2 - 2 N \pi_{0}~.
\end{align}
As the phase space path integral is dominated by the classical action, which is simply the action evaluated for this classical solution, using \ref{eq:class_sol_inf_neu_diri} we obtain
\begin{align}\label{eq:inf_cl_action_neu_diri}
\mathcal{A}_{\rm cl} = \frac{3V_{3}}{8 \pi G_{\rm N}} \left[\frac{\Lambda^2}{27} N^3 + \frac{\Lambda}{3} \pi_{0} N^2 + \left(\frac{\Lambda}{3} q_{1} - \mathcal{K}_{c} - \pi_{0}^2 \right) N - \pi_{0} q_{1}\right]~.
\end{align}
\begin{figure}[ht!]
    \centering
    \includegraphics[width=\textwidth]{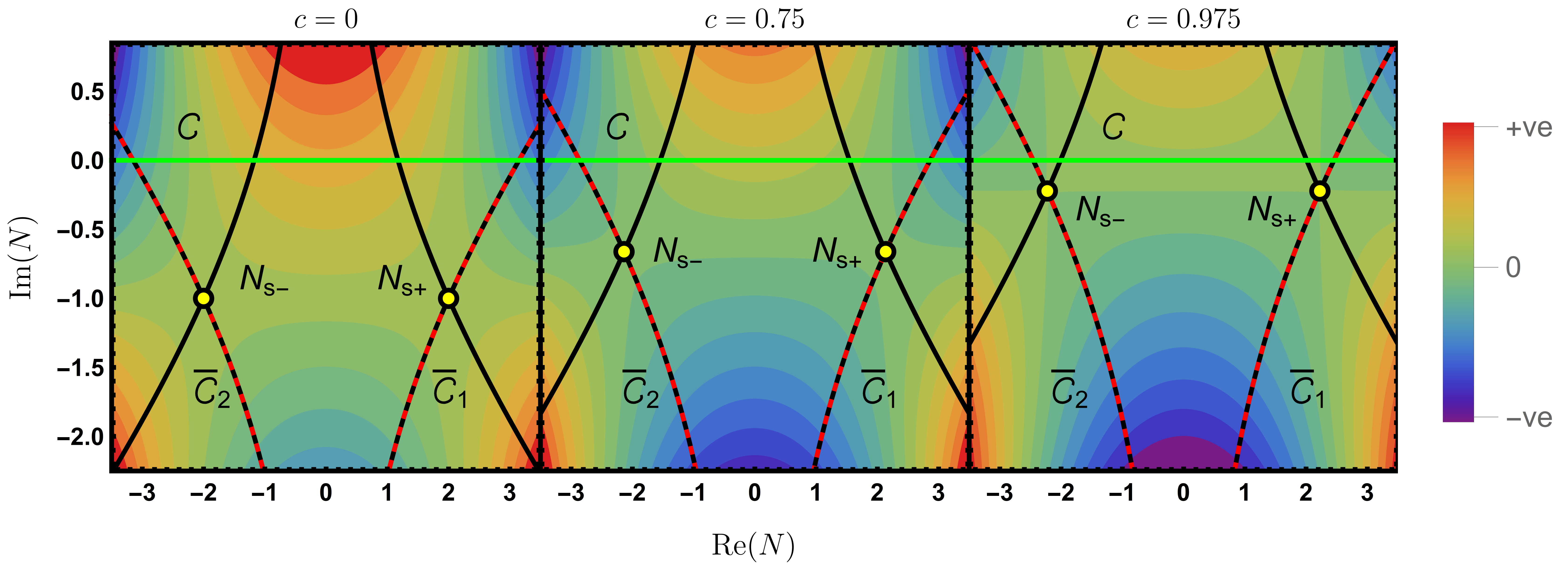}
    \caption{We show the contour plot of ${\rm Re}[i\mathcal{A}_{\rm cl}]$ in the plane of the complex lapse function for different choices of the torsion parameter $c$. In all the plots, the regions where ${\rm Re}[i\mathcal{A}_{\rm cl}]$ is negative(positive) have been shown in shades of blue(red). The regions of convergence asymptotically remain within the wedges ${\rm Arg}\left[N+i\left(3/\Lambda\right)\sqrt{\mathcal{K}_{c}}\right] \in \left(0,\pi/3\right)\cup\left(2\pi/3,\pi\right)\cup\left(4\pi/3,5\pi/3\right)$. The two saddle points $(N_{\rm s\pm})$ have been presented as yellow dots. The black curves are the steepest descent/ascent flow lines passing through the saddle points. $\mathcal{C}$ corresponds to the original integration contour (green line), which is deformed into $\bar{\mathcal{C}}_{2}\cup\bar{\mathcal{C}}_{1}$ (red dashed curves) such that these pass through the saddle points $N_{\rm s\pm}$ and runs along the Lefschetz thimbles starting and ending in the regions of convergence.}
    \label{fig:inf_Neumann_Dirichlet}
\end{figure}
Now, as for the integration over the lapse function, we shall again use the steepest descent approximation, wherein we extend the lapse function from real to a complex variable and then choose a suitable integration contour, along which the ${\rm Re}[\mathcal{A}_{\rm cl}]$ remains constant, whereas ${\rm Im}[\mathcal{A}_{\rm cl}]$ becomes a positive quantity, signaling exponential decay, away from the saddle point (see \ref{fig:inf_Neumann_Dirichlet}). Therefore, along this new contour, the integral is approximated by the saddle points of the action. For the classical action in \ref{eq:inf_cl_action_neu_diri}, there are two saddle points and they are located at,
\begin{align}
\label{eq:saddles_inf_neu_diri}
N_{\rm s\pm} = \frac{3}{\Lambda} \left(-\pi_{0} \pm \sqrt{\frac{\Lambda}{3}q_{1} - \mathcal{K}_{c}} \right)~.
\end{align}
If we require the saddle point geometries to correspond to Hartle-Hawking no-boundary geometry that starts with a zero size, then we must impose the additional condition that $\bar{q}(t=0)\big|_{N_{s\pm}} = 0$. This in turn determines the possible choices for the initial momentum to be
\begin{align}
\label{eq:initial_cond_inf_neu_diri}
\pi_{0} = \pm i\sqrt{\mathcal{K}_{c}}~.
\end{align}
As in the previous example, here also we will work with the positive imaginary solution since the negative imaginary choice for $\pi_0$ leads to unstable perturbations around the saddle point geometries. Due to this choice, the saddle points are located in the lower half complex plane, as evident from \ref{fig:inf_Neumann_Dirichlet}. With a stronger value for the torsion parameter, it follows that $\mathcal{K}_{c}\to0$, and hence $\pi_{0}$ tends to zero as well, thereby shifting the saddle points toward the real $N$ axis. This feature can also be seen by comparing the three plots in \ref{fig:inf_Neumann_Dirichlet}, for different choices of the torsion parameter. Moreover, thimbles at saddle point $N_{\rm s+}$ in the present case satisfy
\begin{align}
{\rm Re} \left[\boldsymbol{|} \delta N \boldsymbol{|}^2 \frac{\Lambda}{3}\sqrt{\frac{\Lambda}{3}q_{1}-\mathcal{K}_{c}} ~ e^{2i\alpha}\right] = 0 ~.
\end{align}
In the classically allowed domain we have $\left(\Lambda/3 \right)q_{1}>\mathcal{K}_{c}$, and thus the expression within the square root is real. Thus the above equation implies $\alpha = \left(4n \pm 1\right)\pi/4,~n\in \mathbb{Z}$, whereas visual inspection reveals the thimble originating at $N_{\rm s+}$ running to the right makes an angle $\pi/4$. On the other hand, towards infinity, the thimble takes an angle $\pi/6$.

Further, as evident from \ref{fig:inf_Neumann_Dirichlet}, unlike previous scenarios, here both the saddle points will contribute, and hence the real integration contour $\mathcal{C}$ has to be deformed into $\bar{\mathcal{C}}_2 \cup \bar{\mathcal{C}}_1$, for which the convergence of the integrand is ensured. Thus the wave function in the coordinate space, approximated by the behavior of the classical action at the saddle points become
\begin{align}
\Psi^{\rm I}(q_{1}) & \sim \sum_{j=s+,s-} e^{i\frac{\pi}{4}{\rm sign}\left(\partial^2_{N}\mathcal{A}_{\rm cl}\big|_{N_{j}}\right)}\sqrt{\frac{2\pi\hbar}{\left|\partial^2_{N}\mathcal{A}_{\rm cl}\big|_{N_{j}}\right|}} e^{\frac{i}{\hbar} \mathcal{A}_{\rm cl}(N_{j})}
    \nonumber\\
    &\propto \frac{e^{+\frac{2}{\mathfrak{h}\Lambda}\mathcal{K}_{c}^{\frac{3}{2}}}}{\sqrt[4]{\frac{\Lambda}{3}q_{1}-\mathcal{K}_{c}}}\cos\left[\left(\frac{2}{\mathfrak{h} \Lambda}\right) \left(\frac{\Lambda}{3}q_{1} - \mathcal{K}_{c}\right)^{\frac{3}{2}} -\frac{\pi}{4}\right]~.
\end{align}
The cosine function arises since the above expression involves a sum over both saddle points. The above wave function exactly corresponds to the semi-classical limit of the coordinate space wave function in \ref{eq:HH_wave_function_WDW}, derived from the Wheeler-DeWitt equation. Thus in all these cases, there is a direct one-to-one correspondence between the wave functions obtained from the semi-classical limit of the Path-integral formulation with those derived from the Wheeler-DeWitt equation.

\subsection{Wave function from path integral in Bouncing scenario}
\label{sec:PI_quantization_bouncing_scenario}

Having demonstrated the path integral quantization for torsionful inflationary cosmology, we now turn our attention to the torsionful bouncing scenario. Since the path integral quantization for the bouncing model, as we shall see, proceeds along the lines of the inflationary case, we choose to demonstrate the path integral quantization for only one suitable boundary condition --- an initial Neumann and final Dirichlet condition. This should lead us to the coordinate space bouncing wave function in its semi-classical limit. Let us first recall that the action for the $s$-parametrized bouncing models has the following form
\begin{align}
\label{eq:bounce_action_PI}
\mathcal{A}_{\rm Bounce}=\frac{3 s V_{3}}{4\pi G_{\rm N}}\int_{0}^{1} {\rm d} t \left[-\dot{q}\mathcal{B} - N \left(- \frac{1}{2s} \mathcal{B}^2 + \frac{1}{2s} c^2 - \frac{4\pi G_{\rm N}}{3s}\rho_{0}\left(\Omega - q\right) \right) \right] + \frac{3V_{3}}{8 \pi G_{\rm N}} q \mathcal{B}\Big|_{\rm Bondary}~.
\end{align}
Like before, here also we construct the final action by adding a boundary term to the above action, $\mathcal{A}_{\rm final}=\mathcal{A}_{\rm Bounce}+\mathcal{A}_{\rm Boundary}$, such that the variational problem consistently supports an initial Neumann and final Dirichlet boundary condition. For the action in \ref{eq:bounce_action_PI}, the boundary term has the following form
\begin{align}
\mathcal{A}_{\rm Boundary}=\left(1-2s\right)\frac{3V_{3}}{8\pi G_{\rm N}}q(0) \mathcal{B}(0) - \frac{3V_{3}}{8\pi G_{\rm N}} q(1) \mathcal{B}(1)~,
\end{align}
and we notice that the boundary term depends on the value of $s$, that is, on the particular bouncing model being considered. With this boundary term, the variation of the action $\mathcal{A}_{\rm final}$ yields
\begin{align}
\delta\mathcal{A}_{\rm final}&=\frac{3 s V_{3}}{4\pi G_{\rm N}}
\int {\rm d}t \Bigg[\left(\dot{\mathcal{B}}-\frac{4\pi G_{N}}{3s}N\rho_{0}\right)\delta q +\left(-\dot{q}+\frac{N}{s}\mathcal{B} \right)\delta \mathcal{B}
\nonumber
\\
&\qquad -\delta N \left(- \frac{1}{2s} \mathcal{B}^2 + \frac{1}{2s} c^2 - \frac{4\pi G_{\rm N}}{3s}\rho_{0}\left(\Omega - q\right)\right) \Bigg]
-\frac{3 s V_{3}}{4\pi G_{\rm N}}\Big[\mathcal{B}(1)\delta q(1) + q(0)\delta \mathcal{B}(0)\Big]~,
\end{align}
such that, we can now set the desired boundary conditions for the variational problem
\begin{align}
\label{eq:bounce_boundary}
q(1) = q_{1}~;
\qquad 
\mathcal{B}(0)=\pi_{0}~.
\end{align}
Moreover, from the variations with respect to $q$, $\mathcal{B}$, and $N$, respectively, we recover the dynamical, as well as constraint equations for the bouncing scenario,
\begin{align}
\dot{\mathcal{B}}&=N\frac{\sigma}{2s}~,
\label{eq:bounce_dynamical1}
\\
\dot{q}&=\frac{N}{s} \mathcal{B}~,
\label{eq:bounce_dynamical2}
\\
-\frac{1}{2s}\mathcal{B}^2&+\frac{1}{2s} c^2 - \frac{\sigma}{2s} \left(\Omega - q\right)=0~,
\label{eq:bounce_constraint}
\end{align}
where, we have used the definition $\sigma \equiv (8\pi G_{\rm N}/3)\rho_{0}$, which will be the Hubble squared if a torsionless universe had been filled with constant matter energy density $\rho_{0}$. The solution of the dynamical equations, presented in \ref{eq:bounce_dynamical1} and \ref{eq:bounce_dynamical2}, which is consistent with the boundary conditions in \ref{eq:bounce_boundary}, has the following explicit form
\begin{align}
\label{eq:bounce_class_sol}
    \bar{q}(t) = N^2 \frac{\sigma}{4s^2} t^2 + N \frac{\pi_{0}}{s} t + q_{1} - N^2 \frac{\sigma}{4s^2} - N \frac{\pi_{0}}{s}~.
\end{align}
Again, the path integral over the phase space is dominated by the classical solution to the dynamical equations and hence the coordinate space wave function is an integration over the lapse function with the integrand being $\exp\left[(i/\hbar)\mathcal{A}_{\rm cl}(N)\right]$. Here $\mathcal{A}_{\rm cl}(N)$ is the action evaluated at the above classical solution and is given by
\begin{align}
\mathcal{A}_{\rm cl}(N)=\frac{3sV_{3}}{4\pi G_{\rm N}}\left[N^3\frac{\sigma^2}{24s^3}+N^2\frac{\sigma}{4s^2}\pi_{0} 
+\frac{N}{2s}\Big(-c^2+\pi_{0}^2+\sigma(\Omega - q_{1})\Big)-\pi_{0}q_{1}\right]~.
\end{align}
The integration of the lapse function along the real line has to be converted to an integration over the plane of the complex lapse function to ensure convergence. Subsequently, in the semi-classical limit, one employs the method of steepest descent wherein the most contribution to the integral comes from the relevant saddle points through which the Lefschetz thimbles pass. Thus it is important to determine the saddle points of the classical action $\mathcal{A}_{\rm cl}(N)$ in the complex $N$ plane and these are given by, 
\begin{align}
\label{eq:bounce_saddle}
N_{\rm s\pm}=\frac{-\pi_{0}\pm\sqrt{\sigma\left(q_{1}-\Omega\right)+c^2}}{(\sigma/2s)}~.
\end{align}
Since we are working in a bouncing model, the universe can never reach zero size, classically. Therefore, we demand the saddle point geometries to also start from a finite size, and hence we impose the following initial condition: $\bar{q}(t=0)\big|_{N_{\rm s\pm}} =q_{0}$, for a finite real number $q_{0}>0$, yielding,
\begin{align}
\pi_{0}=\pm i \sqrt{\sigma\left(\Delta - q_{0} \right)}~,
\end{align}
where for convenience, we have defined a quantity $\Delta\equiv\Omega-\left(c^2/\sigma\right)$, which is the value of the scale factor $q$ at the bounce in the classical theory (see, the case of $q$-independent model of torsion in \ref{scale_bounce}, for details). A priory $q_{0}$ is an arbitrary quantity, and we will keep it that way for reasons to be elaborated on later, in connection with the stability analysis. It may appear tempting to equate $q_{0}$ with the scale associated with the bouncing scenario, that is, for a $q$-independent model using the condition: $\Delta=q_{0}$. Since $\Omega$ and $c$ are parameters of the problem, the above condition can always be imposed (however, if one elevates the torsion parameter $c$ to be an operator, the above classical constraint may not be satisfied). This in turn demands $\pi_{0}=0$. This is because $\pi_{0}$ is related to the Hubble parameter and at the bouncing scale the Hubble parameter identically vanishes. For vanishing $\pi_{0}$ and for the final size $q_{1}$ of the universe is such that, $q_{1}>\Delta$, it follows that the saddle points $N_{\rm s\pm}$ are both situated on the real axis in the plane of complex lapse function. However, it will turn out that the above saddle points are unstable under external scalar perturbations. The other possibility will be to consider $q_{0}=0$, which would provide a formal similarity of the wave function in the bouncing model with the Hartle-Hawking no-boundary proposal. However, we avoid this case as the saddle point geometry encounters an initial singularity (see \cite{PhysRevD.103.106008}). Thus, we will neither assume $q_{0}=0$ nor shall we take it to be equal to the scale of the bounce, rather we will keep $q_{0}$ arbitrary for the moment being, but smaller than the scale of the bounce. In which case, $\pi_{0}$ is a purely imaginary quantity and the saddle points are complex. The saddle points, as we shall see, are stable only in the case of positive imaginary $\pi_{0}$.

\begin{figure}[ht!]
\centering
\includegraphics[width=\textwidth]{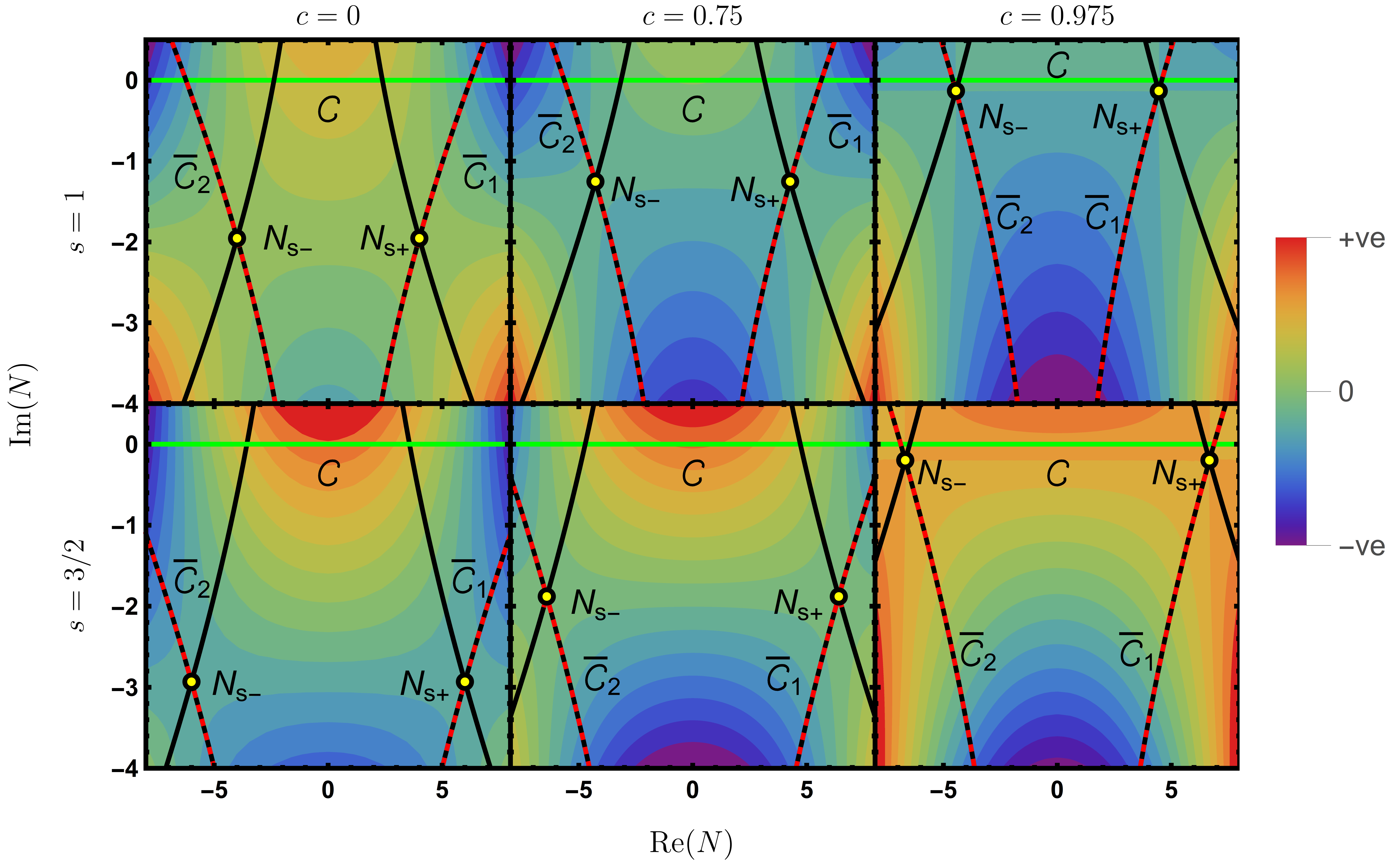}
\caption{We present the saddle points, as well as the steepest descent and the steepest ascent contours. For this plot, we have chosen the initial scale factor of the saddle point geometry to be $q_{0}=0.045$. Note that the initial size is taken to be smaller than the scale of classical bounce. In this case, there are two saddle points and both lie in the lower half of the complex plane. The existence of two complex saddle points is generic and exists irrespective of the bouncing model under consideration, e.g., whether we consider the matter bounce scenario ($s=1$) or otherwise (for, e.g., $s=3/2$). The steepest descent contours (depicted by red and black dashed lines) asymptote to the dark blue regions, where ${\rm Im}[\mathcal{A}_{\rm cl}]>0$ and the path integral along them converges, while the steepest ascent contours (black lines) asymptote to dark red regions with ${\rm Im}[\mathcal{A}_{\rm cl}] < 0$ and the path integral along them does not converge. The original integration contour $\mathcal{C}$, shown by a green line, needs to be deformed to the steepest descent contours $\bar{\mathcal{C}}_{1}\cup\bar{\mathcal{C}}_{2
}$ to obtain the wave function of a bouncing universe from the path integral.}
\label{fig:bounce_Neumann_Dirichlet}
\end{figure}

The corresponding situation involving the complex saddle points as well as the steepest descent and steepest ascent contours has been depicted in the complex lapse function plane, in \ref{fig:bounce_Neumann_Dirichlet}. As evident, the original contour of integration $\mathcal{C}$ (demonstrated by the Green line) can be deformed into the union of two steepest descent contours $\mathcal{C}_{1}$ and $\mathcal{C}_{2}$, passing through the saddle points $N_{\rm s+}$ and $N_{\rm s-}$, respectively. This deformation of contour works for all choices of the torsion parameter $c$ (see \ref{fig:bounce_Neumann_Dirichlet} for the integration contour involving two different choices of the torsion parameter). Thus the coordinate space wave function depends on the contribution from both saddle points, and hence it is straightforward to compute the coordinate space wave function of the universe in the bouncing scenario as,
\begin{align}
\label{eq:bounce_wave_function_path_integral}
\Psi^{\rm B}(q_{1}) & \sim \sum_{j=s+,s-} e^{i\frac{\pi}{4}{\rm sign}\left(\partial^2_{N}\mathcal{A}_{\rm cl}\big|_{N_{j}}\right)}\sqrt{\frac{2\pi\hbar}{\left|\partial^2_{N}\mathcal{A}_{\rm cl}\big|_{N_{j}}\right|}} e^{\frac{i}{\hbar} \mathcal{A}_{\rm cl}(N_{j})}
\nonumber
\\
&\qquad \propto \frac{\exp\left[\frac{2\sqrt{\sigma}}{3\mathfrak{h}_{s}}\left(\Delta+\frac{q_{0}}{2}\right)\sqrt{\Delta-q_{0}}\right]}{\sqrt[4]{\sigma \left(q_{1}-\Delta\right) }} \cos\left[\frac{2\sqrt{\sigma}}{3\mathfrak{h}_{s}}\left(q_{1} - \Delta \right)^{\frac{3}{2}}-\frac{\pi}{4}\right]~.
\end{align}
The above wave function corresponds to the solution of the Wheeler-DeWitt equation in the semi-classical limit, as expected, and can be seen from \ref{eq:bounce_wavefunction}. Thus we have demonstrated that the wave functions arising from the solution of the Wheeler-DeWitt equation, also arise from the semi-classical limit of the path integral, even in the context of a bouncing scenario, when the appropriate contour(s) of integration has been used. It is to be emphasized that, even though the lapse integral can either be performed over the whole real line, leading to a solution of the Hamiltonian constraint, or over the half-line, leading to Green's function, here we have chosen to calculate the former because then the obtained result can be readily checked against the known and expected solution
to the Wheeler-DeWitt equation. We will now determine the stability of these saddle points under scalar field perturbation and shall determine the corresponding power spectrum.

\subsubsection{Interpretation of the bouncing wave function}
\label{sec:interpretation_bounce_wave_function}

In quantum cosmology, given a wave function, computing the probability and assigning physical interpretation to it can be a challenging task in its own right and thus have engendered numerous suggestions of how an appropriate probability measure can be defined (see, for example, \cite{PhysRevD.39.1116,Page:2006hr,PhysRevLett.100.201301,PhysRevD.80.124032,PhysRevD.99.123531,Lehners:2023yrj}). In the present context we may define a relative probability between any two given configurations of the early universe in the following manner. We start with the conserved current associated with the Wheeler-DeWitt equation in the coordinate space representation, which yields,
\begin{align}
J = - \frac{i\hbar}{2}\left(\Psi^{*\rm B}\partial_{q_{1}}\Psi^{\rm B}-\Psi^{\rm B}\partial_{q_{1}}\Psi^{*\rm B}\right)~,
\end{align}
where $\Psi^{\rm B}$ is the wavefunction, dependent on the final scale factor $q_{1}$, in the bouncing scenario. For the real wave function of the bouncing universe, as in \ref{eq:bounce_wave_function_path_integral}, the above current is identically zero, because the WKB modes corresponding to the expanding and contracting branches of the bouncing universe cancel each other. However, for a single WKB mode of the universe, say the one corresponding to the saddle point $N_{\rm s+}$, the current is given by, 
\begin{align}
J_{\rm \Delta} \propto \exp\left[ \frac{4\sqrt{\sigma}}{3\mathfrak h_{s}}\left(\Delta+\frac{q_{0}}{2}\right)\sqrt{\Delta - q_{0}} \right]~.
\end{align}
Then the probability measure of the universe, for a given value of $\Delta$ is ${\rm d}P_{\Delta} = J_{\Delta} {\rm d}V$, where ${\rm d}V$ is a volume element of the minisuperspace. $J_{\Delta}$ being a constant (that is the whole point of the existence of a conserved current), the total probability is obtained by integrating over the entire minisuperspace, which becomes proportional to its volume, and is infinite. Since the above does not provide a sensible description for probabilities, rather than talking about absolute probabilities here, we can define a relative probability  of any configuration of the universe characterized by a value of $\Delta$ with respect to a fiducial state as
\begin{align}\label{prob_torsion}
P\left(\Delta|\Delta_{\rm fiducial}\right) & \equiv \lim_{{\rm vol}\to\infty} \frac{J_{\Delta} \cdot\left({\rm vol}\right)}{J_{\Delta_{\rm fiducial}}\cdot \left({\rm vol}\right)} = \frac{J_{\Delta}}{J_{\Delta_{\rm fiducial}}}  \nonumber\\ & = \exp\left[\frac{4\sqrt{\sigma}}{3\mathfrak{h}_{s}}\left\{\left(\Delta+\frac{q_{0}}{2}\right)\sqrt{\Delta-q_{0}}-\left(\Delta_{\rm fiducial}+\frac{q_{0}}{2}\right)\sqrt{\Delta_{\rm fiducial}-q_{0}}\right\}\right]~.
\end{align}
Without the loss of any generality, this fiducial state can be taken to be the torsionless state of the universe. Then given a value of $\Omega$, which refers to the relative density of the bounce-enabling matter compared to the standard matter field present, we always have $\Delta_{\rm fiducial}\geq\Delta$, with the equality holding true only when the torsion vanishes. Therefore, we see that a state with non-zero torsion has a lower probability associated with it compared to a state with zero torsion. This can be understood as the quantum version of the classical secondary constraint which implies that torsion vanishes in the case of vanishing spin current coming from the fermionic degrees of freedom, thereby, rendering the first and the second-order formalism of the gravity theory equivalent. In quantum theory, then, this classical condition becomes a statement regarding the probability of small torsion being large, i.e., zero torsion is a favorable state of the universe. 

Moreover, it has been observed that whenever the solution of the following form $\Psi \sim e^{\left(A+iP\right)/\hbar}$ is substituted to the Wheeler-DeWitt equation and is expanded in the powers of $\hbar$, the first few leading order terms in the equation reduce to the classical Hamilton-Jacobi equation, if the amplitude of the wave function varies considerably slower than the phase of the wave function. In other words, if $(\partial A/\partial q_{1})\ll(\partial P/\partial q_{1})$ (see, \cite{PhysRevD.77.123537,PhysRevD.91.083525} for details). For the case of the saddle point $N_{\rm s+}$, in the context of bouncing cosmologies we obtain,
\begin{align}
\frac{\partial A/\partial q_{1}}{\partial P/\partial q_{1}} \propto \frac{1}{\left(q_{1}-\Delta\right)^{\frac{3}{2}}} \sim a_{1}^{-\frac{3}{2s}} \to 0 ~,
\end{align}
where the limit corresponds to a large final size for the universe, i.e., $q_{1}\gg1$ and we have used the relation $a_{1} = q_{1}^s$. We can immediately see that the universe attains classicality as its size increases at a rate $1/V^{1/2s}$, where $V$ is the volume of the universe. For the case of an inflationary universe, $s=1/2$, and the rate of classicalization corresponds exactly to the volume of the universe. However, in a bouncing model, say in the case of matter bounce (with $s=1$), the classicality is obtained at a rate $1/\sqrt{V}$, much slower than the inflationary case. This may imply that for such bouncing scenarios the quantum nature of the universe is retained longer than the inflationary case as the universe continues to grow. We will investigate this further in a separate work.

We would like to emphasize that, there is no clear prescription for the choice of the initial boundary condition. Demanding that the saddle point geometry be the Hartle-Hawking (HH) no-boundary geometry, and that the perturbations around the saddle is stable, uniquely determine the free-parameter of the initial Robin condition, or, the value of the momentum for the initial Neumann boundary condition. Intriguingly, the power spectrum predicted by the no-boundary saddle point geometry with both initial Neumann and Robin conditions is exactly the same. Hence from the power spectrum, it is almost impossible to distinguish the initial conditions for our universe, if the HH no-boundary wavefunction describes the initial quantum state of our universe. For the bouncing scenario, on the other hand, there can be some differences. We wish to investigate this point --- the effect of initial boundary conditions on the power spectrum in a bouncing scenario --- in our future work.

\section{Quantum perturbations against the saddle point geometry}
\label{sec:quantized_perturbations}

In the path integral formalism, it is possible to impose various boundary conditions on the geometries at the initial and the final hypersurfaces. Each of these boundary conditions leads to different wave functions for our universe, in the semi-classical limit. But for these wave functions to represent a valid quantum state of the universe, they must be stable under external perturbations. Such a stability analysis will also enable us to determine the power spectrum of these primordial fluctuations, which can have a direct observational signature in the Cosmic Microwave Background.


In what follows, we first provide a general setup regarding the scalar perturbations, before specializing to inflationary and bouncing scenarios.

\subsection{General setup}

In this section, we consider a real scalar field, described by $\phi(x)$, on the background spacetime defined by the saddle point geometries that dominate the mini-superspace path integral, as we have discussed above. The perturbation due to the scalar field will be considered to be small so that the back reaction of the scalar field on the background geometry can be ignored. Since the perturbing scalar field lives on the classical background geometry, we may take its action to be $\mathcal{A}_{\rm pert}[\phi,\bar{q},N_{\rm s}]$, defined on the background spacetime characterized by the scale factor $\bar{q}(t)$, which is the classical solution to the Einstein's dynamical equations and then evaluating the same at the saddle points $N_{\rm s}$ of the (gravity+matter) action. Explicitly, the action for the scalar field reads
\begin{align}
\label{eq:action_scalar}
\mathcal{A}_{\rm pert}=-\frac{1}{2}\int {\rm d}^4 x \sqrt{-g}\,g^{\mu \nu} \partial_{\mu} \phi \partial_{\nu} \phi ~,
\end{align}
which corresponds to the minimal coupling of a scalar field with gravity. It is to be emphasized that the spacetime torsion does not directly couple to the scalar degree of freedom. Moreover, as far as the background spacetime is concerned, the classical solution $\bar{q}(t)$ is also independent of torsion. Therefore the information regarding torsion enters only through the saddle points $N_{\rm s}$, i.e., through Einstein's constraint equation.

Given the symmetries of the background spacetime, it is possible to decompose the scalar field into time-dependent and space-dependent parts, such that,
\begin{align}
\label{phix}
\phi(x)=\sum_{n \ell m} v_{n \ell m}(t) Q_{n \ell m}(\bm{x})~.
\end{align}
Here $v_{n\ell m}(t)$ is the time-dependent part of the scalar field and $Q_{n \ell m}(\bm{x})$ is the space-dependent part. We will come back to the time-dependent part in a moment, while the spatial part satisfies the Helmholtz equation
\begin{align}
D^{2}Q_{n \ell m}(r,\theta,\phi) = -k^2 Q_{n \ell m} (r,\theta,\phi)~,
\end{align}
where $D^{2}=\gamma^{ij}D_{i}D_{j}$, with $D_{i}$ being the spatial covariant derivative defined by the induced 3-metric $\gamma_{ij}$ on the spatial hypersurface. The co-moving wavenumber $k$ is related to the eigenmode index $n$, but the relationship is different in spaces with different spatial curvature index $\mathcal{K}$. These are expressed as follows
\begin{equation}\label{k-n-connection}
    k^2 = \begin{cases}
    \begin{aligned}
       & n^2, & n & \geq 0, & \mathcal{K} & = 0~, \\
       & n^2 - 1 , & n & = 1,2,3,\dots, & \mathcal{K} & = +1~,
    \end{aligned}
    \end{cases}
\end{equation}
where, for $\mathcal{K}=0$, $n$ is a continuous quantity, while for $\mathcal{K}=1$, $n$ can only take integer values. The spatial eigenfunction $Q_{n\ell m}(r,\theta,\phi)$ can further be decomposed into radial and angular parts, such that the radial part satisfies the radial harmonic equation in spaces with curvature $\mathcal{K}$ and the angular parts are given by spherical harmonics. Moreover, $\ell$ takes values from $0$ to $n-1$ in integer steps, and for each $\ell$, the number $m$ takes values from $-\ell$ to $+\ell$ (for further details of this decomposition, see \cite{PhysRevD.31.1777,Kiefer:2021iko}). Here, we will only require the following orthonormality condition:
\begin{align}\label{space_norm}
\int {\rm d}^3 \bm{x} \sqrt{\gamma} Q^*_{n \ell m}(\bm{x})Q_{p\ell'm'}(\bm{x}) = \delta\left(n,p\right)\delta_{\ell \ell'}\delta_{mm'}~,
\end{align}
where $\gamma$ is the determinant of the three-metric $\gamma_{ij}$, and the symbol $\delta\left(n,p\right)$ represents the Dirac delta function in the case of flat space and Kronecker delta in the case of closed space, such that,
\begin{align}
\delta(n,p) = \begin{cases}
\begin{aligned}
& \delta(n-p)=\delta(k-k'), & \mathcal{K} & = 0~, \\
& \delta_{np}, & \mathcal{K} & = +1~.
\end{aligned}
\end{cases}
\end{align}
In what follows the angular indices $(\ell,m)$ are not of much significance because of the degeneracies associated with them, rather the index $n$, arising out of the radial behavior of the mode functions of the scalar field is of much importance. Thus we will suppress all the angular indices for notational convenience and hence the action in \ref{eq:action_scalar} can be decomposed as a sum of individual eigenmode components, as,
\begin{align}
\mathcal{A}_{\rm pert} = \sum_{n}\mathcal{A}_{{\rm pert}, n}~.
\end{align}
In the above expression, we will have summation if the spatial sections of the universe are closed, while integration will replace it if the spatial sections are flat. Each of the decomposed parts of the scalar field action can be written as an integration over the coordinate time, such that,
\begin{align}\label{eq:action_scalar_eigen} 
\mathcal{A}_{{\rm pert}, n}[v,\bar{q},N_{\rm s}] = \frac{1}{2}\int_{0}^{1} {\rm d}t \, N_{\rm s} \left( \bar{q}^2 \frac{{\dot{v}_n}^2}{N_{\rm s}^2} -  k^2 \bar{q}^{(4s-2)} v^2_n\right)~, \end{align}
where, the quantity $k$ is related to $n$ through \ref{k-n-connection}, and $\bar{q}$ is the classical solution of Einstein's dynamical equations at the saddle points $N_{\rm s}$. The equation of motion corresponding to the above action in \ref{eq:action_scalar_eigen}, can be expressed in terms of the time-dependent part of the scalar field, namely $v_{n}(t)$, which reads
\begin{align} \label{eq:perturbation_eom} 
\ddot{v}_{n} + 2 \frac{\dot{\bar{q}}}{\bar{q}} \dot{v}_{n} + N_{\rm s}^2 k^2 \bar{q}^{(4s-4)} v_n = 0~. 
\end{align}
Note, from the above equation of motion, it is clear that the eigenmodes $v_{n}(t)$ do not explicitly depend on $(\ell,m)$, thereby justifying the suppression of these indices earlier. We would like to emphasize that similar perturbations can also arise in the gravitational sector. The gravitational perturbation, in the transverse-traceless gauge, leads to two tensor modes $h_{+}$ and $h_{\times}$, both of which \textit{individually} satisfy equations of motion identical to that of a massless scalar field with no potential, as in \ref{eq:perturbation_eom}. The only difference between the tensor and the scalar perturbations is that the action for the tensor perturbations is weighted with $(1/16\pi G)$, which can be captured by re-scaling the scalar field. Thus it follows that, even though we have only shown computations for a scalar perturbation, it can either represent the metric fluctuations or may represent some putative matter field, possibly of quantum origin and perturbing the isotropic and homogeneous universe. This in turn implies that such perturbations are natural in going beyond the minisuperspace approximation and their stability is essential to assess the initial condition for our universe. Therefore, it is
justified to take the point of view that the initial boundary
condition should be such that the background geometry remains stable
against the perturbations, be it scalar or, gravitational.

As the path integral for any dynamical degree of freedom is dominated by the action evaluated at the classical solution, the classical action for the scalar perturbation must be computed. To do this, we rewrite the action for the scalar field in \ref{eq:action_scalar_eigen} by employing integration by parts as
\begin{align} 
\mathcal{A}_{{\rm pert}, n}= & \frac{1}{2}\int_{0}^{1} {\rm d}t \left[ -\frac{\bar{q}^2}{N_{\rm s}} v_n \left(\ddot{v}_{n} + 2 \frac{\dot{\bar{q}}}{\bar{q}} \dot{v}_{n} + N_{\rm s}^2 k^2 \bar{q}^{(4s-4)} v_n \right) + \frac{{\rm d}}{{\rm d}t}\left(\frac{\bar{q}^2}{N_{\rm s}}\dot{v}_n v_n \right) \right]~. 
\end{align}
Note that the terms inside the first round brackets involving $\ddot{v}_{n}$ identically vanish when the classical equations of motion are used, thanks to \ref{eq:perturbation_eom}. Therefore, only the term involving the total time derivative remains in the action for the perturbation scalar field. Expressing the time-dependent part of the scalar perturbation as, $v_{n}(t)\equiv\varphi_{1n}\{F_n(t)/F_n(1)\}$, the classical action for the scalar field can be written as
\begin{align} 
\mathcal{A}_{{\rm pert}, n}=\frac{\bar{q}^2(1)}{2N_{\rm s}}\frac{\dot{F}_n(1)}{F_n(1)}\varphi_{1n}^2~,
\end{align}
where, we have imposed the initial conditions that $v_{n}(0)=0$. Since the time-dependence in $v_{n}$ translates into the time-dependence of the function $F_{n}$, it follows that $F_{n}$ also satisfies the same equation of motion as $v_{n}$, in particular, 
\begin{align}\label{eq:mode_differential_equation} 
\ddot{F}_{n} + 2 \frac{\dot{\bar{q}}}{\bar{q}} \dot{F}_{n} + N_{\rm s}^2 k^2 \bar{q}^{(4s-4)} F_n = 0~. 
\end{align}
Thus, we have decomposed the scalar field into individual mode functions, which are decoupled from each other and can be determined independently using the second-order differential equation, presented in \ref{eq:mode_differential_equation}. The next step is to consider the wave function of the background, along with the perturbation scalar field.  

The wave functional for the (gravity+matter+scalar field) system, in the momentum space, can be represented as $\Phi_{\rm total} [\mathcal{B}_1,\varphi_1] \approx \Phi(\mathcal{B}_1)\chi[\varphi_1(\bm{x})]$, i.e., the total wave functional is a product of the wave function for the unperturbed universe in the momentum space, given by $\Phi(\mathcal{B}_1)$ and the wave functional for the perturbation, denoted by $\chi[\varphi_1(\bm{x})]$. A similar description exists in the coordinate space representation as well, where $\Psi_{\rm total}[q_1,\varphi_1]\approx \Psi(q_1)\chi[\varphi_1(\bm{x})]$. In both of these cases, we have introduced the notation: $\phi(t_{\rm f}=1,\bm{x})\equiv \varphi_{1}(\bm{x})$. The wave functions for the universe, in both coordinate and momentum representations, have already been discussed in detail in the last section, here we concentrate on the wave functional $\chi[\varphi_1(\bm{x})]$. Since the scalar field can be decomposed into a large number of decoupled modes, it follows that, the wave functional $\chi[\varphi_1(\bm{x})]$ can be expressed as
\begin{align}\label{decomp_scalar}
\chi[\varphi_1(\bm{x})]=\prod_{n}\chi_{n}(\varphi_{1n})~, 
\end{align}
where the wave function $\chi_{n}(\varphi_{1n})$, associated with each of the decoupled modes is defined as a sum over histories of that mode between its initial and final values $\varphi_{0n}$ to $\varphi_{1n}$, such that,
\begin{align}\label{eq:scalar_path_integral}
\chi_{n}(\varphi_{1n})=\int_{-\infty}^{\infty} {\rm d} \varphi_{0n} \int_{v_{n}(0)=\varphi_{0n}}^{v_{n}(1) = \varphi_{1n}} \mathcal{D} [v_{n}] e^{\frac{i}{\hbar}\mathcal{A}_{\rm pert,n}[v_{n},\bar{q},N_{\rm s}]} \chi_{0n}(\varphi_{0n})~.
\end{align}
Here, $\chi_{0n}(\varphi_{0n})$ is a suitably chosen initial wave function for the $n$-th mode of the scalar field $\phi$. It is conventional to choose the initial value of all the mode functions to be zero, i.e., $\varphi_{0n} = 0$, which in turn implies $F_{n}(0)=0$ as well. Then the initial wave function for the scalar field perturbation can be simply chosen as $\chi_{0n}(\varphi_{0n})=\delta(\varphi_{0n})$, and hence the path integral turns out to be
\begin{align}\label{eq:perturbation_wavefunction}
\chi_{n}(\varphi_{1n}) \propto \exp\left[\frac{i}{\hbar}\mathcal{A}_{{\rm pert}, n}\right]=A_{n}\exp\left[i \frac{\bar{q}^2(1)}{2\hbar N_{\rm s}}\frac{\dot{F}_n(1)}{F_n(1)}\varphi_{1n}^2 \right]~, 
\end{align}
where $A_{n}$ is a suitable normalization. Note that the real part of the coefficient of $\varphi_{1n}^2$ in the exponent of the above wave function has to be negative, that is,
\begin{align}\label{stability_index}
\mathbb{E}_{\rm s}(n) \equiv {\rm Re}\left[i \frac{\bar{q}^2(1)}{2\hbar N_{\rm s}}\frac{\dot{F}_n(1)}{F_n(1)}\right]<0,
\end{align}
to describe a Gaussian distribution and be physically acceptable. On the other hand, if the coefficient in the exponent $\mathbb{E}_{\rm s}(n)$ is positive then the perturbations assume an \textit{inverse} Gaussian distribution, favoring larger and larger values of $v_{1n}$. Such a distribution then leads to instability and is said to be unphysical. When $\mathbb{E}_{\rm s}(n)$ is identically zero, the wave function of that particular mode becomes pure phase and hence is non-normalizable.

Assuming that around a particular saddle point geometry we get physically viable perturbations with Gaussian distribution, that is with $\mathbb{E}_{\rm s}(n)<0$, then the normalization constant of the mode functions $A_{n}$ can be determined by imposing the condition that the integral of $\chi^*_{n}(\varphi_{1n})\chi_{n}(\varphi_{1n})$ over all possible final configurations $\varphi_{1n}$ of the $n$-th mode of the scalar field should be unity, yielding
\begin{align}
|A_{n}|^2 = \sqrt{\frac{2}{\pi} \Big|\mathbb{E}_{\rm s}(n)\Big|}~. 
\end{align}
The modulus takes care of the fact that $\mathbb{E}_{\rm s}(n)$ is negative for stable configurations of the scalar field. Given the wave functions associated with each individual mode of the scalar field, we can calculate the power spectrum associated with the scalar perturbations. The computations will be different depending on the values of $\mathcal{K}$, here we will present the computation for a closed universe, with $\mathcal{K} = +1$. The power spectrum is obtained by computing the expectation value of the squared scalar perturbation in the state described by the wave functional $\chi[\varphi_{1}(\bm{x})]$, and then averaged over all space at the final hypersurface (where we make our observations), which yields,
\begin{align}
\left\langle \chi[\varphi_{1}(\bm{x})]|\varphi_{1}^2(\bm{x})|\chi[\varphi_{1}(\bm x)]\right\rangle_{\rm avg} 
&=\frac{1}{V_{3}}\int {\rm d}^3 \bm{x}\sqrt{\gamma}\sum_{\substack{n \ell m \\ n' \ell' m'}}  \prod_{p} \left\langle \chi_{p}(\varphi_{1p})| v_{n}(1) v^{*}_{n'}(1)|\chi_{p}(\varphi_{1p}) \right\rangle Q_{n \ell m}(\bm{x}) Q^*_{n' \ell' m'}(\bm{x})
\nonumber
\\
& = \sum_{n \ell m} \frac{1}{2\pi^2} \left\langle \chi_{n}(\varphi_{1n})| v_{n}(1) v^*_{n}(1)|\chi_{n}(\varphi_{1n}) \right\rangle \prod_{p \neq n}\left\langle \chi_{p}(\varphi_{1p})| \chi_{p}(\varphi_{1p}) \right\rangle \nonumber \\ &
= \sum_{n}\frac{n^2}{2\pi^2} \left\langle \chi_{n}(\varphi_{1n})| v_{n}(1) v^*_{n}(1)|\chi_{n}(\varphi_{1n}) \right\rangle~.
\end{align}
Here, we have used the result that for the spatial sector with $\mathcal{K}=1$, the three-volume is $V_{3}=2\pi^{2}$. Moreover, the last equality follows from the fact that we are working with Gaussian wave functions $\chi_{p}(\varphi_{1p})$ which are normalized to unity, and the sum over all the degenerate $\ell$ and $m$ indices yields the factor of $n^{2}$. Thus, we finally obtain
\begin{align}
\left\langle \chi[\varphi_{1}(\bm{x})]|\varphi_{1}^2(\bm{x})|\chi[\varphi_{1}(\bm x)]\right\rangle_{\rm avg} 
=\sum_{n} \frac{n}{n^2-1} \mathcal{P}(n)~,
\end{align}
where the power spectrum $\mathcal{P}(n)$, associated with the mode $n$, is defined as
\begin{align}
\mathcal{P}(n)\equiv\frac{n(n^2-1)}{2\pi^2} \left\langle \chi_{n}(\varphi_{1n})|v_{n }(1)v^*_{n}(1)|\chi_{n}(\varphi_{1n})\right\rangle~.
\end{align}
The above definition of the power spectrum at a given scale $n$ is motivated by the fact that the logarithmic integration in the continuum is equivalent in the closed universe with the following discrete sum
\begin{align}
\int \frac{{\rm d}k}{k} \to \sum_{n}\frac{n}{n^2-1}~,
\end{align}
along with the fact that the power spectrum is usually defined as simply the square of the eigenmode amplitude per logarithmic interval. Further, recalling that $v_{n}(1)=\varphi_{1n}$, we can readily calculate the expectation value of $v_{n}(1)v_{n}^{*}(1)$ in the state described by the wave function $\chi_{n}(\varphi_{1n})$, which yields,
\begin{align}
\left\langle \chi_{n}(\varphi_{1n})|v_{n }(1)v^*_{n}(1)|\chi_{n}(\varphi_{1n})\right\rangle
&=\int_{-\infty}^{\infty} {\rm d}\varphi_{1n} \varphi_{1n}^2 \Big|A_{n} \exp\left[i \frac{\bar{q}^2(1)}{2\hbar N_{\rm s}}\frac{\dot{F}_n(1)}{F_n(1)}\varphi_{1n}^2 \right]\Big|^2 
\nonumber
\\
&=\frac{1}{4\left|\mathbb{E}_{\rm s}(n)\right|}~.
\end{align}
Therefore, the power spectrum becomes,
\begin{align}\label{power_spectrum_discrete}
\mathcal{P}(n) = \frac{n(n^2-\mathcal{K})}{8\pi^2 \left|\mathbb{E}_{\rm s}(n)\right|}~,
\end{align}
where, we have inserted the term involving $\mathcal{K}$, to simply provide a transition from the closed to the flat universe. For the flat universe, with $\mathcal{K}=0$, one can simply replace any summation by integration and the discrete index $n$ need to be replaced by the continuum index $k$, which yields (see, \cite{PhysRevD.86.103524} for more details), 
\begin{align}\label{power_spectrum_continuous}
\mathcal{P}(k)=\frac{k^3}{8\pi^2 \left|\mathbb{E}_{\rm s}(k)\right|} ~.
\end{align}
The definition of $\mathbb{E}_{\rm s}(k)$ is identical to \ref{stability_index}, with the discrete index $n$ being replaced by the continuum index $k$. Thus we have determined the power spectrum for both closed and flat models of the universe, and it depends on the modes as --- $n^{2}(n-1)$ for the closed universe and $k^{3}$ for the flat universe. In addition, it depends on the classical solution $\bar{q}$, the saddle point $N_{\rm s}$, and the mode functions $F_{n}$ and its derivative at the observing hypersurface located at $t_{\rm f}=1$. The determination of the mode function from \ref{eq:mode_differential_equation} depends on the choice of $s$ and hence on the bouncing and inflationary models of the early universe. In what follows we depict the stability of perturbations around the saddle point geometry and its implication for boundary conditions in the path integral approach. 

\subsection{Stability of perturbations around saddle point geometry}
\label{sec:stability_perturbation}

Saddle point geometries dominate the path integral for the unperturbed universe and depend heavily on the boundary conditions. We would like to see whether the scalar perturbations, described above, are stable around the saddle points picked up by the Picard-Lefschetz theory. In cases, when the perturbations around a saddle point geometry grow uncontrollably, the wave function computed with such saddle point(s), even if deemed mathematically correct, cannot be considered physical. In this section, we scrutinize the stability of various saddle points we discovered in previous sections while discussing the path integral problem set up with different boundary conditions. Since the stability depends crucially on the quantity $\mathbb{E}_{\rm s}(n)$, defined in \ref{stability_index}, which in turn depends on the mode function $F_{n}$ and its derivative, the analysis will differ from inflation and bouncing scenarios.  
\subsubsection{Inflationary scenario}
\label{sec:inf_perturbation_stability_Dir_Neu}

In order to study the stability of the saddle points, as well as to evaluate the power spectrum, we first need the mode functions $F_{n}(t)$, which satisfies the differential equation in \ref{eq:mode_differential_equation}. In the case of inflationary scenario, that is when $s=1/2$, the two linearly independent solutions of \ref{eq:mode_differential_equation} read
\begin{align}\label{eq:perturbation_solutions} 
f_n(t),g_n(t) = & \frac{1}{\sqrt{\bar{q}(t)}}\left[\frac{t - \delta}{t - \gamma}\right]^{\pm \frac{\mu}{2}} \left\{ \left[1 \mp \mu \right]\left(\gamma - \delta\right) + 2 (t - \gamma) \right\}~, 
\end{align}
where $f_{n}$ corresponds to the `+'ve sign in the exponent and `-'ve sign inside the curly bracket and $g_{n}$ is given by the opposite signs. The constants $\gamma$ and $\delta$ appearing in the above solutions are the two roots of the equation $\bar{q}(t)=0$, where $\bar{q}(t)$ is the classical solution, which can be represented as,
\begin{align}
\bar{q}(t) = N_{\rm inf}^2H^2 (t-\gamma)(t-\delta)~,
\end{align}
where $N_{\rm inf}$ corresponds to the saddle points associated with the inflationary scenario. In arriving at the above equation, we have set $H^2 = \Lambda/3$, and have defined the constant $\mu$ as,
\begin{align}
\mu^2 \equiv 1 - \frac{4k^2}{(\gamma-\delta)^2 N_{\rm inf}^2 H^4}~,
\end{align}
which holds for the flat universe. Recalling the relation between the principle eigenmode number $n$ and the comoving wavenumber $k$, the above quantity $\mu$ for the closed universe is obtained by simply making the transformation: $k^2\rightarrow (n^2 - \mathcal{K})$. In what follows, we shall set $\mathcal{K}=+1$, i.e., we will discuss the stability of saddle points corresponding to a closed universe. We will now evaluate the quantities $\gamma$, $\delta$, and $\mu$, appearing above, for the saddle points corresponding to different boundary conditions associated with the inflationary paradigm, so that we can get the solutions for the scalar perturbation explicitly and inspect whether the saddle points are stable. Subsequently, in \ref{sec:power_spectrum} using these results we shall calculate the power spectrum.
\paragraph{Dirichlet-Neumann boundary condition}

The path integral prescription and the associated saddle points depend on the choice of boundary conditions. In the inflationary scenario, we employed three possible boundary conditions and the first one being the Dirichlet boundary condition at the initial time and the Neumann boundary condition at the final time. In which case, as we have depicted in \ref{sec:Inf_Diri_Neu}, the Picard-Lefshetz theory picks up the saddle point located on the upper-half complex plane, which becomes,
\begin{align}  
N_{\rm inf+} = \frac{3}{\Lambda}\left(\mathcal{B}_{1} + i \sqrt{ \mathcal{K}_{c}} \right)~. 
\end{align}
Here, $\mathcal{B}_{1}$ corresponds to the momentum at the final time, and $\mathcal{K}_{c}=\mathcal{K}-c^{2}$, where $\mathcal{K}$ is the spatial curvature in the absence of torsion and $c$ is the completely antisymmetric part of the torsion tensor. Evaluation of the quantities $\gamma$, $\delta$ and $\mu$ for the above saddle point yields,
\begin{align}\label{eq:gamma_delta_mu_inf_diri_neu}
\gamma= 0~, 
\quad 
\delta= \frac{2i\mathcal{K}_{c}}{\mathcal{B}_{1}+i\mathcal{K}_{c}}~, 
\quad 
\mu= \sqrt{\frac{\mathcal{K}_{c}+k^2}{\mathcal{K}_{c}}}~,
\end{align}
where, for a closed inflationary universe $k^2 = n^2 - 1$, with $n$ being discrete. Having determined the constants involved in the problem, consider the small-time ($t\ll 1$) behavior of the mode functions,
\begin{align} 
f_n(t),g_n(t) & \propto t^{\mp \frac{1}{2}\left(\mu \pm 1\right)}~, 
\end{align}
and hence, for $\mu \pm 1 > 0$, the mode $f_{n}$ will not be well-behaved for small $t$ and hence cannot be considered to be describing the observable universe at early times. It turns out that if we consider the case when $1\geq \mathcal{K}_{c}>0$, then from \ref{eq:gamma_delta_mu_inf_diri_neu}, it follows that $\mu \pm 1 > 1$, and hence the mode $f_{n}(t)\sim t^{-(\mu \pm 1)/2}$, diverges at $t=0$, and has to be rejected on physical grounds. On the other hand, $g_{n}(t)\sim t^{+(\mu \pm 1)/2}$ for the above choice of the parameter $\mu$ and goes to zero at $t=0$. Therefore, this is the allowable physical solution to be considered.

Having discussed the allowable model function, let us check whether the perturbations are stable around the saddle point $N_{\rm inf+}$ for the initial Dirichlet and final Neumann boundary conditions. For this purpose, we must calculate the quantity $\mathbb{E}_{\rm s}(n)$ with the physically allowable solution $g_{n}(t)$, which for large $\mathcal{B}_{1}$, corresponding to the classical limit, yields, 
\begin{align}
\mathbb{E}_{\rm inf+}(n) & =\textrm{Re}\left[i \frac{\bar{q}^2(1)}{2\hbar N_{\rm inf+}} \left\{\frac{\dot{F}_n(1)}{F_n(1)}\right\}_{g_{n}}\right]
\nonumber
\\
&=\textrm{Re}\left[-i\frac{\mathcal{B}_{1} (n^2-1)}{\frac{2\Lambda}{3}}+\frac{(n^2-1) \sqrt{n^2-c^2}}{\frac{2\Lambda}{3}}+\mathcal{O}\left(\frac{1}{\mathcal{B}_{1}}\right)\right]
\simeq +\frac{(n^2-1)\sqrt{n^2-c^2}}{\frac{2\Lambda}{3}}+\mathcal{O}\left(\frac{1}{\mathcal{B}^2_{1}}\right)~.
\end{align}
As evident, $\mathbb{E}_{\rm inf+}(n)>0$ for any $n>1$ and from \ref{eq:perturbation_wavefunction} it follows that for $\mathbb{E}_{\rm inf+}(n)>0$ the perturbation has an \textit{inverse} Gaussian distribution. Therefore the perturbations grow and the system is unstable. Thus even though the unperturbed wave function, as in \ref{eq:path_Dirichlet_Neumann_inflation}, is mathematically consistent with the Picard-Lefschetz theory, such a wave function cannot describe the initial moments of our universe. Therefore, an initial Dirichlet and final Neumann boundary condition cannot give rise to a stable universe.  

Before moving forward to other boundary conditions, however, let us consider the case where the torsion dominates the curvature scale $\mathcal{K}$, yielding a negative or a vanishing value for the quantity $\mathcal{K}_{c}$. When $\mathcal{K}_{c}<0$, there will always be modes for which the quantity $\mu$ becomes imaginary (as $n$ can correspond to any arbitrary eigenmode, and for a given value of $c^2$, we can always find a mode such that $\mathcal{K}_{c}+k^2>0$). As a result, both the solutions $f_{n},g_{n}$ behave in the following manner,
\begin{align}
f_n(t),g_n(t) & \propto t^{\mp i \frac{|\mu|}{2} - \frac{1}{2}}~.
\end{align}
Note that for small $t$, the absolute value of the complex modes become, $|f_{n}|,|g_{n}| \sim t^{-\frac{1}{2}}$, and hence diverges. Thus neither of these modes can be considered as perturbations, and hence we find no solutions to the equation satisfied by the mode functions, which can be considered physical. Therefore, it is only legitimate to constrain the torsion degree of freedom such that we have always $\mathcal{K}_{c}>0$.

Finally, we come back to the case where $\mathcal{K}_{c}=0$, i.e., the torsion and $\mathcal{K}$ cancel each other completely. In this case, the mode functions of the perturbing scalar field take the following form
\begin{align}
f_n(t),g_{n}(t)&=\frac{2}{\mathcal{B}_{1}}\sqrt{\frac{\Lambda}{3}} e^{\mp \frac{ik}{\mathcal{B}_{1}t}}\left(1\pm \frac{ik}{\mathcal{B}_{1}t}\right)~.
\end{align}
Furthermore, in the limit $\mathcal{K}_{c} \to 0$, the quantity $\{N_{\rm inf+}/\bar{q}\}$ simplifies to $(\mathcal{B}_{1}t^2)^{-1}$ and hence one can identify the conformal time in this limit as the integral of the above quantity over the coordinate time $t$, yielding $\eta=-(\mathcal{B}_{1}t)^{-1}$. Intriguingly, the above eigenmodes in the conformal time resemble the familiar Bunch-Davies solutions
\begin{align}
\left(\frac{\mathcal{B}_{1}}{2}\sqrt{\frac{3}{\Lambda}}\right)f_k(t),g_{k}(t)& = e^{\pm ik\eta}\left(1\mp ik\eta\right)~.
\end{align}
Then calculating the stability index $\mathbb{E}_{\rm inf+}(k)$ with the positive frequency solution $g_{k}$, we obtain
\begin{align}
\mathbb{E}_{\rm inf+}(k) \approx + \frac{k^3}{\frac{2\Lambda}{3}}~.
\end{align}
Again we see that the wave function for the individual modes with positive frequency behaves as an inverse Gaussian function and hence is unphysical. Therefore, we see that we must restrict ourselves to the criteria that there is a bound on the torsional degree of freedom $1 > c^2\geq 0$. This conclusion turns out to be true for all the different boundary conditions considered for the inflationary scenario as well.
\paragraph{Robin-Neumann boundary condition}

Having demonstrated the instability associated with the Dirichlet-Neumann boundary conditions, let us deal in this section with the stability of staddle point geometry under the Robin-Neumann boundary condition, as discussed in \ref{sec:Robin_Neumann_inflation}. Again, starting from \ref{eq:class_sol_robin_neumann} and evaluating it at the saddle point \ref{eq:saddle_robin_neumann}, we can rewrite the classical solution as
\begin{align}
\bar{q}(t) = \frac{\Lambda N_{\rm inf}^2}{3}\left(t - \frac{\left(\sqrt{\frac{\Lambda}{3}}a_{0} + i \sqrt{\mathcal{K}_{c}}\right)^2}{\sqrt{\frac{\Lambda}{3}}a_{0}\left(\sqrt{\frac{\Lambda}{3}}a_{0} - 2 \mathcal{B}_{1}\right)-\mathcal{K}_{c}}\right)\left(t - \frac{\left(\sqrt{\frac{\Lambda}{3}}a_{0} - i \sqrt{\mathcal{K}_{c}}\right)^2}{\sqrt{\frac{\Lambda}{3}}a_{0}\left(\sqrt{\frac{\Lambda}{3}}a_{0} - 2 \mathcal{B}_{1}\right)-\mathcal{K}_{c}}\right),
\end{align}
and hence we can identify the quantities $\gamma$ and $\delta$ with,
\begin{align} 
\gamma= \frac{\left(\sqrt{\frac{\Lambda}{3}}a_{0} + i \sqrt{\mathcal{K}_{c}}\right)^2}{\sqrt{\frac{\Lambda}{3}}a_{0}\left(\sqrt{\frac{\Lambda}{3}}a_{0} - 2 \mathcal{B}_{1}\right)-\mathcal{K}_{c}}~, 
\qquad 
\delta= \frac{\left(\sqrt{\frac{\Lambda}{3}}a_{0} - i \sqrt{\mathcal{K}_{c}}\right)^2}{\sqrt{\frac{\Lambda}{3}}a_{0}\left(\sqrt{\frac{\Lambda}{3}}a_{0} - 2 \mathcal{B}_{1}\right)-\mathcal{K}_{c}}~. 
\end{align}
In tune with the no-boundary proposal, if we set $\bar{q}(t=0)=0$, then it turns out that we must set the parameter $a_{0}$ to the following value: $a_{0} = \pm i \sqrt{3/\Lambda} \sqrt{\mathcal{K}_{c}}$. We first present the case where the imaginary part of $a_{0}$ is positive and in this case for the parameters $\gamma$, $\delta$ and $\mu$, we obtain, 
\begin{align} 
\gamma=\frac{-2i\sqrt{\mathcal{K}_{c}}}{\mathcal{B}_{1}-i\sqrt{\mathcal{K}_{c}}}~, 
\qquad 
\delta= 0~, 
\qquad
\mu= \frac{\sqrt{\mathcal{K}_{c}+k^2}}{\sqrt{\mathcal{K}_{c}}}~.
\end{align}
We see that the quantity $\mu$, in this case, is identical to that in \ref{eq:gamma_delta_mu_inf_diri_neu}. Therefore, the conclusion regarding the bound on the parameter $c^2$ still holds true in this case as well, i.e., we will have the following inequality $0 < \mathcal{K}_{c} \leq 1$ to be true. Moreover, from \ref{eq:perturbation_solutions}, it follows that, with the above inequality satisfied, only the solution $f_{n}(t)$ is regular at $t=0$. For this solution the stability index $\mathbb{E}_{\rm inf}(n)$, appearing in the exponent of the perturbation wave function \ref{eq:perturbation_wavefunction} becomes
\begin{align}
\mathbb{E}_{\rm inf}(n)&={\rm Re} \left[i\frac{\bar{q}^2(1)}{2\hbar N_{\rm inf}}\left\{\frac{\dot{F}_n(1)}{F_n(1)}\right\}_{f_{n}}\right] 
\nonumber
\\
&={\rm Re}\left[-i\frac{\mathcal{B}_{1} (n^2-1)}{\frac{2\Lambda}{3}} - \frac{(n^2-1) \sqrt{n^2-c^2}}{\frac{2\Lambda}{3}} +\mathcal{O}\left(\frac{1}{\mathcal{B}_{1}}\right)\right] \simeq - \frac{(n^2-1) \sqrt{n^2-c^2}}{\frac{2\Lambda}{3}} + \mathcal{O}\left(\frac{1}{\mathcal{B}^2_{1}}\right)~.
\end{align}
Therefore, the perturbation has a Gaussian distribution and the system is stable. On the other hand, it can be shown that with the negative imaginary choice for the parameter $a_{0}=-i\sqrt{3/\Lambda}\sqrt{\mathcal{K}_{c}}$, the perturbation has an inverse Gaussian distribution and this case cannot be considered physical. Therefore, the saddle point associated with the path integral involving the Robin-Neumann boundary condition is stable for a certain choice of parameters. 

\paragraph{Neumann-Dirichlet boundary condition} In this final scenario involving an inflationary paradigm, we consider an initial Neumann and a final Dirichlet boundary condition. As in the previous case, here also we start with the classical solution corresponding to the above boundary condition, presented in \ref{eq:class_sol_inf_neu_diri}. In this case, there are two saddle points, and evaluating the classical solution for both of these saddle points, as in \ref{eq:saddles_inf_neu_diri}, we can obtain the parameters $\gamma$, $\delta$, and $\mu$. The determination of these parameters also requires fixing the initial momenta, which we consider to have the following value $\pi_{0} = +i\sqrt{\mathcal{K}_{c}}$, and that yields,
\begin{align}
\gamma_{\pm}=\frac{2\sqrt{\mathcal{K}_{c}}}{\pm i\sqrt{\frac{\Lambda}{3}q_{1}-\mathcal{K}_{c}} + \sqrt{\mathcal{K}_{c}}}~, 
\qquad 
\delta_{\pm}=0~, 
\qquad 
\mu_{\pm}=\sqrt{\frac{\mathcal{K}_{c}+k^2}{\mathcal{K}_{c}}}~.
\end{align}
Again, the conclusion regarding the bound on the value of the torsion parameter, namely $0<\mathcal{K}_{c}\leq 1$ still holds in this case. Using the above expressions for the parameters $\gamma_{\pm}$, $\delta_{\pm}$ and $\mu_{\pm}$, from \ref{eq:perturbation_solutions} we realize that only $f_{n}(t)$ is well behaved at $t=0$, among the two linearly independent solutions for the differential equation satisfied by the mode function. Therefore, the stability parameter, defined in \ref{stability_index} becomes, 
\begin{align}
\mathbb{E}_{\pm}(n)&=\textrm{Re}\left[i\frac{\bar{q}^2(1)}{2N_{\textrm{inf}\pm}}\left\{\frac{\dot{F}_n(1)}{F_n(1)}\right\}_{f_{n}} \right]
\nonumber
\\
&=\textrm{Re}\left[\mp i\frac{\sqrt{q_{1}} (n^2-1)}{2\sqrt{\frac{\Lambda}{3}}} - \frac{(n^2-1) \sqrt{n^2-c^2}}{\frac{2\Lambda}{3}} +\mathcal{O}\left(\frac{1}{\sqrt{q_{1}}}\right)\right]
\simeq- \frac{(n^2-1) \sqrt{n^2-c^2}}{\frac{2\Lambda}{3}} +\mathcal{O}\left(\frac{1}{{q_{1}}}\right)~,
\end{align}
which is negative for all modes with $n>1$ and hence the wave functions are Gaussian in nature. Thus we observe that the perturbation around both the saddle points for the Neumann-Dirichlet boundary condition is deemed relevant by the Picard-Lefschetz theory and is stable and therefore, the corresponding wave function can be considered physical.

\subsubsection{Bouncing scenarios}

In the case of a bouncing scenario, there is a free parameter in the problem, namely the initial size of the saddle point universe $q_{0}$. As we have discussed before, there are two tempting possibilities in this regard, we can take $q_{0}$ to be identical to the scale of classical bounce, or, we can take $q_{0}=0$, following the no-boundary scenario. However, for $q_{0}$ equal to the classical scale of bounce, the stability index $\mathbb{E}_{\rm s}(k)$ becomes ill-behaved, often positive for certain values of comoving wavenumber and hence the modes become inverse Gaussian, signaling instability. The incompatibility of the bouncing wave function with the no-boundary proposal can be motivated along the following lines --- the saddle point geometry in the bouncing scenario is singular when $q_{0}\rightarrow 0$ and hence should be avoided (see, \cite{PhysRevD.103.106008}). Thus the bouncing wave function can neither have any correspondence with the no-boundary wave function, nor the initial size of the saddle point universe in the path integral approach be equal to the classical scale of bounce. 

In the context of bouncing scenarios, for reasonable values of $s$, the differential equation satisfied by the mode functions, namely \ref{eq:mode_differential_equation}, cannot be solved analytically. Thus, we need to find out these mode functions by numerically solving the associated differential equation. We proceed to do this exercise for the specific case $s=1$, whereas similar computations can be performed for other $s$-models as well. In addition, we also incorporate the classical solution, derived in \ref{eq:bounce_class_sol}, which reads,
\begin{align}
\bar{q}_{s\pm}(t) = N_{s\pm}^2\frac{\sigma}{4s^2}\left(t-\frac{\pi_{0}+i\sqrt{\sigma \Delta}}{\pi_{0}\mp\sqrt{\sigma \left(q_{1}-\Delta\right)}}\right)\left(t-\frac{\pi_{0}-i\sqrt{\sigma \Delta}}{\pi_{0}\mp\sqrt{\sigma \left(q_{1}-\Delta\right)}}\right)~.
\end{align}
and then the classical solution is evaluated at the two saddle points associated with the path integral in the bouncing cosmology, located at \ref{eq:bounce_saddle}. Inclusion of all these results provides us an estimation for the stability index $\mathbb{E}_{\rm s}(k)$ and one can check if this index is positive or, negative, thereby pointing out the instability or, the stability of the saddle points, respectively, under scalar perturbation.

\begin{figure}[H]
    \centering
    \begin{minipage}{0.48\textwidth}
    \includegraphics[width=\textwidth]{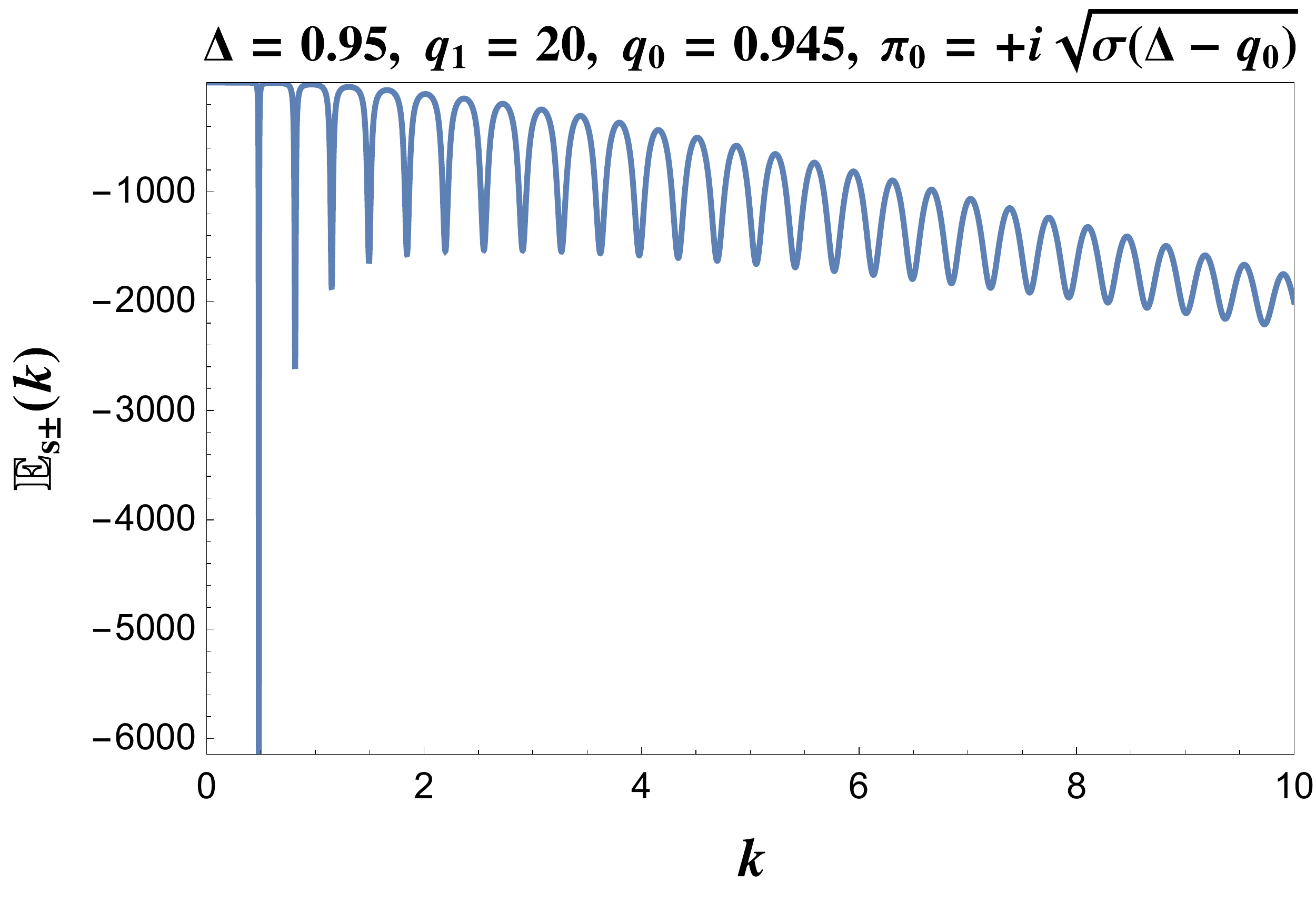}
    \end{minipage}
    \hfill
    \begin{minipage}{0.48\textwidth}
    \includegraphics[width=\textwidth]{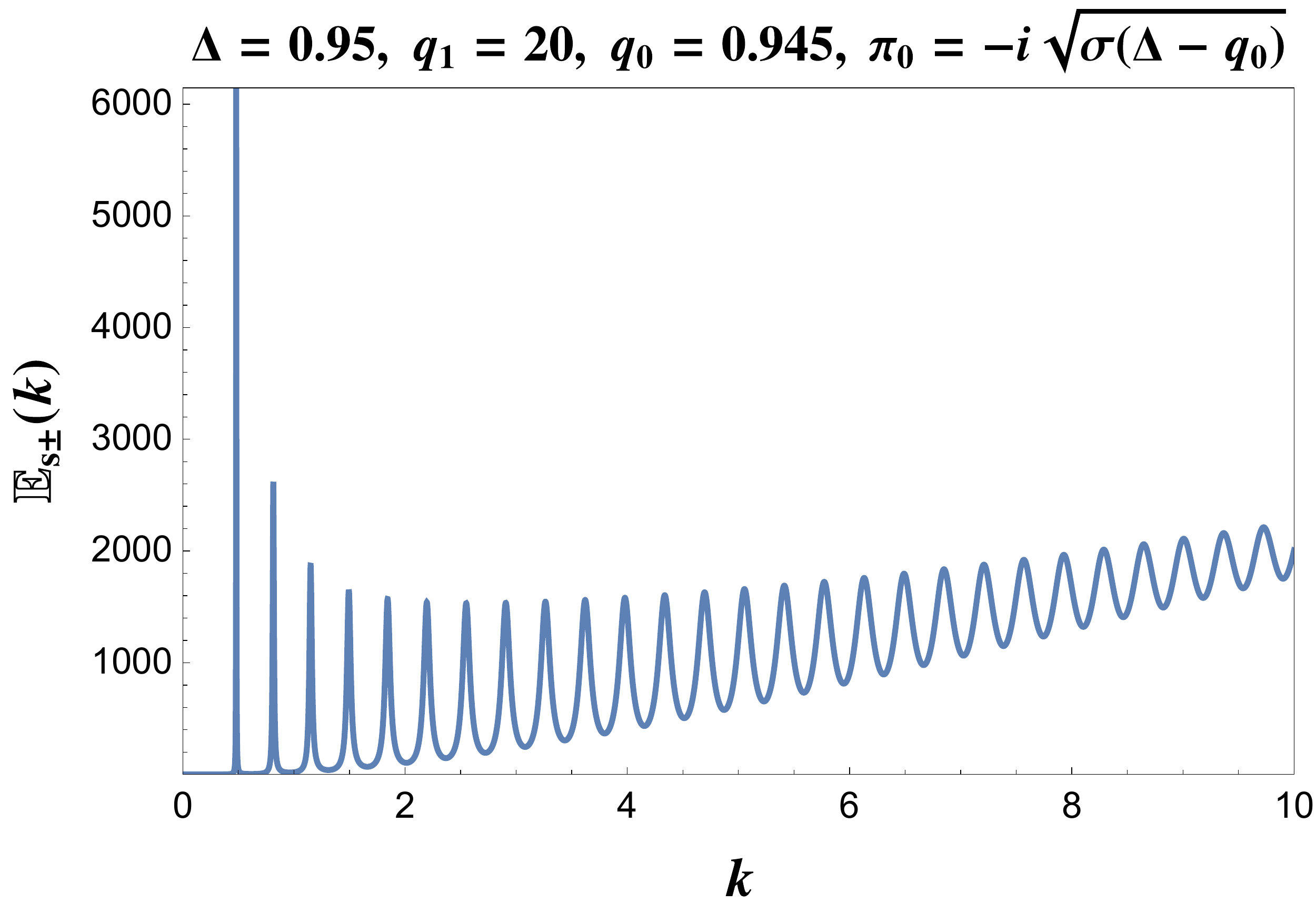}
    \end{minipage}
    \caption{We have plotted the stability index $\mathbb{E}_{\rm s}$ with the wave number associated with various mode functions for different choices of $\pi_{0}$, and for the situation where the initial size of the universe is very close to the classical scale of bounce but smaller than the same. The left hand plot is for $\textrm{Im}\,\pi_{0}>0$, while the right hand plot depicts the case where $\textrm{Im}\,\pi_{0}<0$. As evident $\mathbb{E}_{\rm s}<0$ for all $k$ if and only if $\textrm{Im}\,\pi_{0}>0$ and hence the configurations corresponding to the left-hand plot is stable, while the configurations associated with the right-hand plot is unstable.}
    \label{fig:stable_01}
\end{figure}

\begin{figure}[H]
    \centering
    \begin{minipage}{0.48\textwidth}
    \includegraphics[width=\textwidth]{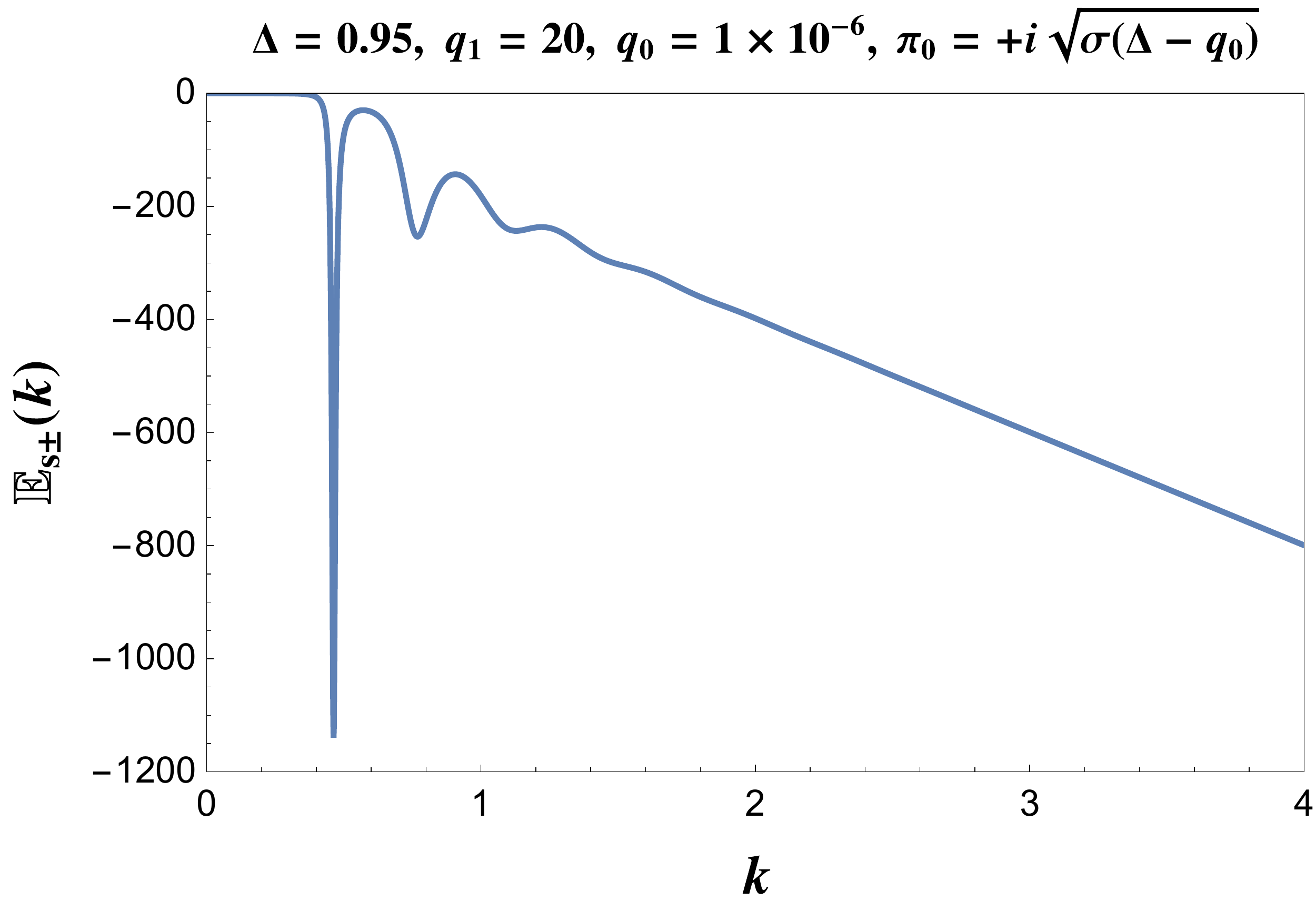}
    \end{minipage}
    \hfill
    \begin{minipage}{0.48\textwidth}
    \includegraphics[width=\textwidth]{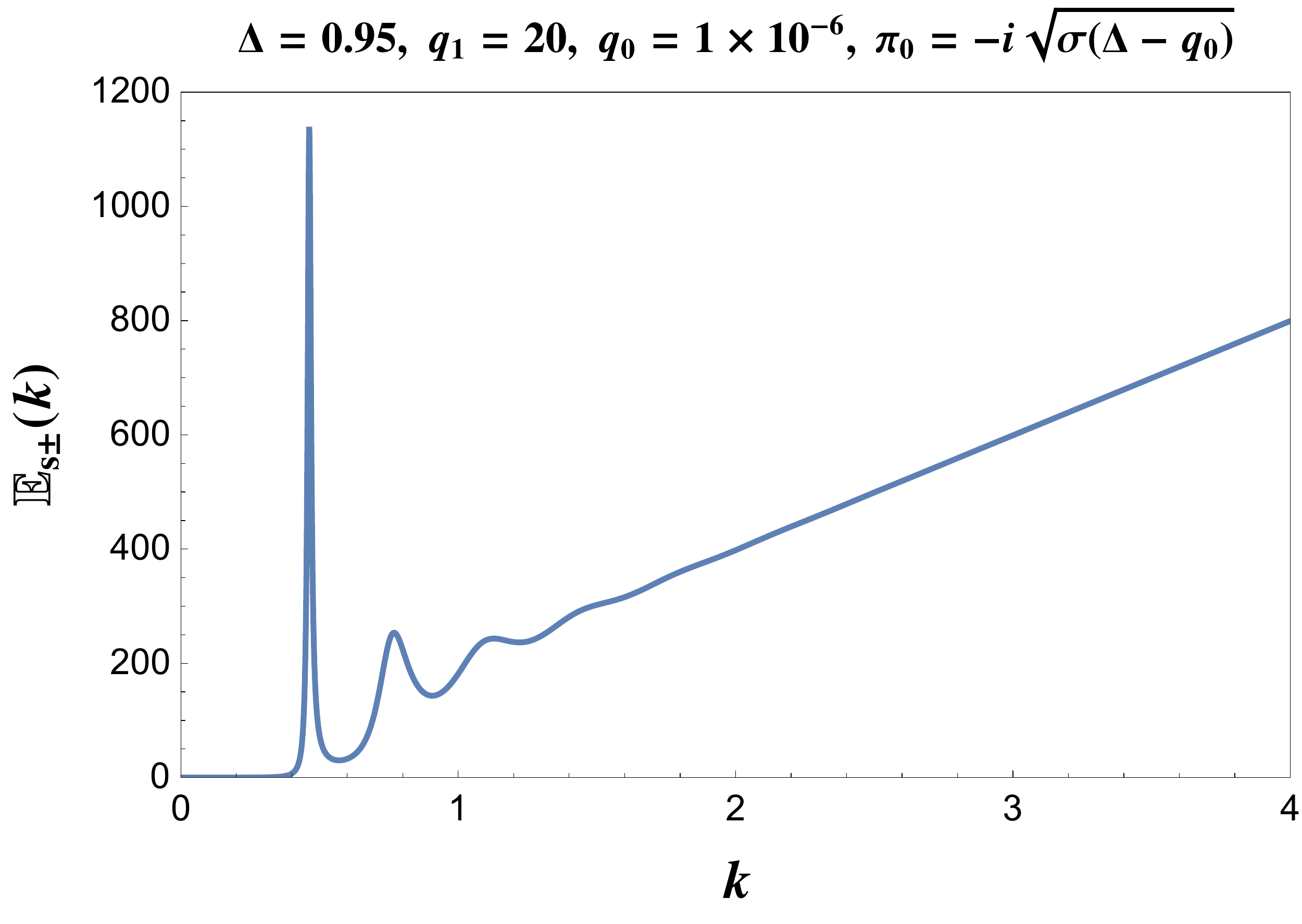}
    \end{minipage}
    \caption{The stability index $\mathbb{E}_{\rm s}$ has been plotted with the wave number $k$ for a universe whose initial size is very small compared to the classical scale of the bounce. In this case also, only for $\textrm{Im}\,\pi_{0}>0$, the $\mathbb{E}_{\rm s}<0$ and hence stability is achieved for such a configuration of the universe. Note that this result holds good for both the saddle points of the path integral in the bouncing scenario.}
    \label{fig:stable_02}
\end{figure}

\begin{figure}[H]
    \centering
    \begin{minipage}{0.48\textwidth}
    \includegraphics[width=\textwidth]{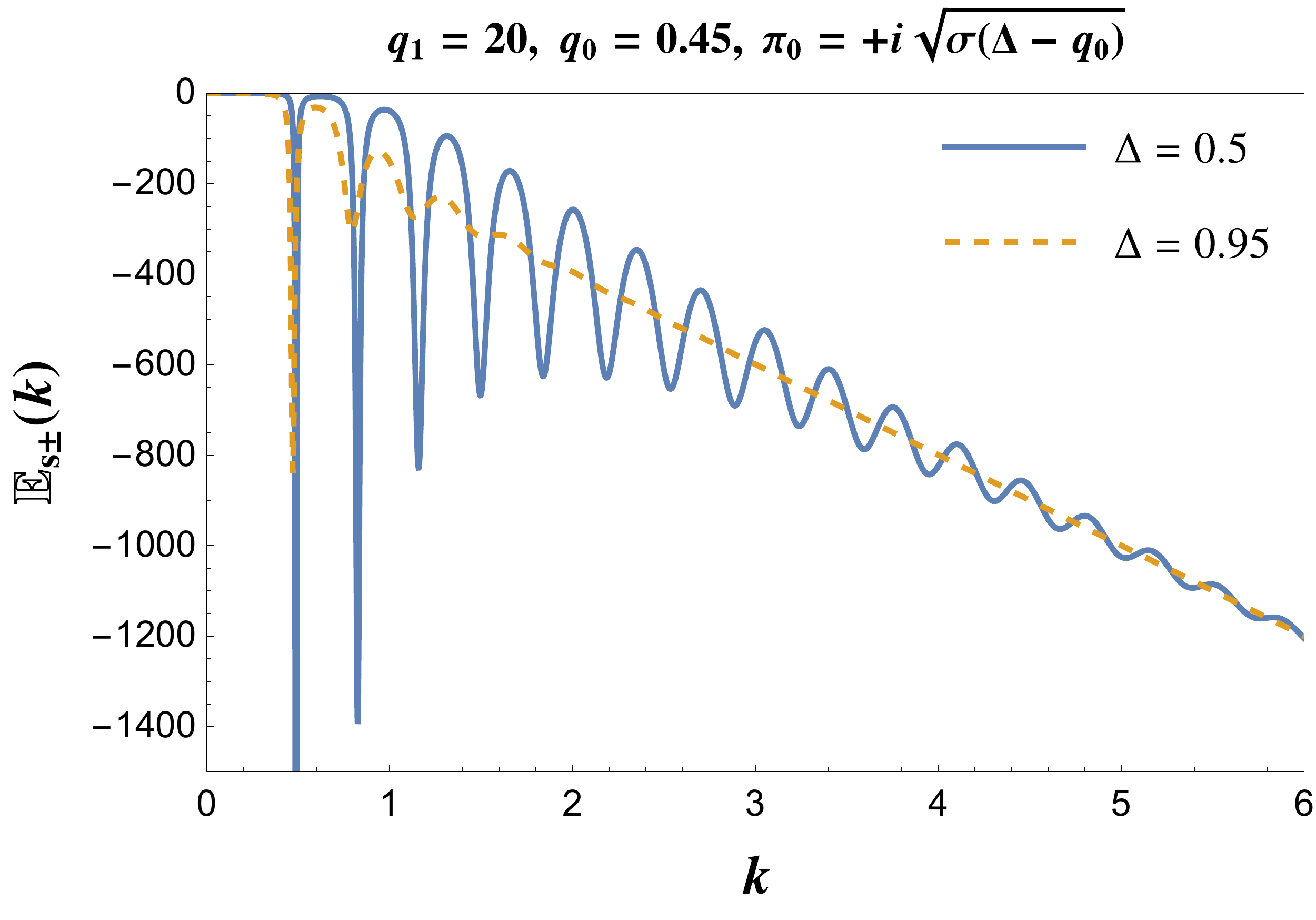}
    \end{minipage}
    \hfill
    \begin{minipage}{0.48\textwidth}
    \includegraphics[width=\textwidth]{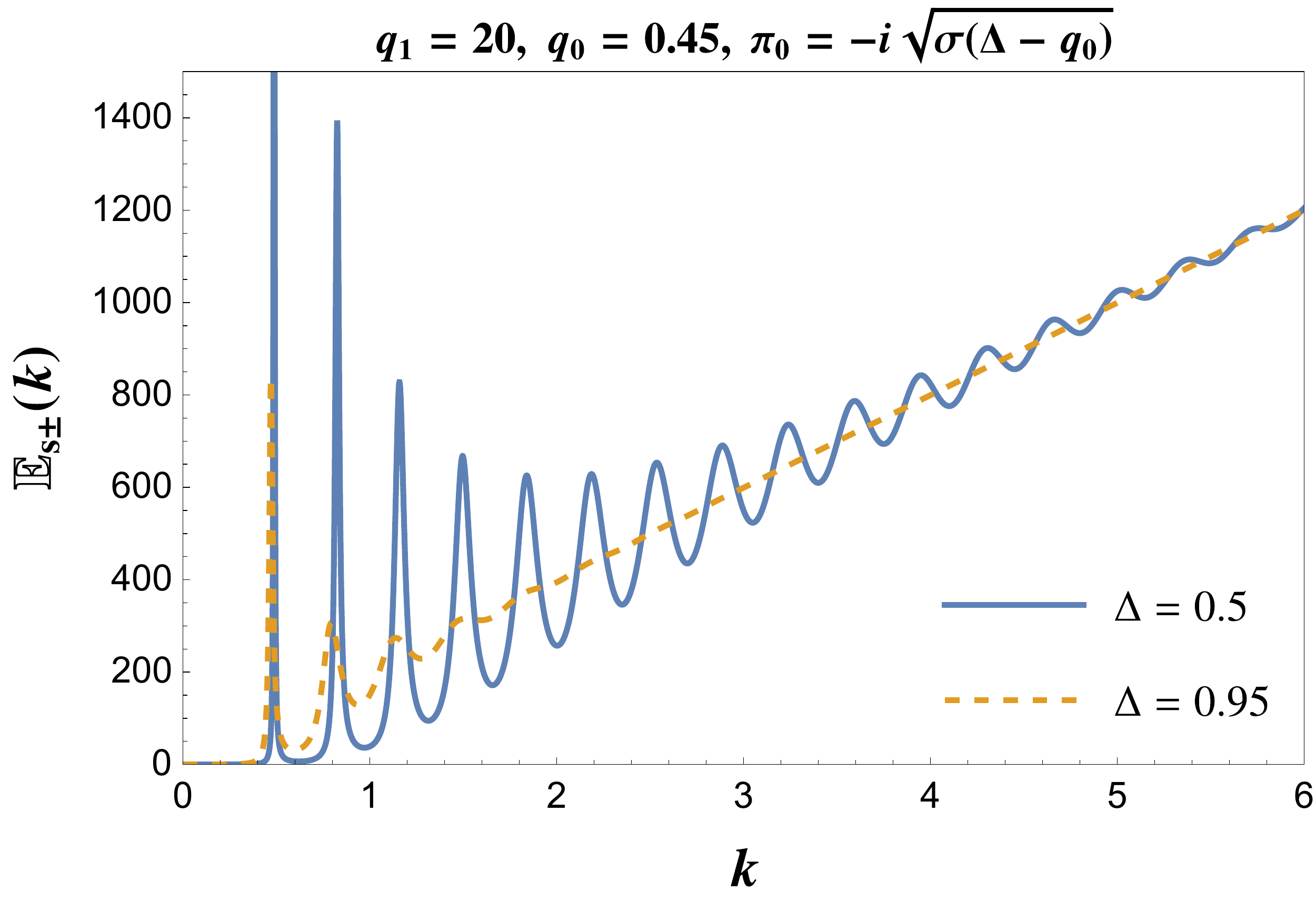}
    \end{minipage}
    \caption{The variation of the stability index $\mathbb{E}_{\rm s}$ with the wave number $k$ is presented for two saddle point universes whose classical scales of the bounce are different. Again, universes with a positive imaginary part for $\pi_{0}$ have a negative stability index and hence are stable. This is true irrespective of the value of torsion, as it holds for different choices of the torsion parameter $c$.}
    \label{fig:stable_03}
\end{figure}

Following this approach, we have considered two scenarios, one in which $q_{0}\ll \Delta$, and another with $q_{0}\lesssim \Delta$, and have plotted the stability index $\mathbb{E}_{\rm s}(k)$ against the wave number $k$, for the matter bounce scenario (which corresponds to $s=1$), for two possible choices of the quantity $\pi_{0}$, appearing in the saddle point geometries. The result of this analysis has been presented in \ref{fig:stable_01} to \ref{fig:stable_03}. As evident, universes having imaginary part of $\pi_{0}$ as positive, provide negative values of the stability index $\mathbb{E}_{\rm s}$ for all wave modes $k$ and hence are stable. On the other hand, universes with the imaginary part of $\pi_{0}$ as negative have the stability index $\mathbb{E}_{\rm s}$ to be positive and hence are unstable. These results are generic in nature, e.g., holds when the initial size of the universe is very close to the classical scale of bounce but smaller than the same (see \ref{fig:stable_01}). Similar results hold true for universes whose initial size is small \emph{but}, not zero, as depicted in \ref{fig:stable_02}. Finally, it turns out that the above result is also independent of the value of the torsion, as long as it is smaller than the relative abundance of bounce-enabling matter. This is evident from \ref{fig:stable_03}, where the stability index $\mathbb{E}_{\rm s}$ is negative for universes having imaginary part of $\pi_{0}$ as positive as well as larger and smaller values of torsion (corresponding to smaller and larger values of $\Delta$, respectively). Thus, we conclude that, as long as the relative abundance of bounce-enabling exotic matter is greater than the torsion, and for all possible choices of the initial size of the universe, which is greater than zero but smaller than the classical scale of bounce, the saddle points in the path integral of bouncing wave function are stable for positive values of the imaginary part of $\pi_{0}$.

\begin{figure}[H]
    \centering
    \begin{minipage}{0.48\textwidth}
    \includegraphics[width=\textwidth]{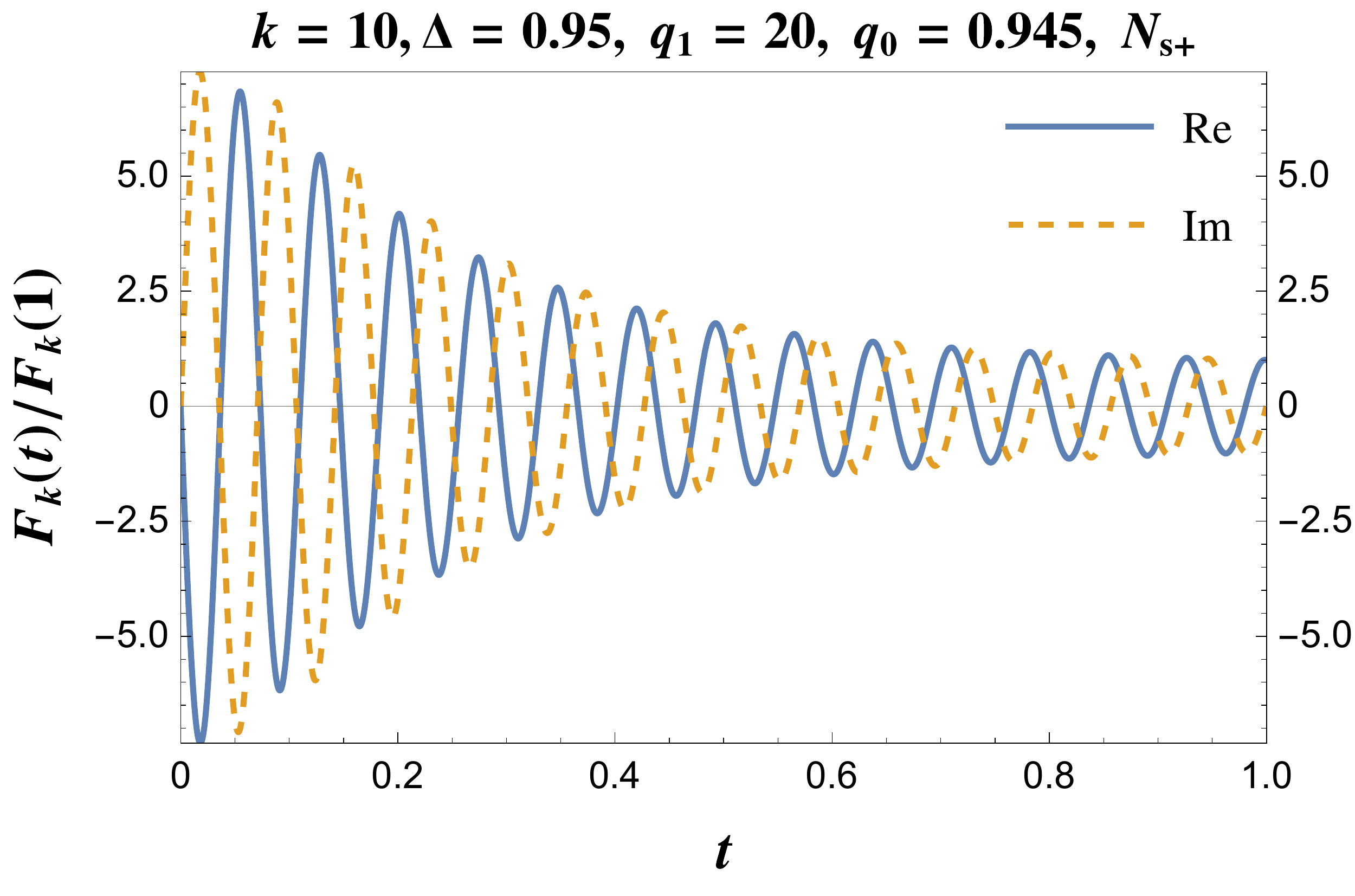}
    \end{minipage}
    \hfill
    \begin{minipage}{0.48\textwidth}
    \includegraphics[width=\textwidth]{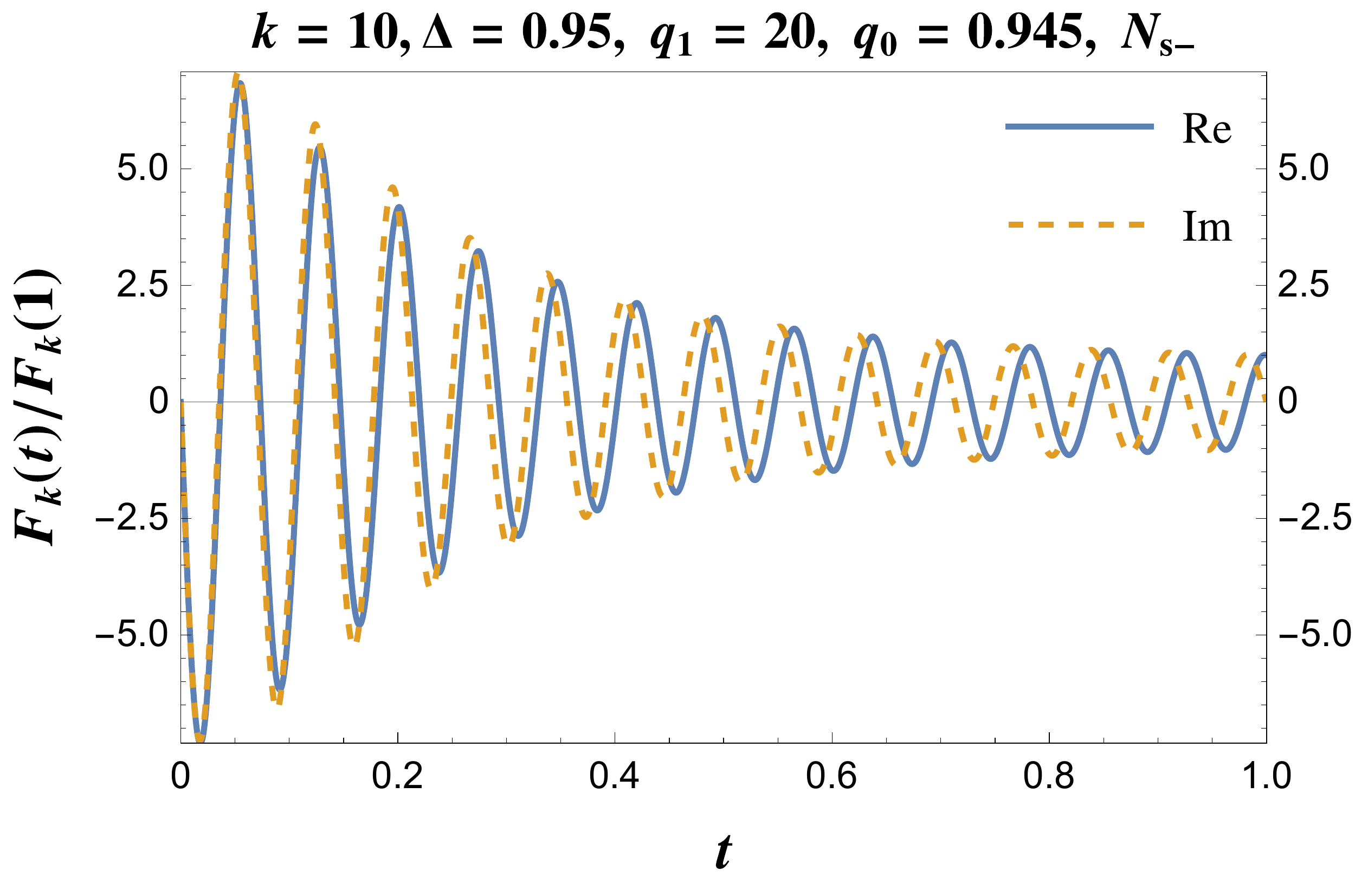}
    \end{minipage}
    \caption{The variation of the real and imaginary parts of the mode function with time has been depicted for wave number $k=10$. In both of these plots, we have taken the initial size of the universe to be very close to the classical scale of bounce, but smaller than it. The left-hand plot depicts the variation of the mode function associated with perturbation about the saddle point $N_{\rm s+}$, while the right-hand side plot describes the variation of the $k=10$ mode for scalar perturbation about the other saddle point $N_{\rm s-}$.}
    \label{fig:scalar_var_bounce_01}
\end{figure}

\begin{figure}[ht!]
    \centering
    \begin{minipage}{0.48\textwidth}
    \includegraphics[width=\textwidth]{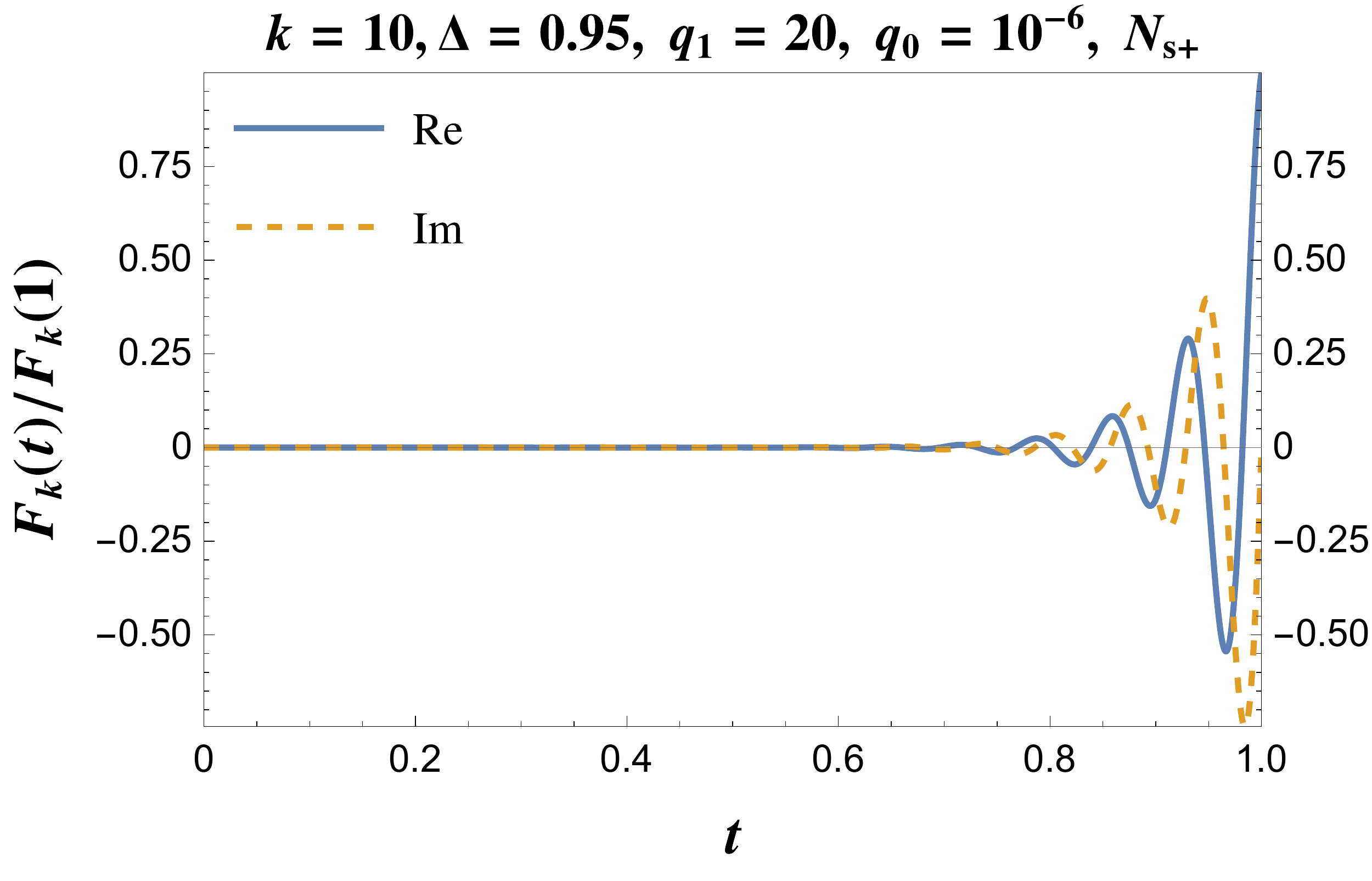}
    \end{minipage}
    \hfill
    \begin{minipage}{0.48\textwidth}
    \includegraphics[width=\textwidth]{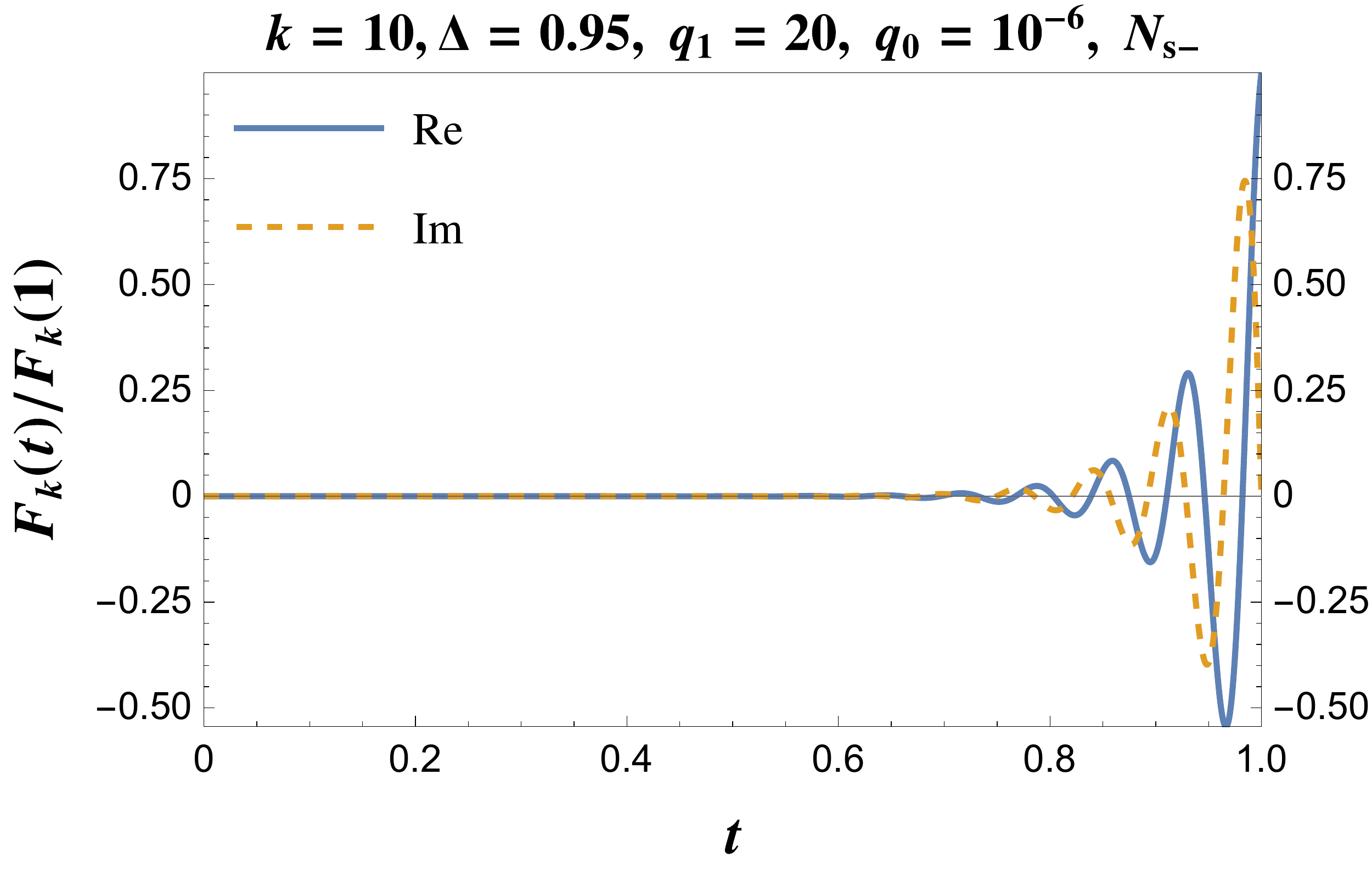}
    \end{minipage}
    \caption{We have depicted the time variation of the real and imaginary parts of the mode function with wave number $k=10$ perturbing a universe whose initial size is very small. The left-hand plot depicts the variation of a specific scalar mode around the saddle point $N_{\rm s+}$, while the right-hand side plot describes the variation of the same about the other saddle point $N_{\rm s-}$.}
    \label{fig:scalar_var_bounce_02}
\end{figure}

Besides studying the stability of saddle point geometries under scalar perturbations, we have also demonstrated the time variation of the mode function associated with the scalar perturbation around both saddle points for stable configurations of the background spacetime. As evident from \ref{fig:scalar_var_bounce_01}, for the initial size of the universe very close to the classical scale of bounce (but smaller), it follows that both the real and imaginary parts of the mode function oscillate in time, with the amplitude decaying. On the other hand, if the initial size of the universe is very close to zero, then also the real and imaginary part of the mode function oscillates in time, but the amplitude grows (see \ref{fig:scalar_var_bounce_02}). In both of these cases, the real and imaginary parts of the mode functions are out of phase for scalar perturbation around the saddle point $N_{\rm s+}$, while for scalar perturbation around $N_{\rm s-}$, the real and imaginary parts of the scalar perturbation are mostly in phase. Having discussed the conditions for stability of scalar perturbation around the saddle point, along with the time variation of the mode function, we will concentrate on the power spectrum of the perturbing scalar field in the next section.   

\subsection{Power spectrum}\label{sec:power_spectrum}

\begin{figure}[H]
    \centering
    \includegraphics[width=0.5\textwidth]{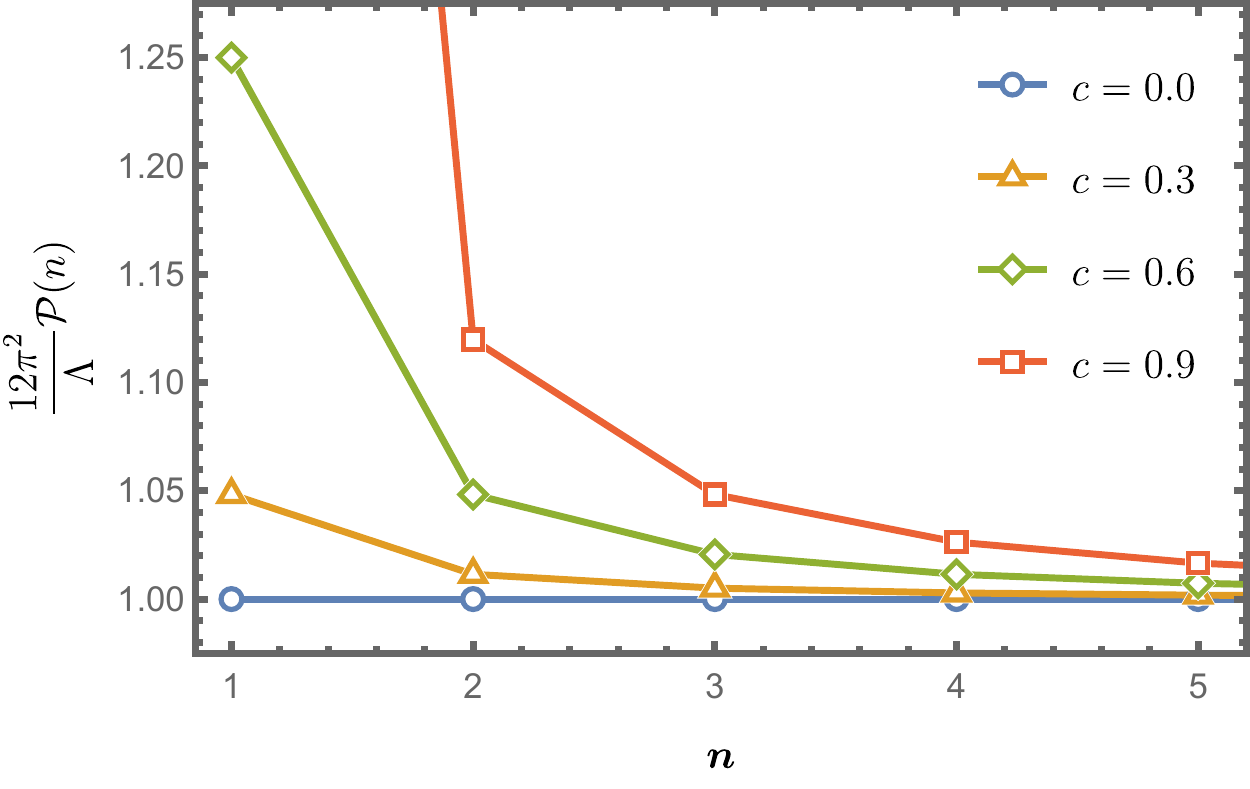}
    \caption{We have depicted the variation of the power spectrum, scaled by $(12\pi^{2}/\Lambda)$, with the mode number $n$, for different choices of the torsion parameter $c$. As evident, for zero torsion, the power spectrum is scale invariant, while for finite torsion there is more power at large physical length scales.}
    \label{fig:power_inflation}
\end{figure}

In this section, we discuss the power spectrum associated with the scalar perturbations for both inflationary and bouncing scenarios. As evident from \ref{power_spectrum_discrete} and \ref{power_spectrum_continuous}, the power spectrum depends on the stability index, which in turn depends on the classical solution, the saddle points, and the mode functions. For an inflationary universe, the mode functions can be exactly solved for and hence the power spectrum can be explicitly computed. Considering a closed universe (with $\mathcal{K}=1$), the power spectrum yields, 
\begin{align}
\mathcal{P}(n) = \left(\frac{\Lambda}{12\pi^2}\right)\frac{n}{\sqrt{n^2-c^2}} ~.
\end{align}
As evident, for $c=0$, i.e., for a torsion-free universe, the power spectrum does not depend on the wave number $n$, leading to a scale-invariant power spectrum. On the other hand, for non-zero values of torsion, the power spectrum becomes scale non-invariant. Moreover, since the torsion appears as a subtracting quantity in the denominator, it follows that for finite $c$, there will be larger power at large physical length scales. This feature is explicitly presented in \ref{fig:power_inflation}. Thus the power spectrum derived from the saddle points in the Lorentzian path integral approach leads to a scale-invariant power spectrum at large comoving wavenumber, i.e., at small scales, which is consistent with the Bunch-Davies vacuum in the inflationary paradigm. It turns out that the feature involving enhanced power at large scale (equivalently, for small wave number) has already been observed in the context of Euclidean path integral with a massless scalar perturbation \cite{Chen:2017aes} as well as in the context of Starobinsky model of inflation \cite{Chen:2019mbu}. Thus, enhanced power at a large physical length scale is quite common in the context of quantum cosmology within the Euclidean approach. Here we have shown that the presence of torsion can also lead to the enhancement of power at such large scales but within the Lorentzian approach to quantum cosmology. This is a distinct signature and can in principle lead to observable consequences regarding the possible existence of torsion. Moreover, in \cite{Chen:2017aes} it was demonstrated that the inclusion of mass for the scalar perturbation can suppress the power spectrum at large scales, rather than enhancing the same, and it would be interesting to see if the same holds true in the presence of torsion as well, which we wish to explore in a future work.  

\begin{figure}[H]
    \centering
    \begin{minipage}{0.48\textwidth}
    \includegraphics[width=\textwidth]{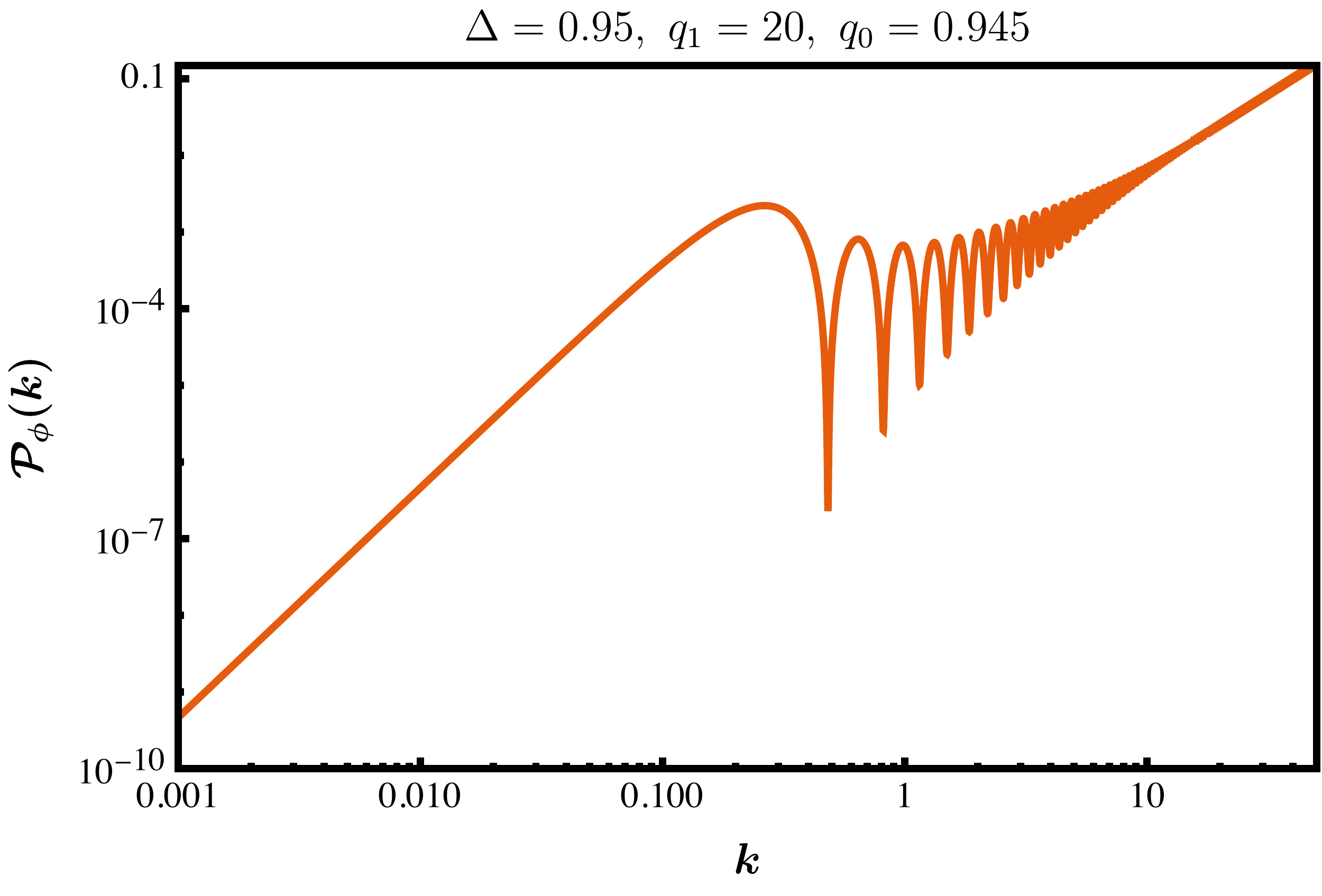}
    \end{minipage}
    \hfill
    \begin{minipage}{0.48\textwidth}
    \includegraphics[width=\textwidth]{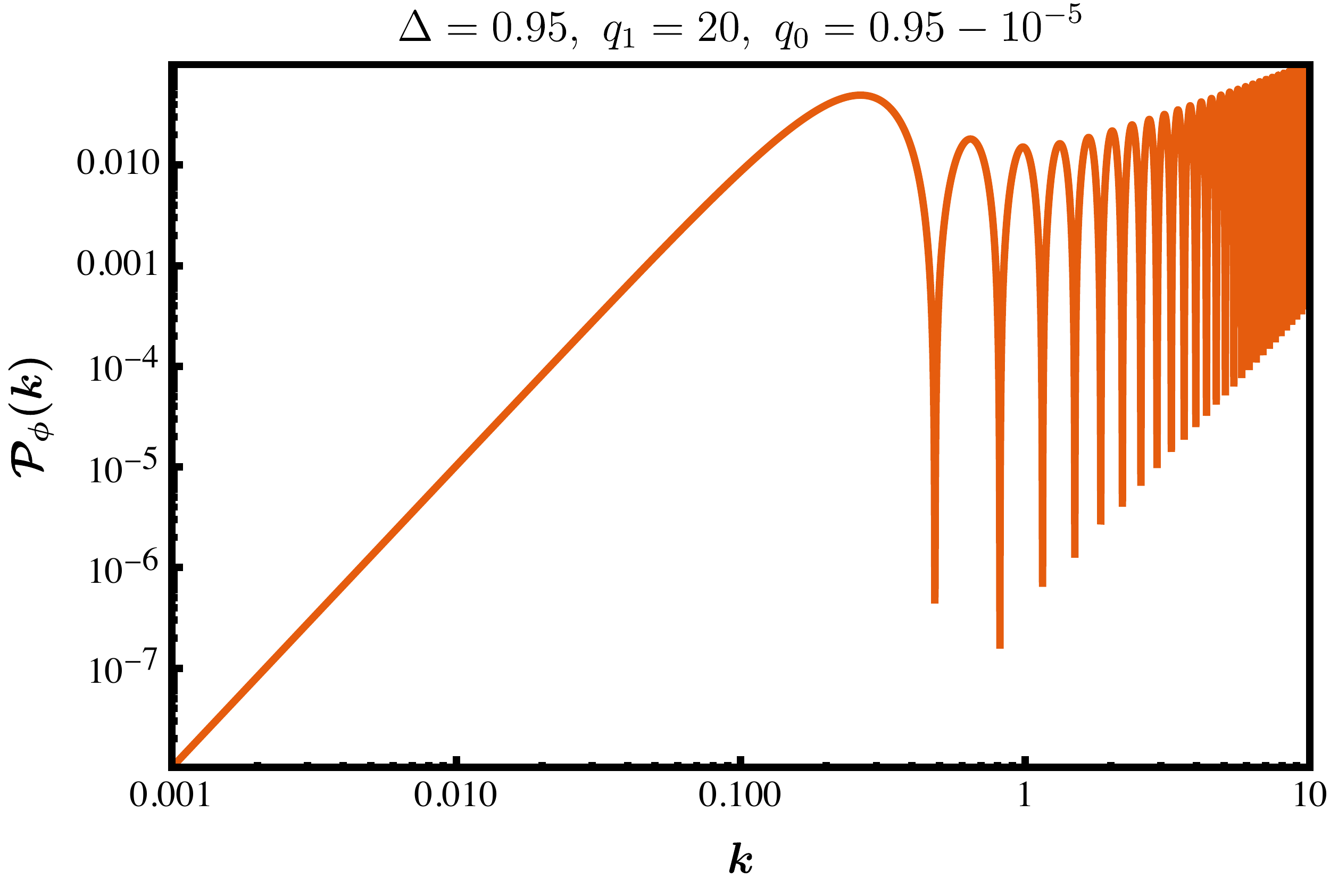}
    \end{minipage}
    \caption{The power spectrum of the perturbing scalar field has been presented in the context of bouncing cosmology. The left-hand side plot is for the power spectrum of the perturbing field when the initial size of the saddle point universe is slightly smaller than the classical scale of bounce. See text for more discussion.}
    \label{fig:power_bounce_01}
\end{figure}

Let us now discuss the corresponding situation in the context of bouncing cosmology. Since in this case the mode functions cannot be determined analytically, we can not provide a closed-form expression for the power spectrum. Rather, we solve for the mode functions numerically (as previously discussed) and hence determine the power spectrum. In the bouncing context, we have taken $\mathcal{K}=0$, and hence the power spectrum will be given by \ref{power_spectrum_continuous}. The corresponding power spectra for the matter bounce model ($s=1$) have been depicted in \ref{fig:power_bounce_01}. The structure of the power spectrum is such that the power spectrum initially grows, then oscillates, and again finally it grows again. For $q_{0}\lesssim \Delta$, the oscillations are smaller and they exist for a smaller range of the comoving wavenumber. On the other hand, when $q_{0}\to\Delta$, the oscillations are larger and also happen over a longer range of the comoving wavenumber.   

\begin{figure}[ht!]
    \centering
    \begin{minipage}{0.48\textwidth}
    \includegraphics[width=\textwidth]{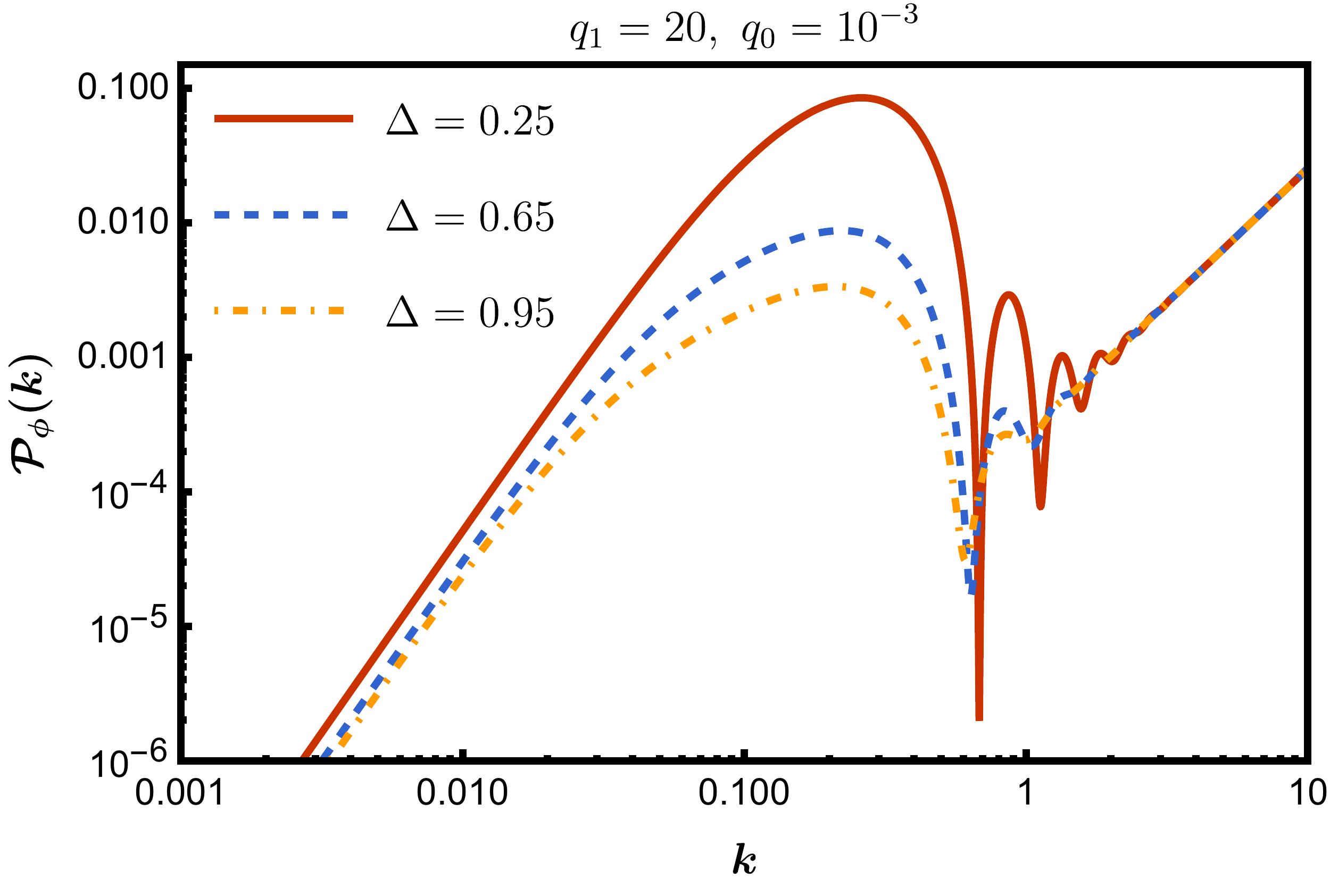}
    \end{minipage}
    \hfill 
    \begin{minipage}{0.48\textwidth}
    \includegraphics[width=\textwidth]{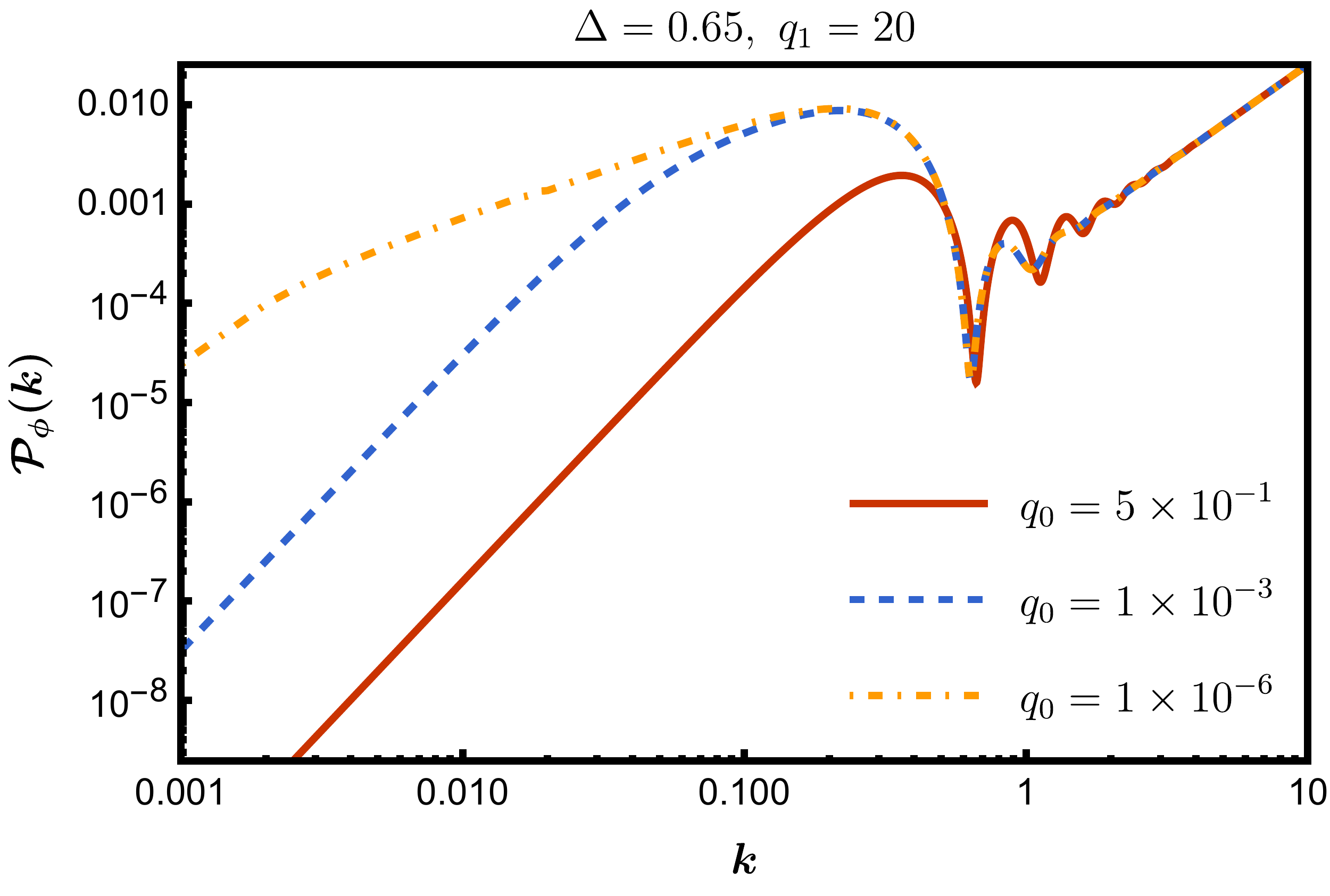}
    \end{minipage}\\
    \centering
    \begin{minipage}{0.48\textwidth}
    \includegraphics[width=\textwidth]{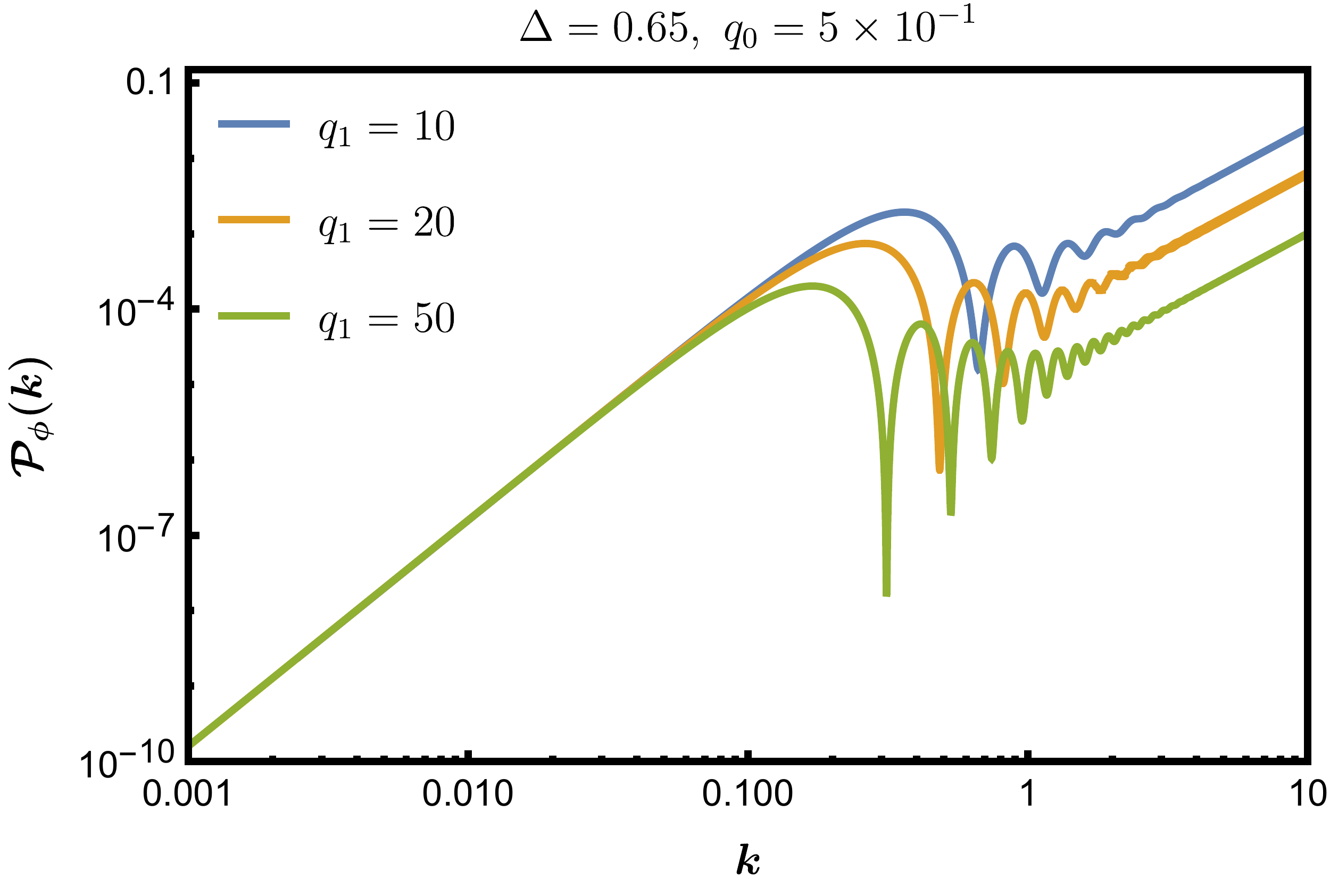}
    \end{minipage}
    \caption{We have depicted the power spectrum of the perturbing scalar field in the bouncing universe for different choices of the torsion parameter (the left-hand plot in the upper row), for various initial size of the universe (the right-hand side plot in the upper row) and lastly for different final sizes of our universe (the plot in the last row). The torsion and the initial size of the saddle point universe affect the power spectrum at large length scales, while the final size of the universe affects the power spectrum at smaller length scales. See text for discussion.}
    \label{fig:power_bounce_02}
\end{figure}

The above features in the bouncing power spectrum have been observed in various quantum models of gravity predicting a bouncing scenario for our universe \cite{PhysRevD.93.023531,PhysRevD.102.126025,PhysRevD.101.086004,ElizagaNavascues:2018bgp}. The power spectrum is divided into three regimes, the infra-red regime, which corresponds to the initial growth, then there is the intermediate regime, where the power spectrum oscillates, but the amplitude remains almost constant, and finally, the ultra-violet regime, where again the power spectrum grows. The growth in the ultra-violet regime is expected, since in the bouncing cosmology, during the phase of bounce the presence of exotic matter can amplify the perturbations. The oscillatory nature of the perturbation is expected too and is a common feature in most of the early universe models. Finally, the infra-red contribution depends on the choice of the vacuum state and other details, e.g., the normalization. We have further elaborated on the variation of the power spectrum with the various parameters of interest, e.g., torsion, and the initial and final size of the saddle point universe. The results of such an analysis have been depicted in \ref{fig:power_bounce_02}. As torsion increases the quantity $\Delta$ decreases, which in turn enhances the oscillatory behavior and the growth in the infra-red sector, while the ultra-violet sector remains identical. This feature we have observed in the context of the inflationary paradigm as well, where also torsion modified the structure at large length scales. Thus we can conclude that irrespective of the early universe physics, the presence of torsion modifies the power spectrum at large length scales, in particular, it enhances the power. The enhancement of power at a large length scale can also be due to the smaller initial size of the universe. As evident from \ref{fig:power_bounce_02}, the smaller the initial size of the universe, the larger is the power at large length scales, however, the ultra-violet behavior of the power spectrum remains unchanged. Finally, we also depict the behavior of the power spectrum with a change in the final size of the universe. This does not affect the infrared behavior but affects the ultraviolet behavior significantly. \ref{fig:power_bounce_02} also demonstrates that a universe with a smaller final size will have larger power at smaller length scales, than a universe with a larger final size. Thus we observe that the power spectrum of the perturbing scalar field in the bouncing scenario is very similar to the previous results in the literature arising from quantum gravity models. In our work, we have explored the matter bounce scenario, with $s=1$, in detail. It will be interesting to explore other possibilities for the parameter $s$ and investigate how the power spectrum is modified. We hope to return to these issues in a future work.

\section{Conclusion and discussions}

The existence of spacetime torsion is a tantalizing possibility and appears in the most natural extension of general relativity, namely in the Einstein-Cartan theory. The presence of spacetime torsion would prohibit the formation of spacetime singularity, by introducing a divergent term in Raychaudhuri's equation. However, it had remained elusive, despite several experimental searches. Classically, due to the absence of macroscopic fermionic currents, the spacetime torsion is not expected to exist (actually, Einstein's equations predict vanishing torsion in the absence of any macroscopic fermionic current). The corresponding statement cannot not be made in the quantum domain, due to the uncertainty relation. Since the gravitational Hamiltonian does not involve the time derivative of the torsional degree of freedom, the momentum conjugate to the torsional degree of freedom must vanish and hence the torsion parameter cannot be set to zero in the quantum domain. Thus the early universe cosmology must have a non-zero value of torsion. Following this argument we have studied the minisuperspace quantum cosmology with a non-zero value of spacetime torsion. Intriguingly, the gravitational Hamiltonian only depends on the completely anti-symmetric part of the torsion tensor, while the other non-trivial contributions from the torsion tensor in the minisuperspace have been absorbed in the momentum conjugate to the scale factor $q(t)$. This also brings out two inequivalent quantization schemes in the presence of torsion. In this work, we have considered the tetrad and the spin-connection components to be more fundamental so that torsion components are derived from these. Thus in our approach, only the completely antisymmetric part of the torsion appears in the Hamiltonian, while another component appears in the momentum $\mathcal{B}$. On the other hand, we could have considered the tetrad and the torsion to be fundamental and the spin connections as derived entities. In that case, however, the wave function would explicitly depend on the scale factor and all the torsion components. Therefore, these two approaches are inequivalent. This inequivalency also arises in the path integral formalism, since in the former one performs the `phase-space' path integral over the Liouville's measure $\mathcal{D}[\mathcal{B}] \mathcal{D}[q]$, while in the latter one reexpress the momentum $\mathcal{B}$ in terms of the scale factor and torsion and then perform a `coordinate-space' path integral over the measure $\mathcal{D}[q]$. In such a case the two torsion degrees of freedom appear explicitly and are treated similarly. Here we have used the phase-space path integral and it would be interesting to perform the coordinate space path integral and determine the wave function to exactly pinpoint the in-equivalences. We hope to return to this in a future work.

In this work, we have explicitly demonstrated the equivalence between the solutions of the Wheeler-DeWitt equation and the semi-classical wave function derived using the path integral framework. This equivalence holds true for both inflationary and bouncing scenarios. In particular, for the inflationary scenario, we have determined the saddle points in the Lorentzian path integral using initial Dirichlet as well as initial Neumann and Robin boundary conditions. It turns out that the saddle point associated with the initial Dirichlet boundary condition is unstable under scalar perturbation, while the saddle points with initial Robin or Neumann boundary conditions are stable. 

For both bouncing and inflationary scenarios, the associated semi-classical wave function depicts two interesting behavior. First of all, it seems possible to answer the question, of why spacetime torsion seems to be small in the universe. It turns out that if one calculates the relative probability between two configurations of the universe with different values for the torsion component, then one finds that the universe with a lower value of torsion has a higher relative probability to occur. Thus it is more natural for a universe to have a smaller value of torsion than a large value. Moreover, it turns out that classicalization happens fastest in an inflationary universe and the rate depends on the volume. On the other hand, for the bouncing scenarios, the classicalization is slower and hence the bouncing scenarios are expected to retain more quantum nature with the evolution of the universe.  

The perturbation analysis, presented in this work, also suggests interesting behavior. For example, in the context of an inflationary paradigm with torsion, for Robin or Neumann boundary conditions \cite{PhysRevD.100.123543,PhysRevD.106.023511}, stable saddle points exist, if and only if the torsion parameter never becomes larger than the spatial curvature index $\mathcal{K}=+1$. Therefore, the effective curvature of space has to be bounded within $0<\mathcal{K}_{c}\leq 1$. This seems to indicate that there is no transition between the positive curvature space to flat or, negative curvature space through quantum dynamics. Similarly, we find, for the bouncing scenarios, the stability of perturbation requires that the energy density of the torsion component does not overwhelm that of the bounce-enabling matter. Thus in both inflation and bounce, the torsion is supposed to contribute by a small amount and hence cannot change the overall nature of the classical geometry. 

We have also answered the following question in this work, how does the presence of torsion affect the power spectrum? For this purpose, we have computed the power spectrum of the perturbing scalar field around the saddle point geometries for both inflationary and bouncing scenarios. From such exercises, we find that for an inflationary universe with a positive spatial curvature index ($\mathcal{K}=+1$), the stronger the value of the torsion, the more is the enhancement of the power spectrum in the large length scales (equivalent, for low co-moving wave number). For the bouncing scenarios, we have computed the power spectrum numerically for the case of a universe with vanishing spatial curvature index ($\mathcal{K}=0$) and filled with two fluids --- ordinary matter (fall-off $\propto a^{-3}$) and an energy condition violating `phantom' radiation (fall-off $\propto a^{-4}$). In this case, which corresponds to the case of matter bounce, at large physical length scales, the power spectrum is suppressed and is characterized by a region of initial growth as the co-moving wavenumber $k$ increases, then an intermediate region of oscillations, and finally a region of growth again. The shape of the power spectrum in the initial and oscillatory regions is sensitive to the difference in the abundance of the bouncing matter and torsion, and the initial quantum state chosen for the universe, while the smaller length scale part of the power spectrum depends crucially on the final size. Note that, we have considered the matter bounce model as a simple example, but the formalism developed in this work, including the numerical treatment of perturbations and thereby obtaining the power spectrum, can be readily extended to the case of other bouncing models as well. Also following the discussion around \ref{prob_torsion}, as our analysis clearly shows, the probability of a universe with a small torsion parameter is indeed large, compared to a universe with a large torsion parameter. This allows us to expect, albeit in a qualitative manner that torsion is going to have very little influence on the structure formation. A more detailed discussion is necessary to make a quantitative statement, which will be presented in a subsequent work. Finally, implications for the allowable complex matrices (as in \cite{PhysRevD.105.026022, Jonas:2022uqb}) with the inclusion of torsion, and also in the context of the bouncing scenario needs to be explored, which we wish to perform in a separate work in the future. 

\section*{acknowledgements}

Research of S.C. is funded by the INSPIRE Faculty fellowship from DST, Government of India (Reg. No. DST/INSPIRE/04/2018/000893). The research of VM is funded by the INSPIRE fellowship from the DST, Government of India (Reg. No. DST/INSPIRE/03/2019/001887). VM acknowledges Karthik Rajeev, Rathul N. Raveendran, and Deep Ghosh for helpful discussions during the preparation of this manuscript.

\appendix
\labelformat{section}{Appendix #1} 
\labelformat{subsection}{Appendix #1}
\section{Derivation of the components of the Spin-connection}\label{AppA}

Given the tetrad one-forms in \ref{tetrad}, we obtain the following relations, 
\begin{align}
\boldsymbol{d}\boldsymbol{e}^{0}&=\frac{N}{q^{p}}\boldsymbol{d}^{2}t+\dfrac{d}{dt}\left(\frac{N}{q^{p}}\right)\boldsymbol{d}t\wedge \boldsymbol{d}t=0~,
\\
\boldsymbol{d}\boldsymbol{e}^{1}&=\frac{q^{s}}{K(r)}\boldsymbol{d}^{2}r
-\frac{q^{s}K'(r)}{K^{2}(r)}\boldsymbol{d}r\wedge \boldsymbol{d}r+\frac{sq^{s-1}\dot{q}}{K(r)}\boldsymbol{d}t\wedge \boldsymbol{d}r
=\frac{sq^{p-1}\dot{q}}{N}\boldsymbol{e}^{0}\wedge \boldsymbol{e}^{1}~,
\\
\boldsymbol{d}\boldsymbol{e}^{2}&=q^{s}r\boldsymbol{d}^{2}\theta
+q^{s}\boldsymbol{d}r\wedge \boldsymbol{d}\theta+sq^{s-1}\dot{q}r\boldsymbol{d}t\wedge \boldsymbol{d}\theta
=\frac{sq^{p-1}\dot{q}}{N}\boldsymbol{e}^{0}\wedge \boldsymbol{e}^{2}+\frac{K(r)}{q^{s}r}\boldsymbol{e}^{1}\wedge \boldsymbol{e}^{2}~,
\\
\boldsymbol{d}\boldsymbol{e}^{3}&=q^{s}r\sin \theta\boldsymbol{d}^{2}\phi
+q^{s}\sin \theta\boldsymbol{d}r\wedge \boldsymbol{d}\phi
+q^{s}r\cos \theta \boldsymbol{d}\theta \wedge \boldsymbol{d}\phi 
+sq^{s-1}\dot{q}r\sin \theta\boldsymbol{d}t\wedge \boldsymbol{d}\phi
\nonumber
\\
&=\frac{sq^{p-1}\dot{q}}{N}\boldsymbol{e}^{0}\wedge \boldsymbol{e}^{3}
+\frac{K(r)}{q^{s}r}\boldsymbol{e}^{1}\wedge \boldsymbol{e}^{3}
+\frac{\cot \theta}{q^{s}r}\boldsymbol{e}^{2}\wedge \boldsymbol{e}^{3}
\end{align}
Thus from the Cartan's structure equation in \ref{cartan_first_structure} and the expression for the torsion tensor in the minisuperspace, as described by \ref{torsion}, we obtain,
\begin{align}
\boldsymbol{\omega}^{0i}\wedge \boldsymbol{e}^{i}&=\boldsymbol{T}^{0}-\boldsymbol{d}\boldsymbol{e}^{0}=0
\label{omega1}
\\
\boldsymbol{\omega}^{0i}\wedge \boldsymbol{e}^{0}+\boldsymbol{\omega}^{ij}\wedge \boldsymbol{e}^{j}&=\boldsymbol{T}^{i}-\boldsymbol{d}\boldsymbol{e}^{i}=\mathcal{T}(t) \boldsymbol{e}^0 \wedge \boldsymbol{e}^i + \mathcal{C}(t) \epsilon^i_{\,\,\, jk} \boldsymbol{e}^j \wedge \boldsymbol{e}^k-\boldsymbol{d}\boldsymbol{e}^{i}
\label{omega2}
\end{align}
where, we have used the result that $\boldsymbol{\omega}^{ab}$ is antisymmetric and the indices are raised by the flat spacetime Minkowski metric. From \ref{omega1}, we obtain, $\boldsymbol{\omega}^{0i}=f(t)\boldsymbol{e}^{i}$, where $f(t)$ is an arbitrary function of the cosmological time $t$, as defined in \ref{minimetric}. This suggests the following expressions for the other components of the spin connection from \ref{omega2}, 
\begin{align}
\boldsymbol{\omega}^{12}\wedge \boldsymbol{e}^{2}+\boldsymbol{\omega}^{13}\wedge \boldsymbol{e}^{3}&=\left[f(t)+\mathcal{T}(t)-\frac{sq^{p-1}\dot{q}}{N}\right]\boldsymbol{e}^{0}\wedge \boldsymbol{e}^{1}+2\mathcal{C}(t) \boldsymbol{e}^2 \wedge \boldsymbol{e}^3
\\
-\boldsymbol{\omega}^{12}\wedge \boldsymbol{e}^{1}+\boldsymbol{\omega}^{23}\wedge \boldsymbol{e}^{3}&=\left[f(t)+\mathcal{T}(t)-\frac{sq^{p-1}\dot{q}}{N}\right]\boldsymbol{e}^{0}\wedge \boldsymbol{e}^{2}+2\mathcal{C}(t) \boldsymbol{e}^3 \wedge \boldsymbol{e}^1-\frac{K(r)}{q^{s}r}\boldsymbol{e}^{1}\wedge \boldsymbol{e}^{2}~,
\\
-\boldsymbol{\omega}^{13}\wedge \boldsymbol{e}^{1}-\boldsymbol{\omega}^{23}\wedge \boldsymbol{e}^{2}&=\left[f(t)+\mathcal{T}(t)-\frac{sq^{p-1}\dot{q}}{N}\right]\boldsymbol{e}^{0}\wedge \boldsymbol{e}^{3}+2\mathcal{C}(t) \boldsymbol{e}^1 \wedge \boldsymbol{e}^2-\frac{K(r)}{q^{s}r}\boldsymbol{e}^{1}\wedge \boldsymbol{e}^{3}-\frac{\cot \theta}{q^{s}r}\boldsymbol{e}^{2}\wedge \boldsymbol{e}^{3}~,
\end{align}
These equations can be satisfied provided the spin connections have the following expressions,
\begin{align}
\boldsymbol{\omega}^{12}&=-\frac{K(r)}{q^{s}r}\boldsymbol{e}^{2}-\mathcal{C}(t) \boldsymbol{e}^3~,
\\
\boldsymbol{\omega}^{13}&=-\frac{K(r)}{q^{s}r}\boldsymbol{e}^{3}+\mathcal{C}(t) \boldsymbol{e}^2~,
\\
\boldsymbol{\omega}^{23}&=-\frac{\cot \theta}{q^{s}r}\boldsymbol{e}^{3}-\mathcal{C}(t) \boldsymbol{e}^1~,
\end{align}
along with the relation, $\mathcal{T}(t)+f(t)=(sq^{p-1}\dot{q}/N)$. These results have been used in the main text. 

\section{Derivation of the curvature components} \label{AppB}

In the previous section we have derived the relevant components of the tetrad and the spin connections in synchronization with the homogeneity and isotropy of the background cosmological spacetime. For our purpose, we need to determine the gravitational action in the presence of torsion in the minisuperspace, and this requires determining the curvature two forms. Owing to the symmetries of the spacetime, determination of only two components of the curvature two-form will suffice for our purpose and these two components are $\boldsymbol{R}^{01}$ and $\boldsymbol{R}^{12}$. Using the structure of the tetrad and the spin-connection derived above, we obtain,
\begin{align} 
\boldsymbol{R}^{01}&=\boldsymbol{ d} \boldsymbol{\omega}^{01} + \boldsymbol{\omega}^0_{\,\,\, 2}\wedge \boldsymbol{\omega}^{21} + \boldsymbol{\omega}^0_{\,\,\, 3}\wedge \boldsymbol{\omega}^{31} \nonumber
\\ 
&=\boldsymbol{d}\left(\mathcal{B} (t) \frac{q^{-2s+1}}{K(r)}\boldsymbol{d}r\right) - \left(q^{-3s+1} \mathcal{B} (t) \boldsymbol{e}^2\right) \wedge \left(-\frac{K(r)}{r} q^{-s} \boldsymbol{e}^2 -  q^{d} c(t) \boldsymbol{e}^3\right) 
\nonumber
\\ 
&\qquad-\left(q^{-3s+1} \mathcal{B} (t) \boldsymbol{e}^3\right) \wedge \left(-\frac{K(r)}{r} q^{-s} \boldsymbol{e}^3 + q^{d} c(t) \boldsymbol{e}^2 \right)  
\nonumber 
\\ 
&=\frac{q^{-2s}}{K(r)}\left(q \dot{\mathcal{B}}(t) + \left(1-2s\right) \dot{q} \mathcal{B}(t) \right) \boldsymbol{d} t \wedge \boldsymbol{d}r + 2 q^{-3s+d+1} \mathcal{B} (t) c(t) \boldsymbol{e}^2 \wedge \boldsymbol{e}^3 
\nonumber 
\\  
&=\frac{q^{-3s+p}}{N}\left(q \dot{\mathcal{B}}(t) + \left(1-2s\right) \dot{q} \mathcal{B}(t) \right) \boldsymbol{e}^0 \wedge \boldsymbol{e}^1 + 2 q^{-3s+d+1} \mathcal{B} (t) c(t) \boldsymbol{e}^2 \wedge \boldsymbol{e}^3~,
\end{align}
and, 
\begin{align} 
\boldsymbol{R}^{12}&=\boldsymbol{d} \boldsymbol{\omega}^{12} + \boldsymbol{\omega}^1_{\,\,\, 0} \wedge \boldsymbol{\omega}^{02} + \boldsymbol{\omega}^1_{\,\,\, 3} \wedge \boldsymbol{\omega}^{32} \nonumber 
\\ 
&=\boldsymbol{d} \left(-{K(r)} \boldsymbol{d} \theta -  q^{s+d} c(t) r \sin \theta \boldsymbol{d}\phi \right) + q^{-6s+2}\mathcal{B}^2(t) \boldsymbol{e}^1 \wedge \boldsymbol{e}^2 \nonumber\\ & - \left(-K(r) \sin\theta \boldsymbol{d}\phi + q^{s+d} c(t) r \boldsymbol{d}\theta \right) \wedge \left(-\cos\theta \boldsymbol{d} \phi - \frac{q^{s+d}}{K(r)} c(t) \boldsymbol{d}r \right)  
\nonumber 
\\
&=\mathcal{K} \frac{r}{K(r)} \boldsymbol{d}r \wedge \boldsymbol{d}\theta - \left(q^{s+d} \dot{c}(t) + \left(s+d\right) q^{s+d-1} \dot{q} c(t)\right) r \sin\theta \boldsymbol{d}t \wedge \boldsymbol{d} \phi + q^{-6s+2} \mathcal{B}^2(t) \boldsymbol{e}^1 \wedge \boldsymbol{e}^2 
\nonumber 
\\ 
&\qquad+\frac{q^{2(s+d)}}{K(r)} c^2(t) r \boldsymbol{d}\theta \wedge \boldsymbol{d} r  
\nonumber
\\
&=q^{-2s} \left(\mathcal{K} + \mathcal{B}^2(t) q^{-4s+2} - c^2(t) q^{2(s+d)} \right)\boldsymbol{e}^1 \wedge \boldsymbol{e}^2 - \frac{q^{p+d}}{N}\left(\dot{c}(t) + \left(s+d\right) \frac{\dot{q}}{q} c(t)\right) \boldsymbol{e}^0 \wedge \boldsymbol{e}^3~. 
\end{align}
The other non-zero components of the curvature two-form can be derived given the above. The gravitational action, on the other hand depends on the curvature two-form and a term dependent on the cosmological constant. Using the above expressions, we obtain these two contributions to the gravitational action to read,
\begin{align} 
&\epsilon_{abcd} \boldsymbol{e}^a \wedge  \boldsymbol{e}^b \wedge \boldsymbol{R}^{cd} 
\nonumber
\\ 
&=2 \epsilon_{ab0i} \boldsymbol{e}^a \wedge \boldsymbol{e}^b \wedge \boldsymbol{R}^{0i} +  \epsilon_{ab i j} \wedge \boldsymbol{e}^a \wedge \boldsymbol{e}^b \wedge \boldsymbol{R}^{i j } 
\nonumber
\\ 
&=2 \epsilon_{0 i j k } \boldsymbol{R}^{0 i} \wedge \boldsymbol{e}^j \wedge \boldsymbol{e}^k +  \epsilon_{a b i j} \boldsymbol{e}^a \wedge \boldsymbol{e}^b \wedge \boldsymbol{R}^{i j} 
\nonumber
\\ 
&=2\epsilon_{0 i j k } \left(\frac{q^{-3s+p}}{N}\left(q \dot{\mathcal{B}}(t) + \left(1-2s\right) \dot{q} \mathcal{B}(t) \right) \boldsymbol{e}^0 \wedge \boldsymbol{e}^i + \epsilon^i_{\,\,\, mn} q^{-3s+d+1} \mathcal{B} (t) c(t) \boldsymbol{e}^m \wedge \boldsymbol{e}^n\right) \wedge \boldsymbol{e}^j \wedge \boldsymbol{e}^k 
\nonumber
\\ 
&\qquad+\epsilon_{a b i j} \boldsymbol{e}^a \wedge \boldsymbol{e}^b \wedge \Big[q^{-2s} \left(\mathcal{K} + \mathcal{B}^2(t) q^{-4s+2} - c^2(t) q^{2(s+d)} \right)\boldsymbol{e}^i \wedge \boldsymbol{e}^j 
\nonumber
\\
&\qquad- \epsilon^{ij}_{\,\,\, k}\frac{q^{p+d}}{N}\left(\dot{c}(t) + \left(s+d\right) \frac{\dot{q}}{q} c(t)\right) \boldsymbol{e}^0 \wedge \boldsymbol{e}^k\Big] 
\nonumber
\\ 
&=2\epsilon_{0 i j k } \left(\frac{q^{-3s+p}}{N}\left(q \dot{\mathcal{B}}(t) + \left(1-2s\right) \dot{q} \mathcal{B}(t) \right) \boldsymbol{e}^0 \wedge \boldsymbol{e}^i\right) \wedge \boldsymbol{e}^j \wedge \boldsymbol{e}^k \nonumber
\\ 
&\qquad+\epsilon_{a b i j} \boldsymbol{e}^a \wedge \boldsymbol{e}^b \wedge \left(q^{-2s} \left(\mathcal{K} + \mathcal{B}^2(t) q^{-4s+2} - c^2(t) q^{2(s+d)} \right)\boldsymbol{e}^i \wedge \boldsymbol{e}^j\right) 
\nonumber
\\
&=2\times 3!\left[\frac{q^{-3s+p}}{N}\left(q \dot{\mathcal{B}}(t) + \left(1-2s\right) \dot{q} \mathcal{B}(t) \right) + q^{-2s} \left(\mathcal{K} + \mathcal{B}^2(t) q^{-4s+2} - c^2(t) q^{2(s+d)} \right) \right]\boldsymbol{e}^0\wedge \boldsymbol{e}^1 \wedge \boldsymbol{e}^2 \wedge \boldsymbol{e}^3 
\nonumber
\\
&=2\times 3! \left[\frac{q^{-3s+p}}{N}\left(q \dot{\mathcal{B}}(t) + \left(1-2s\right) \dot{q} \mathcal{B}(t) \right) + q^{-2s} \left(\mathcal{K} + \mathcal{B}^2(t) q^{-4s+2} - c^2(t) q^{2(s+d)} \right) \right] \frac{q^{3s-p} N r^2 \sin\theta}{\sqrt{1- \mathcal{K} r^2}} {\rm d} t {\rm d} r {\rm d}\theta {\rm d}\phi 
\nonumber 
\\ 
&=2 \times 3! \left[\left(q \dot{\mathcal{B}}(t) + \left(1-2s\right) \dot{q} \mathcal{B}(t) \right) + q^{s-p} N \left(\mathcal{K} + \mathcal{B}^2(t) q^{-4s+2} - c^2(t) q^{2(s+d)} \right) \right] \frac{ r^2 \sin\theta}{\sqrt{1-\mathcal{K} r^2}} {\rm d} t {\rm d} r {\rm d}\theta {\rm d}\phi~,
\end{align}
and,
\begin{align} 
\frac{\Lambda}{6} \epsilon_{abcd} \boldsymbol{e}^a \wedge \boldsymbol{e}^b \wedge \boldsymbol{e}^c \wedge \boldsymbol{e}^d = \frac{4!\Lambda}{6} \frac{q^{3s-p} N r^2 \sin\theta}{\sqrt{1-k r^2}} {\rm d} t {\rm d} r {\rm d}\theta {\rm d}\phi = 4\Lambda \frac{q^{3s-p} N r^2 \sin\theta}{\sqrt{1-k r^2}} {\rm d} t {\rm d} r {\rm d}\theta {\rm d}\phi \end{align}
These are the results we have used in the main text. 

\bibliographystyle{utphys1}
\bibliography{reference}

\providecommand{\href}[2]{#2}\begingroup\raggedright\begin{thebibliography}{10}

\bibitem{Einstein:1916vd}
A.~Einstein, ``{The foundation of the general theory of relativity.},''
  \href{http://dx.doi.org/10.1002/andp.19163540702}{{\em Annalen Phys.}
  {\bfseries 49} no.~7, (1916) 769--822}.

\bibitem{Cartan1923}
E.~Cartan, ``Sur les vari\'et\'es \`a connexion affine et la th\'eorie de la
  relativit\'e g\'en\'eralis\'ee (premi\`ere partie),''
  \href{http://dx.doi.org/10.24033/asens.751}{{\em Annales scientifiques de
  l'\'Ecole Normale Sup\'erieure} {\bfseries 3e s{\'e}rie, 40} (1923)
  325--412}. \url{http://www.numdam.org/articles/10.24033/asens.751/}.

\bibitem{Cartan1925}
E.~Cartan, ``Sur les vari\'et\'es \`a connexion affine, et la th\'eorie de la
  relativit\'e g\'en\'eralis\'ee (deuxi\`eme partie),''
  \href{http://dx.doi.org/10.24033/asens.761}{{\em Annales scientifiques de
  l'\'Ecole Normale Sup\'erieure} {\bfseries 3e s{\'e}rie, 42} (1925) 17--88}.
  \url{http://www.numdam.org/articles/10.24033/asens.761/}.

\bibitem{RevModPhys.48.393}
F.~W. Hehl, P.~von~der Heyde, G.~D. Kerlick, and J.~M. Nester, ``General
  relativity with spin and torsion: Foundations and prospects,''
  \href{http://dx.doi.org/10.1103/RevModPhys.48.393}{{\em Rev. Mod. Phys.}
  {\bfseries 48} (Jul, 1976) 393--416}.
  \url{https://link.aps.org/doi/10.1103/RevModPhys.48.393}.

\bibitem{Chakraborty:2018qew}
S.~Chakraborty and R.~Dey, ``{Noether Current, Black Hole Entropy and Spacetime
  Torsion},'' \href{http://dx.doi.org/10.1016/j.physletb.2018.10.027}{{\em
  Phys. Lett. B} {\bfseries 786} (2018) 432--441},
  \href{http://arxiv.org/abs/1806.05840}{{\ttfamily arXiv:1806.05840 [gr-qc]}}.

\bibitem{Banerjee:2018yyi}
R.~Banerjee, S.~Chakraborty, and P.~Mukherjee, ``{Late-time acceleration driven
  by shift-symmetric Galileon in the presence of torsion},''
  \href{http://dx.doi.org/10.1103/PhysRevD.98.083506}{{\em Phys. Rev. D}
  {\bfseries 98} no.~8, (2018) 083506},
  \href{http://arxiv.org/abs/1802.04150}{{\ttfamily arXiv:1802.04150 [gr-qc]}}.

\bibitem{Ivanov:2016xjm}
A.~N. Ivanov and M.~Wellenzohn, ``{Einstein\textendash{}cartan Gravity with
  Torsion Field Serving as an Origin for the Cosmological Constant or Dark
  Energy Density},'' \href{http://dx.doi.org/10.3847/0004-637X/829/1/47}{{\em
  Astrophys. J.} {\bfseries 829} no.~1, (2016) 47},
  \href{http://arxiv.org/abs/1607.01128}{{\ttfamily arXiv:1607.01128 [gr-qc]}}.

\bibitem{Dey:2017fld}
R.~Dey, S.~Liberati, and D.~Pranzetti, ``{Spacetime thermodynamics in the
  presence of torsion},''
  \href{http://dx.doi.org/10.1103/PhysRevD.96.124032}{{\em Phys. Rev. D}
  {\bfseries 96} no.~12, (2017) 124032},
  \href{http://arxiv.org/abs/1709.04031}{{\ttfamily arXiv:1709.04031 [gr-qc]}}.

\bibitem{Blagojevic:2006jk}
M.~Blagojevic and B.~Cvetkovic, ``{Black hole entropy in 3-D gravity with
  torsion},'' \href{http://dx.doi.org/10.1088/0264-9381/23/14/013}{{\em Class.
  Quant. Grav.} {\bfseries 23} (2006) 4781},
  \href{http://arxiv.org/abs/gr-qc/0601006}{{\ttfamily arXiv:gr-qc/0601006}}.

\bibitem{Barnich:2016rwk}
G.~Barnich, P.~Mao, and R.~Ruzziconi, ``{Conserved currents in the Cartan
  formulation of general relativity},'' in {\em {About Various Kinds of
  Interactions}: {Workshop in honour of Professor Philippe Spindel}}.
\newblock 11, 2016.
\newblock \href{http://arxiv.org/abs/1611.01777}{{\ttfamily arXiv:1611.01777
  [gr-qc]}}.

\bibitem{Hehl:1974cn}
F.~W. Hehl, G.~D. Kerlick, and P.~Von Der~Heyde, ``{General relativity with
  spin and torsion and its deviations from einstein's theory},''
  \href{http://dx.doi.org/10.1103/PhysRevD.10.1066}{{\em Phys. Rev. D}
  {\bfseries 10} (1974) 1066--1069}.

\bibitem{Shapiro:2001rz}
I.~L. Shapiro, ``{Physical aspects of the space-time torsion},''
  \href{http://dx.doi.org/10.1016/S0370-1573(01)00030-8}{{\em Phys. Rept.}
  {\bfseries 357} (2002) 113},
  \href{http://arxiv.org/abs/hep-th/0103093}{{\ttfamily arXiv:hep-th/0103093}}.

\bibitem{Shie:2008ms}
K.-F. Shie, J.~M. Nester, and H.-J. Yo, ``{Torsion Cosmology and the
  Accelerating Universe},''
  \href{http://dx.doi.org/10.1103/PhysRevD.78.023522}{{\em Phys. Rev. D}
  {\bfseries 78} (2008) 023522},
  \href{http://arxiv.org/abs/0805.3834}{{\ttfamily arXiv:0805.3834 [gr-qc]}}.

\bibitem{Poplawski:2010kb}
N.~J. Pop\l{}awski, ``{Cosmology with torsion: An alternative to cosmic
  inflation},'' \href{http://dx.doi.org/10.1016/j.physletb.2010.09.056}{{\em
  Phys. Lett. B} {\bfseries 694} (2010) 181--185},
  \href{http://arxiv.org/abs/1007.0587}{{\ttfamily arXiv:1007.0587
  [astro-ph.CO]}}. [Erratum: Phys.Lett.B 701, 672--672 (2011)].

\bibitem{PhysRevD.103.104008}
J.~Magueijo and T.~Zlosnik, ``Quantum torsion and a {H}artle-{H}awking beam,''
  \href{http://dx.doi.org/10.1103/PhysRevD.103.104008}{{\em Phys. Rev. D}
  {\bfseries 103} (May, 2021) 104008}.
  \url{https://link.aps.org/doi/10.1103/PhysRevD.103.104008}.

\bibitem{PhysRevD.104.026002}
J.~Magueijo, ``Real {C}hern-{S}imons wave function,''
  \href{http://dx.doi.org/10.1103/PhysRevD.104.026002}{{\em Phys. Rev. D}
  {\bfseries 104} (Jul, 2021) 026002}.
  \url{https://link.aps.org/doi/10.1103/PhysRevD.104.026002}.

\bibitem{Alexander_2021}
S.~Alexander, G.~Herczeg, and J.~Magueijo, ``A generalized
  {H}artle{\textendash}{H}awking wave function,''
  \href{http://dx.doi.org/10.1088/1361-6382/abf2f6}{{\em Classical and Quantum
  Gravity} {\bfseries 38} no.~9, (Apr, 2021) 095011}.
  \url{https://doi.org/10.1088%2F1361-6382%2Fabf2f6}.

\bibitem{Albertini:2022yny}
E.~Albertini, S.~Alexander, G.~Herczeg, and J.~Magueijo, ``{Torsion and the
  probability of inflation},''
  \href{http://dx.doi.org/10.1088/1475-7516/2022/11/036}{{\em JCAP} {\bfseries
  11} (2022) 036}, \href{http://arxiv.org/abs/2203.12640}{{\ttfamily
  arXiv:2203.12640 [gr-qc]}}.

\bibitem{Isichei:2022uzl}
R.~Isichei and J.~Magueijo, ``{Minisuperspace quantum cosmology from the
  Einstein-Cartan path integral},''
  \href{http://dx.doi.org/10.1103/PhysRevD.107.023526}{{\em Phys. Rev. D}
  {\bfseries 107} no.~2, (2023) 023526},
  \href{http://arxiv.org/abs/2210.05583}{{\ttfamily arXiv:2210.05583
  [hep-th]}}.

\bibitem{Gielen:2022yez}
S.~Gielen and E.~Nash, ``{Quantum cosmology of pure connection general
  relativity},'' \href{http://arxiv.org/abs/2212.06198}{{\ttfamily
  arXiv:2212.06198 [gr-qc]}}.

\bibitem{martini2023covariant}
R.~Martini, G.~Paci, and D.~Sauro, ``Covariant spin-parity decomposition of the
  torsion and path integrals,'' 2023.

\bibitem{Freedman:2012zz}
D.~Z. Freedman and A.~Van~Proeyen, {\em {Supergravity}}.
\newblock Cambridge Univ. Press, Cambridge, UK, 5, 2012.

\bibitem{Krasnov:2020lku}
K.~Krasnov, \href{http://dx.doi.org/10.1017/9781108674652}{{\em {Formulations
  of General Relativity}}}.
\newblock Cambridge Monographs on Mathematical Physics. Cambridge University
  Press, 11, 2020.

\bibitem{Feng:2021lfa}
J.~C. Feng and S.~Chakraborty, ``{Weiss variation for general boundaries},''
  \href{http://dx.doi.org/10.1007/s10714-022-02953-0}{{\em Gen. Rel. Grav.}
  {\bfseries 54} no.~7, (2022) 67},
  \href{http://arxiv.org/abs/2111.06897}{{\ttfamily arXiv:2111.06897 [gr-qc]}}.

\bibitem{Chakraborty:2020yag}
S.~Chakraborty and T.~Padmanabhan, ``{Eddington gravity with matter: An
  emergent perspective},''
  \href{http://dx.doi.org/10.1103/PhysRevD.103.064033}{{\em Phys. Rev. D}
  {\bfseries 103} no.~6, (2021) 064033},
  \href{http://arxiv.org/abs/2012.08542}{{\ttfamily arXiv:2012.08542 [gr-qc]}}.

\bibitem{PhysRevD.28.2960}
J.~B. Hartle and S.~W. Hawking, ``Wave function of the universe,''
  \href{http://dx.doi.org/10.1103/PhysRevD.28.2960}{{\em Phys. Rev. D}
  {\bfseries 28} (Dec, 1983) 2960--2975}.
  \url{https://link.aps.org/doi/10.1103/PhysRevD.28.2960}.

\bibitem{PhysRevD.95.103508}
J.~Feldbrugge, J.-L. Lehners, and N.~Turok, ``Lorentzian quantum cosmology,''
  \href{http://dx.doi.org/10.1103/PhysRevD.95.103508}{{\em Phys. Rev. D}
  {\bfseries 95} (May, 2017) 103508}.
  \url{https://link.aps.org/doi/10.1103/PhysRevD.95.103508}.

\bibitem{PhysRevLett.119.171301}
J.~Feldbrugge, J.-L. Lehners, and N.~Turok, ``No smooth beginning for
  spacetime,'' \href{http://dx.doi.org/10.1103/PhysRevLett.119.171301}{{\em
  Phys. Rev. Lett.} {\bfseries 119} (Oct, 2017) 171301}.
  \url{https://link.aps.org/doi/10.1103/PhysRevLett.119.171301}.

\bibitem{PhysRevD.97.023509}
J.~Feldbrugge, J.-L. Lehners, and N.~Turok, ``No rescue for the no boundary
  proposal: Pointers to the future of quantum cosmology,''
  \href{http://dx.doi.org/10.1103/PhysRevD.97.023509}{{\em Phys. Rev. D}
  {\bfseries 97} (Jan, 2018) 023509}.
  \url{https://link.aps.org/doi/10.1103/PhysRevD.97.023509}.

\bibitem{PhysRevD.100.063517}
A.~Di~Tucci, J.~Feldbrugge, J.-L. Lehners, and N.~Turok, ``Quantum
  incompleteness of inflation,''
  \href{http://dx.doi.org/10.1103/PhysRevD.100.063517}{{\em Phys. Rev. D}
  {\bfseries 100} (Sep, 2019) 063517}.
  \url{https://link.aps.org/doi/10.1103/PhysRevD.100.063517}.

\bibitem{PhysRevLett.122.201302}
A.~Di~Tucci and J.-L. Lehners, ``No-boundary proposal as a path integral with
  robin boundary conditions,''
  \href{http://dx.doi.org/10.1103/PhysRevLett.122.201302}{{\em Phys. Rev.
  Lett.} {\bfseries 122} (May, 2019) 201302}.
  \url{https://link.aps.org/doi/10.1103/PhysRevLett.122.201302}.

\bibitem{PhysRevD.100.123543}
A.~Di~Tucci, J.-L. Lehners, and L.~Sberna, ``No-boundary prescriptions in
  lorentzian quantum cosmology,''
  \href{http://dx.doi.org/10.1103/PhysRevD.100.123543}{{\em Phys. Rev. D}
  {\bfseries 100} (Dec, 2019) 123543}.
  \url{https://link.aps.org/doi/10.1103/PhysRevD.100.123543}.

\bibitem{PhysRevD.106.023511}
K.~Rajeev, ``Wave function of the universe as a sum over eventually inflating
  universes,'' \href{http://dx.doi.org/10.1103/PhysRevD.106.023511}{{\em Phys.
  Rev. D} {\bfseries 106} (Jul, 2022) 023511}.
  \url{https://link.aps.org/doi/10.1103/PhysRevD.106.023511}.

\bibitem{Lehners:2023yrj}
J.-L. Lehners, ``{Review of the No-Boundary Wave Function},''
  \href{http://arxiv.org/abs/2303.08802}{{\ttfamily arXiv:2303.08802
  [hep-th]}}.

\bibitem{PhysRevD.103.106008}
K.~Rajeev, V.~Mondal, and S.~Chakraborty, ``No-boundary wave function,
  wheeler-dewitt equation, and path integral analysis of the bouncing quantum
  cosmology,'' \href{http://dx.doi.org/10.1103/PhysRevD.103.106008}{{\em Phys.
  Rev. D} {\bfseries 103} (May, 2021) 106008}.
  \url{https://link.aps.org/doi/10.1103/PhysRevD.103.106008}.

\bibitem{Narain_2021}
G.~Narain, ``On gauss-bonnet gravity and boundary conditions in lorentzian
  path-integral quantization,''
  \href{http://dx.doi.org/10.1007/jhep05(2021)273}{{\em Journal of High Energy
  Physics} {\bfseries 2021} no.~5, (May, 2021) }.
  \url{http://dx.doi.org/10.1007/JHEP05(2021)273}.

\bibitem{Narain_2022}
G.~Narain, ``Surprises in lorentzian path-integral of gauss-bonnet gravity,''
  \href{http://dx.doi.org/10.1007/jhep04(2022)153}{{\em Journal of High Energy
  Physics} {\bfseries 2022} no.~4, (Apr., 2022) }.
  \url{http://dx.doi.org/10.1007/JHEP04(2022)153}.

\bibitem{ailiga2023lorentzian}
M.~Ailiga, S.~Mallik, and G.~Narain, ``Lorentzian robin universe,'' 2023.

\bibitem{Brandenberger:2016vhg}
R.~Brandenberger and P.~Peter, ``{Bouncing Cosmologies: Progress and
  Problems},'' \href{http://dx.doi.org/10.1007/s10701-016-0057-0}{{\em Found.
  Phys.} {\bfseries 47} no.~6, (2017) 797--850},
  \href{http://arxiv.org/abs/1603.05834}{{\ttfamily arXiv:1603.05834
  [hep-th]}}.

\bibitem{Raveendran:2018why}
R.~N. Raveendran and L.~Sriramkumar, ``{Viable scalar spectral tilt and
  tensor-to-scalar ratio in near-matter bounces},''
  \href{http://dx.doi.org/10.1103/PhysRevD.100.083523}{{\em Phys. Rev. D}
  {\bfseries 100} no.~8, (2019) 083523},
  \href{http://arxiv.org/abs/1812.06803}{{\ttfamily arXiv:1812.06803
  [astro-ph.CO]}}.

\bibitem{Raveendran:2017vfx}
R.~N. Raveendran, D.~Chowdhury, and L.~Sriramkumar, ``{Viable tensor-to-scalar
  ratio in a symmetric matter bounce},''
  \href{http://dx.doi.org/10.1088/1475-7516/2018/01/030}{{\em JCAP} {\bfseries
  01} (2018) 030}, \href{http://arxiv.org/abs/1703.10061}{{\ttfamily
  arXiv:1703.10061 [gr-qc]}}.

\bibitem{PhysRevD.100.083523}
R.~N. Raveendran and L.~Sriramkumar, ``Viable scalar spectral tilt and
  tensor-to-scalar ratio in near-matter bounces,''
  \href{http://dx.doi.org/10.1103/PhysRevD.100.083523}{{\em Phys. Rev. D}
  {\bfseries 100} (Oct, 2019) 083523}.
  \url{https://link.aps.org/doi/10.1103/PhysRevD.100.083523}.

\bibitem{Will:1993ns}
C.~M. Will, {\em {Theory and experiment in gravitational physics}}.
\newblock {C}ambridge {U}niversity {P}ress, 1993.

\bibitem{Bohmer:2017dqs}
C.~G. B\"ohmer, P.~Burikham, T.~Harko, and M.~J. Lake, ``{Does space-time
  torsion determine the minimum mass of gravitating particles?},''
  \href{http://dx.doi.org/10.1140/epjc/s10052-018-5719-y}{{\em Eur. Phys. J. C}
  {\bfseries 78} no.~3, (2018) 253},
  \href{http://arxiv.org/abs/1709.07749}{{\ttfamily arXiv:1709.07749 [gr-qc]}}.

\bibitem{PhysRevLett.89.121101}
B.~Mukhopadhyaya, S.~Sen, and S.~SenGupta, ``Does a randall-sundrum scenario
  create the illusion of a torsion-free universe?,''
  \href{http://dx.doi.org/10.1103/PhysRevLett.89.121101}{{\em Phys. Rev. Lett.}
  {\bfseries 89} (Aug, 2002) 121101}.
  \url{https://link.aps.org/doi/10.1103/PhysRevLett.89.121101}.

\bibitem{PhysRevD.90.107901}
A.~Das, B.~Mukhopadhyaya, and S.~SenGupta, ``Why has spacetime torsion such
  negligible effect on the universe?,''
  \href{http://dx.doi.org/10.1103/PhysRevD.90.107901}{{\em Phys. Rev. D}
  {\bfseries 90} (Nov, 2014) 107901}.
  \url{https://link.aps.org/doi/10.1103/PhysRevD.90.107901}.

\bibitem{TSAMPARLIS197927}
M.~Tsamparlis, ``Cosmological principle and torsion,''
  \href{http://dx.doi.org/https://doi.org/10.1016/0375-9601(79)90265-2}{{\em
  Physics Letters A} {\bfseries 75} no.~1, (1979) 27--28}.
  \url{https://www.sciencedirect.com/science/article/pii/0375960179902652}.

\bibitem{PhysRevD.39.2206}
J.~J. Halliwell and J.~Louko, ``Steepest-descent contours in the path-integral
  approach to quantum cosmology. i. the de sitter minisuperspace model,''
  \href{http://dx.doi.org/10.1103/PhysRevD.39.2206}{{\em Phys. Rev. D}
  {\bfseries 39} (Apr, 1989) 2206--2215}.
  \url{https://link.aps.org/doi/10.1103/PhysRevD.39.2206}.

\bibitem{Rajeev:2021yyl}
K.~Rajeev, V.~Mondal, and S.~Chakraborty, ``{Bouncing with shear: implications
  from quantum cosmology},''
  \href{http://dx.doi.org/10.1088/1475-7516/2022/01/008}{{\em JCAP} {\bfseries
  01} no.~01, (2022) 008}, \href{http://arxiv.org/abs/2109.08696}{{\ttfamily
  arXiv:2109.08696 [gr-qc]}}.

\bibitem{Jia:2022nda}
D.~Jia, ``{Truly Lorentzian quantum cosmology},''
  \href{http://arxiv.org/abs/2211.00517}{{\ttfamily arXiv:2211.00517 [gr-qc]}}.

\bibitem{Lehners:2018eeo}
J.-L. Lehners, ``{No smooth beginning for spacetime},''
  \href{http://dx.doi.org/10.1142/S0218271819300052}{{\em Int. J. Mod. Phys. D}
  {\bfseries 28} no.~02, (2018) 1930005}.

\bibitem{PhysRevD.98.066003}
A.~Vilenkin and M.~Yamada, ``Tunneling wave function of the universe,''
  \href{http://dx.doi.org/10.1103/PhysRevD.98.066003}{{\em Phys. Rev. D}
  {\bfseries 98} (Sep, 2018) 066003}.
  \url{https://link.aps.org/doi/10.1103/PhysRevD.98.066003}.

\bibitem{Feldbrugge:2018gin}
J.~Feldbrugge, J.-L. Lehners, and N.~Turok, ``{Inconsistencies of the New
  No-Boundary Proposal},''
  \href{http://dx.doi.org/10.3390/universe4100100}{{\em Universe} {\bfseries 4}
  no.~10, (2018) 100}, \href{http://arxiv.org/abs/1805.01609}{{\ttfamily
  arXiv:1805.01609 [hep-th]}}.

\bibitem{PhysRevD.99.066010}
A.~Vilenkin and M.~Yamada, ``Tunneling wave function of the universe. ii. the
  backreaction problem,''
  \href{http://dx.doi.org/10.1103/PhysRevD.99.066010}{{\em Phys. Rev. D}
  {\bfseries 99} (Mar, 2019) 066010}.
  \url{https://link.aps.org/doi/10.1103/PhysRevD.99.066010}.

\bibitem{PhysRevD.99.043526}
J.~J. Halliwell, J.~B. Hartle, and T.~Hertog, ``What is the no-boundary wave
  function of the universe?,''
  \href{http://dx.doi.org/10.1103/PhysRevD.99.043526}{{\em Phys. Rev. D}
  {\bfseries 99} (Feb, 2019) 043526}.
  \url{https://link.aps.org/doi/10.1103/PhysRevD.99.043526}.

\bibitem{PhysRevD.100.043544}
S.~P. de~Alwis, ``Wave function of the universe and cmb fluctuations,''
  \href{http://dx.doi.org/10.1103/PhysRevD.100.043544}{{\em Phys. Rev. D}
  {\bfseries 100} (Aug, 2019) 043544}.
  \url{https://link.aps.org/doi/10.1103/PhysRevD.100.043544}.

\bibitem{Barrow:2020coo}
J.~D. Barrow and J.~Magueijo, ``{A contextual Planck parameter and the
  classical limit in quantum cosmology},''
  \href{http://dx.doi.org/10.1007/s10701-021-00433-0}{{\em Found. Phys.}
  {\bfseries 51} no.~1, (2021) 22},
  \href{http://arxiv.org/abs/2006.16036}{{\ttfamily arXiv:2006.16036 [gr-qc]}}.

\bibitem{PhysRevD.102.044034}
J.~a. Magueijo, ``Equivalence of the chern-simons state and the hartle-hawking
  and vilenkin wave functions,''
  \href{http://dx.doi.org/10.1103/PhysRevD.102.044034}{{\em Phys. Rev. D}
  {\bfseries 102} (Aug, 2020) 044034}.
  \url{https://link.aps.org/doi/10.1103/PhysRevD.102.044034}.

\bibitem{PhysRevD.31.1777}
J.~J. Halliwell and S.~W. Hawking, ``Origin of structure in the universe,''
  \href{http://dx.doi.org/10.1103/PhysRevD.31.1777}{{\em Phys. Rev. D}
  {\bfseries 31} (Apr, 1985) 1777--1791}.
  \url{https://link.aps.org/doi/10.1103/PhysRevD.31.1777}.

\bibitem{PhysRevD.37.888}
A.~Vilenkin, ``Quantum cosmology and the initial state of the universe,''
  \href{http://dx.doi.org/10.1103/PhysRevD.37.888}{{\em Phys. Rev. D}
  {\bfseries 37} (Feb, 1988) 888--897}.
  \url{https://link.aps.org/doi/10.1103/PhysRevD.37.888}.

\bibitem{PhysRevD.25.3159}
C.~Teitelboim, ``Quantum mechanics of the gravitational field,''
  \href{http://dx.doi.org/10.1103/PhysRevD.25.3159}{{\em Phys. Rev. D}
  {\bfseries 25} (Jun, 1982) 3159--3179}.
  \url{https://link.aps.org/doi/10.1103/PhysRevD.25.3159}.

\bibitem{PhysRevD.38.2468}
J.~J. Halliwell, ``Derivation of the wheeler-dewitt equation from a path
  integral for minisuperspace models,''
  \href{http://dx.doi.org/10.1103/PhysRevD.38.2468}{{\em Phys. Rev. D}
  {\bfseries 38} (Oct, 1988) 2468--2481}.
  \url{https://link.aps.org/doi/10.1103/PhysRevD.38.2468}.

\bibitem{PhysRevD.102.086011}
A.~Di~Tucci, M.~P. Heller, and J.-L. Lehners, ``Lessons for quantum cosmology
  from anti--de sitter black holes,''
  \href{http://dx.doi.org/10.1103/PhysRevD.102.086011}{{\em Phys. Rev. D}
  {\bfseries 102} (Oct, 2020) 086011}.
  \url{https://link.aps.org/doi/10.1103/PhysRevD.102.086011}.

\bibitem{PhysRevD.39.1116}
A.~Vilenkin, ``Interpretation of the wave function of the universe,''
  \href{http://dx.doi.org/10.1103/PhysRevD.39.1116}{{\em Phys. Rev. D}
  {\bfseries 39} (Feb, 1989) 1116--1122}.
  \url{https://link.aps.org/doi/10.1103/PhysRevD.39.1116}.

\bibitem{Page:2006hr}
D.~N. Page, ``{Susskind's challenge to the Hartle-Hawking no-boundary proposal
  and possible resolutions},''
  \href{http://dx.doi.org/10.1088/1475-7516/2007/01/004}{{\em JCAP} {\bfseries
  01} (2007) 004}, \href{http://arxiv.org/abs/hep-th/0610199}{{\ttfamily
  arXiv:hep-th/0610199}}.

\bibitem{PhysRevLett.100.201301}
J.~B. Hartle, S.~W. Hawking, and T.~Hertog, ``No-boundary measure of the
  universe,'' \href{http://dx.doi.org/10.1103/PhysRevLett.100.201301}{{\em
  Phys. Rev. Lett.} {\bfseries 100} (May, 2008) 201301}.
  \url{https://link.aps.org/doi/10.1103/PhysRevLett.100.201301}.

\bibitem{PhysRevD.80.124032}
J.~J. Halliwell, ``Probabilities in quantum cosmological models: A decoherent
  histories analysis using a complex potential,''
  \href{http://dx.doi.org/10.1103/PhysRevD.80.124032}{{\em Phys. Rev. D}
  {\bfseries 80} (Dec, 2009) 124032}.
  \url{https://link.aps.org/doi/10.1103/PhysRevD.80.124032}.

\bibitem{PhysRevD.99.123531}
O.~Janssen, J.~J. Halliwell, and T.~Hertog, ``No-boundary proposal in biaxial
  bianchi ix minisuperspace,''
  \href{http://dx.doi.org/10.1103/PhysRevD.99.123531}{{\em Phys. Rev. D}
  {\bfseries 99} (Jun, 2019) 123531}.
  \url{https://link.aps.org/doi/10.1103/PhysRevD.99.123531}.

\bibitem{PhysRevD.77.123537}
J.~B. Hartle, S.~W. Hawking, and T.~Hertog, ``Classical universes of the
  no-boundary quantum state,''
  \href{http://dx.doi.org/10.1103/PhysRevD.77.123537}{{\em Phys. Rev. D}
  {\bfseries 77} (Jun, 2008) 123537}.
  \url{https://link.aps.org/doi/10.1103/PhysRevD.77.123537}.

\bibitem{PhysRevD.91.083525}
J.-L. Lehners, ``Classical inflationary and ekpyrotic universes in the
  no-boundary wavefunction,''
  \href{http://dx.doi.org/10.1103/PhysRevD.91.083525}{{\em Phys. Rev. D}
  {\bfseries 91} (Apr, 2015) 083525}.
  \url{https://link.aps.org/doi/10.1103/PhysRevD.91.083525}.

\bibitem{Kiefer:2021iko}
C.~Kiefer and T.~Vardanyan, ``{Power spectrum for perturbations in an
  inflationary model for a closed universe},''
  \href{http://dx.doi.org/10.1007/s10714-022-02918-3}{{\em Gen. Rel. Grav.}
  {\bfseries 54} no.~4, (2022) 30},
  \href{http://arxiv.org/abs/2111.07835}{{\ttfamily arXiv:2111.07835 [gr-qc]}}.

\bibitem{PhysRevD.86.103524}
J.~Martin, V.~Vennin, and P.~Peter, ``Cosmological inflation and the quantum
  measurement problem,''
  \href{http://dx.doi.org/10.1103/PhysRevD.86.103524}{{\em Phys. Rev. D}
  {\bfseries 86} (Nov, 2012) 103524}.
  \url{https://link.aps.org/doi/10.1103/PhysRevD.86.103524}.

\bibitem{Chen:2017aes}
P.~Chen, Y.-H. Lin, and D.-h. Yeom, ``{Suppression of long-wavelength CMB
  spectrum from the no-boundary initial condition},''
  \href{http://dx.doi.org/10.1140/epjc/s10052-018-6426-4}{{\em Eur. Phys. J. C}
  {\bfseries 78} no.~11, (2018) 930},
  \href{http://arxiv.org/abs/1707.01471}{{\ttfamily arXiv:1707.01471 [gr-qc]}}.

\bibitem{Chen:2019mbu}
P.~Chen, H.-H. Yeh, and D.-H. Yeom, ``{Suppression of the long-wavelength CMB
  spectrum from the Hartle\textendash{}Hawking wave function in the
  Starobinsky-type inflation model},''
  \href{http://dx.doi.org/10.1016/j.dark.2019.100435}{{\em Phys. Dark Univ.}
  {\bfseries 27} (2020) 100435},
  \href{http://arxiv.org/abs/1903.12045}{{\ttfamily arXiv:1903.12045 [gr-qc]}}.

\bibitem{PhysRevD.93.023531}
S.~Schander, A.~Barrau, B.~Bolliet, L.~Linsefors, J.~Mielczarek, and J.~Grain,
  ``Primordial scalar power spectrum from the euclidean big bounce,''
  \href{http://dx.doi.org/10.1103/PhysRevD.93.023531}{{\em Phys. Rev. D}
  {\bfseries 93} (Jan, 2016) 023531}.
  \url{https://link.aps.org/doi/10.1103/PhysRevD.93.023531}.

\bibitem{PhysRevD.102.126025}
B.-F. Li, J.~Olmedo, P.~Singh, and A.~Wang, ``Primordial scalar power spectrum
  from the hybrid approach in loop cosmologies,''
  \href{http://dx.doi.org/10.1103/PhysRevD.102.126025}{{\em Phys. Rev. D}
  {\bfseries 102} (Dec, 2020) 126025}.
  \url{https://link.aps.org/doi/10.1103/PhysRevD.102.126025}.

\bibitem{PhysRevD.101.086004}
B.-F. Li, P.~Singh, and A.~Wang, ``Primordial power spectrum from the dressed
  metric approach in loop cosmologies,''
  \href{http://dx.doi.org/10.1103/PhysRevD.101.086004}{{\em Phys. Rev. D}
  {\bfseries 101} (Apr, 2020) 086004}.
  \url{https://link.aps.org/doi/10.1103/PhysRevD.101.086004}.

\bibitem{ElizagaNavascues:2018bgp}
B.~Elizaga~Navascu\'es, D.~M. de~Blas, and G.~A.~M. Marug\'an, ``{The Vacuum
  State of Primordial Fluctuations in Hybrid Loop Quantum Cosmology},''
  \href{http://dx.doi.org/10.3390/universe4100098}{{\em Universe} {\bfseries 4}
  no.~10, (2018) 98}, \href{http://arxiv.org/abs/1809.09874}{{\ttfamily
  arXiv:1809.09874 [gr-qc]}}.

\bibitem{PhysRevD.105.026022}
J.-L. Lehners, ``Allowable complex metrics in minisuperspace quantum
  cosmology,'' \href{http://dx.doi.org/10.1103/PhysRevD.105.026022}{{\em Phys.
  Rev. D} {\bfseries 105} (Jan, 2022) 026022}.
  \url{https://link.aps.org/doi/10.1103/PhysRevD.105.026022}.

\bibitem{Jonas:2022uqb}
C.~Jonas, J.-L. Lehners, and J.~Quintin, ``{Uses of complex metrics in
  cosmology},'' \href{http://dx.doi.org/10.1007/JHEP08(2022)284}{{\em JHEP}
  {\bfseries 08} (2022) 284}, \href{http://arxiv.org/abs/2205.15332}{{\ttfamily
  arXiv:2205.15332 [hep-th]}}.

\end{thebibliography}\endgroup
\end{document}